\documentclass[aps,prd,onecolumn,amsmath,11pt,superscriptaddress,floatfix,nofootinbib,preprintnumbers]{revtex4-2}

\usepackage{mathrsfs}
\usepackage{amsfonts}
\usepackage{amsmath,amssymb,bm}
\usepackage{array}
\usepackage{verbatim}
\usepackage{epsfig}
\usepackage{graphicx}
\usepackage{hyperref}
\hypersetup{colorlinks, linkcolor = [rgb]{0, 0, 0.5}, citecolor = [rgb]{0,0.0,0.5}, urlcolor = [rgb]{0,0.0,0.5}}
\usepackage[normalem]{ulem}
\usepackage{xcolor}
\usepackage{ulem}
\usepackage{multirow}

\usepackage{graphicx}
\graphicspath{{./figs/}}
\usepackage{dcolumn}
\usepackage{bm}
\usepackage{slashed}
\usepackage{siunitx}
\usepackage{ulem,xpatch}
\usepackage{hyperref}
\usepackage[mathlines]{lineno}
\usepackage{comment}
\usepackage{soul}
\usepackage{xcolor}
\usepackage{xfrac}

\allowdisplaybreaks[4]

\begin{document}

\title{Polarization analysis in $e^+e^-\rightarrow J/\psi\rightarrow$ baryon-antibaryon pairs}

\newcommand*{\SCNT}{Southern Center for Nuclear-Science Theory (SCNT), Institute of Modern Physics, Chinese Academy of Sciences, Huizhou 516000, China}\affiliation{\SCNT}
\newcommand*{\SDU}{Key Laboratory of Particle Physics and Particle Irradiation (MOE), Institute of Frontier and Interdisciplinary Science, Shandong University, Qingdao, Shandong 266237, China}\affiliation{\SDU}
\newcommand*{\ihep}{Institute of High Energy Physics, Chinese Academy of Sciences, Beijing 100049, China}\affiliation{\ihep}
\newcommand*{\ucas}{University of Chinese Academy of Sciences, Beijing 100049, China}\affiliation{\ucas}
\newcommand*{\HTU}{School of Physics, He'nan Normal University, Xinxiang, Henan 453007, China}\affiliation{\HTU}
\newcommand*{\ytu}{College of Nuclear Equipment and Nuclear Engineering, Yantai University, Yantai, Shandong 264005, China}\affiliation{\ytu}

\author{Zhe Zhang}\email{zhangzhe@impcas.ac.cn}\affiliation{\SCNT}
\author{Tianbo Liu}\email{liutb@sdu.edu.cn}\affiliation{\SDU}\affiliation{\SCNT}
\author{Rong-Gang Ping}\email{pingrg@mail.ihep.ac.cn}\affiliation{\ihep}\affiliation{\ucas}
\author{Jiao Jiao Song}\email{songjiaojiao@ihep.ac.cn}\affiliation{\HTU}
\author{Weihua Yang}\email{yangwh@ytu.edu.cn}\affiliation{\ytu}

\begin{abstract}

We present a comprehensive polarization analysis for the process $e^{+}e^{-} \rightarrow J/\psi \rightarrow$ baryon-antibaryon pairs~$B_{1}\bar{B}_{2}$. All possible helicity amplitudes for $e^{+}e^{-} \rightarrow J/\psi$ and $J/\psi \rightarrow B_{1} \bar{B}_{2}$ are provided, along with their transformation properties under parity and $CP$ operations. Taking full account of beam polarization, we analyze the polarization of the $J/\psi$ produced in $e^{+}e^{-}$ annihilation and show that all eight independent polarization components can be nonzero. By introducing a polarization transfer matrix, we propose a novel method to compute the polarization transfer in the decay $J/\psi \rightarrow B_{1} \bar{B}_{2}$. Considering both hadronic and radiative baryon decay models, we derive the joint angular distribution of the final-state particles. Focusing on the process $J/\psi \rightarrow \Lambda \bar{\Sigma}^{0}$, we provide statistical uncertainty estimates for probing $CP$ violation in associated production of different baryons. This work establishes a complete framework for studying the parity and $CP$ violations present in baryon productions and decays within the helicity formalism.
\end{abstract}

\maketitle

\newpage

\section{Introduction\label{sec: Intro}}

Verifying the symmetries of physical laws and their violation across different energy scales has played a central role in advancing our understanding of particle interactions. Parity violation, originating from weak interactions, was first proposed by Lee and Yang~\cite{Lee:1956qn} and soon confirmed by the beta decay experiment of polarized $^{60}$Co nuclei~\cite{Wu:1957my} and subsequent meson decay experiments~\cite{Garwin:1957hc,Friedman:1957mz}. Meanwhile, $CP$ violation, a necessary condition for matter-antimatter asymmetry in the universe as proposed by Sakharov~\cite{Sakharov:1967dj}, still faces many unresolved issues.

The $CP$ violation was first experimentally discovered in meson decays. In 1964, Brookhaven National Laboratory observed $CP$ violation in the decay of neutral $K$ mesons, marking the first observation of $CP$ violation in a process involving the strange quark~\cite{Christenson:1964fg}. Later, $CP$ violation was also observed in the decays of $B$ mesons, which involve $b$-quarks, by the BABAR collaboration at SLAC~\cite{BaBar:2001pki} and the BELLE collaboration at KEK~\cite{Belle:2001zzw}. More recently, the LHCb Collaboration reported $CP$ violation in the decays of $D^{0}$ mesons, which involve $c$-quarks~\cite{LHCb:2019hro}. These $CP$ violations effects can be explained within the standard model (SM) via the weak interaction and the Cabibbo-Kobayashi-Maskawa (CKM) quark mixing matrix~\cite{Cabibbo:1963yz,Kobayashi:1973fv}. However, the amount of $CP$ violation predicted by the SM is insufficient to explain the observed matter-antimatter imbalance in the universe~\cite{Morrissey:2012db}. Theoretical models have also proposed the possibility of $CP$ violation in the strong interaction
~\cite{tHooft:1976rip,tHooft:1976snw,Jackiw:1976pf,Callan:1976je}, but experimental evidence has not yet been found~\cite{ParticleDataGroup:2024cfk}. As a result, searching for signals of strong $CP$ violation and for new sources of $CP$ violation beyond the SM across various energy scales and in different reaction processes remains a key focus in modern particle physics.

The first search for $CP$ violation in hyperon decays was conducted by comparing the branching ratios of the decays of $\Sigma^{+}$ and $\bar{\Sigma}^{-}$~\cite{Okubo:1958zza}. Although no signal of $CP$ violation was observed, this search sparked significant theoretical interest in exploring $CP$ violation through hyperon decays~\cite{Lee:1957qs,Pais:1959zza,Brown:1983wd,Chau:1983ei,Donoghue:1985ww,Donoghue:1986hh}. Recently, the LHCb Collaboration reported the first observation of $CP$ violation with a significance of 3.1$\sigma$ in decay $\Lambda_{b}^{0}\rightarrow\Lambda K^{+}K^{-}$~\cite{LHCb:2024yzj} and 5.2$\sigma$ in decay $\Lambda_{b}^{0}\rightarrow pK^{-}\pi^{+}\pi^{-}$~\cite{LHCb:2025ray}, respectively. However, no evidence of $CP$ violation has been found in ordinary hyperons~\cite{ParticleDataGroup:2024cfk}.

Electron-positron annihilation at the $J/\psi$ mass peak energy benefits from a significant enhancement in the cross section, offering high-statistics opportunities for studying baryon properties through processes such as $J/\psi \rightarrow B_1\bar{B}_2$~\cite{Kopke:1988cs}. Theoretical studies on the polarization correlations of baryon pairs in such processes provide valuable opportunities to probe baryon $CP$ violation~\cite{Donohue:1969fu,Chen:2007zzf,Perotti:2018wxm,Hong:2023soc,Zhang:2023wmd,Zhang:2023box,Zhang:2024rbl,Dubnickova:1992ii,Tomasi-Gustafsson:2005svz,Czyz:2007wi,Faldt:2017kgy}. The BESIII Collaboration, which has accumulated the world’s largest $J/\psi$ sample, has provided precise measurements of the decay parameters for baryon hadronic decays such as $\Lambda$~\cite{BESIII:2018cnd,BESIII:2022qax}, $\Sigma$~\cite{BESIII:2020fqg,BESIII:2025fre}, and $\Xi$~\cite{BESIII:2021ypr,BESIII:2023drj,BESIII:2023jhj}. Recently, BESIII has also reported hints on $CP$ violation measurements in associated production of different baryons, $J/\psi\rightarrow\Lambda\bar{\Sigma}$~\cite{BESIII:2023cvk}, and baryon radiative decays~\cite{BESIII:2022rgl,BESIII:2023fhs,BESIII:2024nif,BESIII:2024lio}. The proposed STCF experiment~\cite{Achasov:2023gey} will significantly expand the data samples, increasing $J/\psi$ statistics by two orders of magnitude in one year compared to BESIII.

Theoretical studies of $CP$ violation in hyperons have been extended to investigations of $CP$ violation in baryon production processes~\cite{He:1992ng,He:2022jjc,Du:2024jfc}, including scenarios with polarized beams~\cite{Bondar:2019zgm,Salone:2022lpt,Zeng:2023wqw,Fu:2023ose,Cao:2024tvz,Guo:2025bfn}. A key focus of these studies is to establish the joint angular distribution of the final-state products. 
As the amount of data grows (for example, in future STCF experiments), precise measurements of $CP$ violation need to account for specific experimental conditions~\cite{Bondar:2019zgm,Salone:2022lpt,Zeng:2023wqw,Fu:2023ose,Cao:2024tvz,Guo:2025bfn}.  In the BEPCII/BESIII experiment, because the electron and positron beams cross at a small angle of 11 milliradians, the polarization of the $J/\psi$ particle is slightly different in the lab frame compared to its center-of-mass frame. This difference must be considered when measuring CP violation. On the other hand, the effects of the electron mass and the structure of the coupling vertex between the electron and the $J/\psi$  have often been neglected. In $CP$ violation studies, these factors can introduce critical corrections and must be carefully considered.

This paper presents a refined polarization analysis for the process $e^{+}e^{-}\rightarrow J/\psi\rightarrow B_{1}\bar{B}_{2}$. We perform a detailed polarization analysis in both the production and decay subprocesses of $J/\psi$, as well as the decay of baryons, within the helicity formalism developed by Jacob and Wick~\cite{Jacob:1959at}. This helicity-based approach is equivalent to describing the production and decay subprocess of $J/\psi$ in terms of particle form factors, and baryon decays in terms of partial wave amplitudes. We explicitly establish the correspondence between helicity amplitudes and particle form factors or partial-wave amplitudes. 

A comprehensive polarization analysis for the 
subprocess $e^{+}e^{-}\rightarrow J/\psi$ is provided with all possible helicity amplitudes, including parity violating and $CP$ violating ones. These amplitudes are connected to the anomalous magnetic moment form factor and the electric dipole moment of the electron. We also take into account the spin-flip helicity amplitude arising from the electron mass. By matching the helicity amplitudes with the electron form factors, we estimate their magnitudes. We further analyze the effects influencing beam polarization, particularly emphasizing the mixing of transverse and longitudinal polarization when transforming from the laboratory frame to the center-of-mass frame. Based on these, we provide a detailed analysis of the resulting polarization of $J/\psi$.

The complete polarization analyses of the subprocess of $J/\psi$ decay to both baryon–antibaryon pairs $B\bar{B}$ and to different baryons $B_{1}\bar{B}_{2}$ are presented. The helicity amplitudes corresponding to parity conservation, parity violation, and $CP$ violation are introduced. For the process $J/\psi\rightarrow B\bar{B}$, the helicity amplitudes are associated with the form factors of the baryon $B$, and for the process $J/\psi\rightarrow B_{1}\bar{B}_{2}$, they correspond to the transition form factors of $B_{1}\bar{B}_{2}$. We establish the correspondence between the helicity amplitudes and the form factors, and provide the method to search for $CP$ violation within the helicity formalism. Furthermore, a novel approach to represent the polarization transfer in the $J\rightarrow B_{1}\bar{B}_{2}$ subprocess is introduced via the polarization transfer matrix.

We provide the complete expression for the joint angular distribution of the final-state particles. The polarization of hyperons can be measured through their decay processes. We focus on the hadronic and radiative decay of hyperons, providing the helicity amplitudes and the polarization transfer matrix for these decay processes. By combining polarization analyzes of $J/\psi$ production subprocess, its decay into baryons, and the decays of baryons, we present the complete expression for the joint angular distribution of all final-state products. Focusing on the process $J/\psi \rightarrow \Lambda \bar{\Sigma}^{0}$, we provide sensitivity predictions for the helicity amplitude parameters and their associated transition form factors. When combined with measurements from the conjugate channel $J/\psi \rightarrow \bar{\Lambda} \Sigma^{0}$, this allows the search for potential signals of $CP$ violation. These isospin-violating processes are particularly sensitive to electroweak interactions, making them valuable for investigating $CP$ violations.

The structure of this paper is as follows. In Sec.~\ref{sec: Jpsi_Pro}, we present the polarization analysis of the subprocess $e^{+}e^{-}\rightarrow J/\psi$. In Sec.~ \ref{sec: Jpsi_decay}, we provide the polarization analysis of the subprocess $J/\psi\rightarrow B_{1}\bar{B}_{2}$. In Sec.~\ref{sec: joint_ang}, we discuss baryon decays and provide the complete joint angular distribution. Focusing on the $J/\psi\rightarrow\Lambda\bar{\Sigma}^{0}$ process, the sensitivities to the transition form factors are provided. Finally, we present a short summary in Sec.~\ref{sec: Sum}.

\section{Polarization analysis in production of $J/\psi$ \label{sec: Jpsi_Pro}}

In high-energy reactions, the polarization state of particles is described by their spin density matrix. We begin with a systematic analysis of the beam polarization and derive the spin density matrix for $J/\psi$ in the helicity formalism. As the spin-1 particle, $J/\psi$ has eight independent polarization components. For the subprocess $e^{+}e^{-}\rightarrow J/\psi$, the polarization of $J/\psi$ is primarily determined by electron (anomalous) form factors and the polarization of the initial beams. Taking both effects into account, we derive the full expression for the polarization components of $J/\psi$.

\subsection{Spin density matrix}

In the helicity formalism, the spin density matrix of the particle with spin $s$ can be expressed as~\cite{Perotti:2018wxm,Zhang:2024rbl}:
\begin{align}
\rho_{s}= & \sum_{\mu=0}^{4s\left(s+1\right)}S_{\mu}\Sigma_{\mu},
\end{align}
where $S_{\mu}$ are the polarization expansion coefficients, and $\Sigma_{\mu}$ are the corresponding polarization projection matrices.

For the spin-1/2 particle, the spin density matrix is given by:
\begin{align}
S_{\mu}= & \left\{ S_{0},S_{1},S_{2},S_{3}\right\} ,\nonumber \\
\Sigma_{\mu}= & \frac{1}{2}\left\{ I,\sigma_{x},\sigma_{y},\sigma_{z}\right\},\label{eq: Sigma_1h}
\end{align}
where $\sigma_{x}$, $\sigma_{y}$, and $\sigma_{z}$ are the Pauli matrices, $S_{0}$ corresponds to the cross-section term, and $S_{i}$ represents the polarization terms. These polarization terms are related to the Cartesian spin components through the following relation:
\begin{align}
\left\{ S_{1},S_{2},S_{3}\right\} /S_{0}= & \left\{ S_{T}^{x},S_{T}^{y},S_{L}\right\} .
\end{align}
Then, the spin density matrix for the initial-state positron and electron can be expressed as:
\begin{align}
\rho^{+}= & \frac{1}{2}\left(\begin{array}{cc}
1+\bar{S}_{L} & \bar{S}_{T}^{x}-i\bar{S}_{T}^{y}\\
\bar{S}_{T}^{x}+i\bar{S}_{T}^{y} & 1-\bar{S}_{L}
\end{array}\right)\quad\text{for}\quad e^{+},\nonumber \\
\rho^{-}= & \frac{1}{2}\left(\begin{array}{cc}
1+S_{L} & S_{T}^{x}-iS_{T}^{y}\\
S_{T}^{x}+iS_{T}^{y} & 1-S_{L}
\end{array}\right)\quad\text{for}\quad e^{-}.
\end{align}

Under realistic experimental conditions, various factors contribute to beam polarization. For circular accelerators, charged particles experience spontaneous polarization due to their motion in magnetic fields, a phenomenon known as the Sokolov-Ternov effect~\cite{Sokolov:1963zn}. The direction and magnitude of the positron and electron spontaneous polarization depend on factors such as the magnetic field strength, beam energy, beam lifetime, and so on. Furthermore, experiments such as the proposed STCF~\cite{Achasov:2023gey} aim to prepare a polarized beam as the key design feature. For collider experiments at circular accelerators, the two beams typically intersect at a small angle at the interaction point~\cite{BESIII:2009fln}. As a result, the laboratory frame generally does not coincide with the center-of-mass frame.  When polarization vectors are transformed between these frames, transverse and longitudinal polarization components can mix.

\begin{figure}
\begin{centering}
\includegraphics[width=0.6\textwidth]{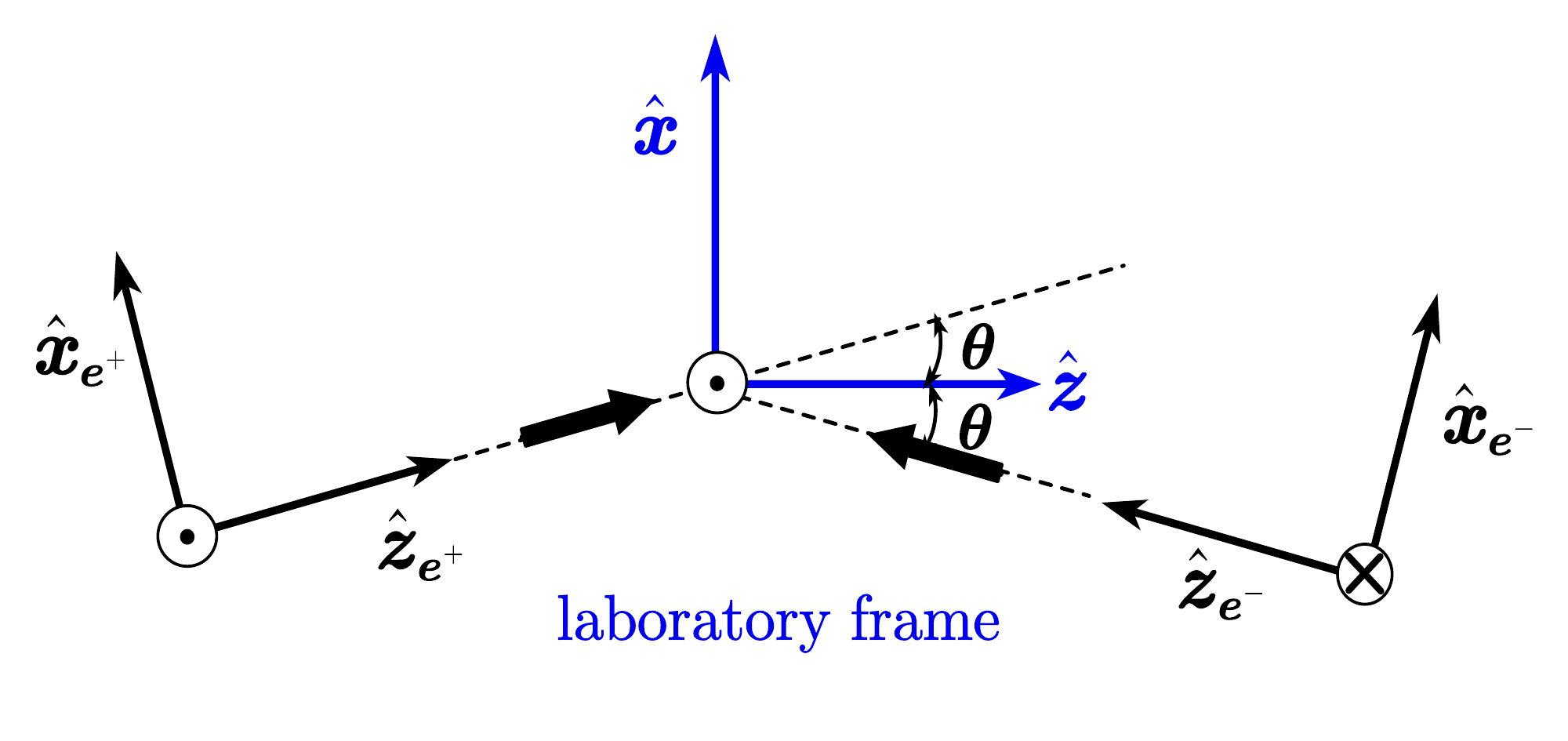}
\par\end{centering}
\begin{centering}
\includegraphics[width=0.6\textwidth]{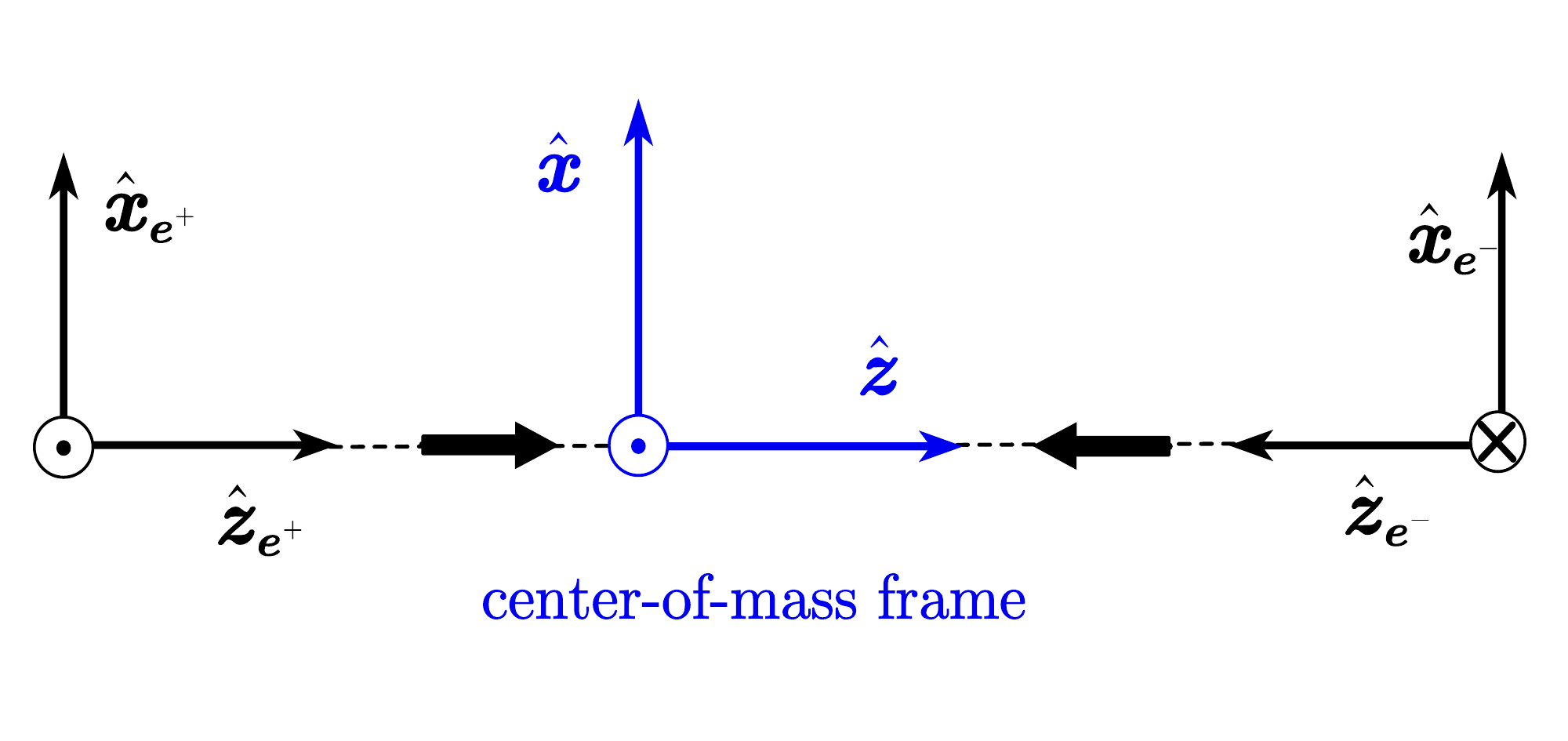}
\par\end{centering}
\caption{\label{fig: lab_and_cm}Comparison of the laboratory frame (upper) and center-of-mass frame (lower) for positron-electron collisions.}
\end{figure}

The coordinate systems of the laboratory and center-of-mass frames are shown in Fig.~\ref{fig: lab_and_cm}. In the laboratory frame, the $z$-axis is defined as the bisector of the angle between the direction of the positron beam and the opposite direction of the electron beam. The plane formed by the two beam directions is taken as the $x\text{-}z$ plane. The momenta of the positron and electron in this frame can be expressed as follows:
\begin{align}
P_{e^{+}}= & \frac{m}{\sqrt{1-\beta^{2}}}\left(1,\beta\sin\theta,0,\beta\cos\theta\right),\nonumber \\
P_{e^{-}}= & \frac{m}{\sqrt{1-\beta^{2}}}\left(1,\beta\sin\theta,0,-\beta\cos\theta\right),
\end{align}
where $\beta$ is the velocity of the beams. The polarization vectors of the positron and electron in the laboratory frame are then given by:
\begin{align}
S_{e^{+}}= & \left(\frac{\beta}{\sqrt{1-\beta^{2}}}\bar{S}_{L,\text{lab}},\frac{\sin\theta}{\sqrt{1-\beta^{2}}}\bar{S}_{L,\text{lab}}+\bar{S}_{T,\text{lab}}^{x}\cos\theta,\bar{S}_{T,\text{lab}}^{y},\frac{\cos\theta}{\sqrt{1-\beta^{2}}}\bar{S}_{L,\text{lab}}-\bar{S}_{T,\text{lab}}^{x}\sin\theta\right),\nonumber \\
S_{e^{-}}= & \left(\frac{\beta}{\sqrt{1-\beta^{2}}}S_{L,\text{lab}},\frac{\sin\theta}{\sqrt{1-\beta^{2}}}S_{L,\text{lab}}+S_{T,\text{lab}}^{x}\cos\theta,-S_{T,\text{lab}}^{y},S_{T,\text{lab}}^{x}\sin\theta-\frac{\cos\theta}{\sqrt{1-\beta^{2}}}S_{L,\text{lab}}\right).
\end{align}
To transform from the laboratory frame to the center-of-mass frame, we apply the following Lorentz boost matrix to the beams:
\begin{align}
\Lambda_{\nu}^{\mu}= & \left(\begin{array}{cccc}
\frac{1}{\sqrt{1-\beta^{2}\sin^{2}\theta}} & -\frac{\beta\sin\theta}{\sqrt{1-\beta^{2}\sin^{2}\theta}} & 0 & 0\\
-\frac{\beta\sin
\theta}{\sqrt{1-\beta^{2}\sin^{2}\theta}} & \frac{1}{\sqrt{1-\beta^{2}\sin^{2}\theta}} & 0 & 0\\
0 & 0 & 1 & 0\\
0 & 0 & 0 & 1
\end{array}\right).
\end{align}
Using this transformation, the momenta of the positron and electron in the center-of-mass frame become
\begin{align}
P_{e^{+}}= & \frac{m}{\sqrt{1-\beta^{2}}}\left(\sqrt{1-\beta^{2}\sin^{2}\theta},0,0,\beta\cos\theta\right),\nonumber \\
P_{e^{-}}= & \frac{m}{\sqrt{1-\beta^{2}}}\left(\sqrt{1-\beta^{2}\sin^{2}\theta},0,0,-\beta\cos\theta\right).
\end{align}
Correspondingly, the polarization vectors of the positron and electron in the center-of-mass frame are:
\begin{align}
S_{e^{+}}=&\left(\frac{\beta\cos\theta}{\sqrt{1-\beta^{2}\sin^{2}(\theta)}}\left(\frac{\bar{S}_{L,\text{lab}}}{\sqrt{1-\beta^{2}}}\cos\theta-\bar{S}_{T,\text{lab}}^{x}\sin\theta\right),\right.\nonumber\\&\left.\frac{\bar{S}_{T,\text{lab}}^{x}\cos\theta+\sqrt{1-\beta^{2}}\bar{S}_{L,\text{lab}}\sin\theta}{\sqrt{1-\beta^{2}\sin^{2}\theta}},\bar{S}_{T,\text{lab}}^{y},\frac{\bar{S}_{L,\text{lab}}\cos\theta}{\sqrt{1-\beta^{2}}}-\bar{S}_{T,\text{lab}}^{x}\sin\theta\right),\nonumber\\S_{e^{-}}=&\left(\frac{\beta\cos\theta}{\sqrt{1-\beta^{2}\sin^{2}\theta}}\left(\frac{S_{L,\text{lab}}}{\sqrt{1-\beta^{2}}}\cos\theta-S_{T,\text{lab}}^{x}\sin\theta\right),\right.\nonumber\\&\left.\frac{S_{T,\text{lab}}^{x}\cos\theta+\sqrt{1-\beta^{2}}S_{L,\text{lab}}\sin\theta}{\sqrt{1-\beta^{2}\sin^{2}(\theta)}},-S_{T,\text{lab}}^{y},S_{T,\text{lab}}^{x}\sin\theta-\frac{S_{L,\text{lab}}\cos\theta}{\sqrt{1-\beta^{2}}}\right).
\end{align}
When further transformed into the helicity frame of the positron and electron, one can extract their polarization components as
\begin{align}
\bar{S}_{T}^{x}= & \frac{\bar{S}_{T,\text{lab}}^{x}\cos\theta+\sqrt{1-\beta^{2}}\bar{S}_{L,\text{lab}}\sin\theta}{\sqrt{1-\beta^{2}\sin^{2}\theta}},\nonumber \\
\bar{S}_{T}^{y}= & \bar{S}_{T,\text{lab}}^{y},\nonumber \\
\bar{S}_{L}= & \frac{\bar{S}_{L,\text{lab}}\cos\theta-\sqrt{1-\beta^{2}}\bar{S}_{T,\text{lab}}^{x}\sin\theta}{\sqrt{1-\beta^{2}\sin^{2}\theta}},\nonumber \\
S_{T}^{x}= & \frac{S_{T,\text{lab}}^{x}\cos\theta-\sqrt{1-\beta^{2}}S_{L,\text{lab}}\sin\theta}{\sqrt{1-\beta^{2}\sin^{2}\theta}},\nonumber \\
S_{T}^{y}= & S_{T,\text{lab}}^{y},\nonumber \\
S_{L}= & \frac{S_{L,\text{lab}}\cos\theta+\sqrt{1-\beta^{2}}S_{T,\text{lab}}^{x}\sin\theta}{\sqrt{1-\beta^{2}\sin^{2}\theta}}.
\end{align}
Since the initial beam polarization is generated in the laboratory frame, while theoretical polarization analysis is typically performed in the center-of-mass frame, experimental observables must also be transformed into the center-of-mass frame for consistency. This transformation leads to a mixing of longitudinal and transverse polarization components, which ultimately affects the joint angular distribution of the final-state particles. Therefore, such effects can manifest themselves in experimental measurements and should not be neglected.

For spin-1 particles, such as $J/\psi$, their spin density matrix can be expressed as: 
\begin{align}
\rho_{1}= & \sum_{\mu=0}^{8}S_{\mu}\Sigma_{\mu}.
\end{align}
The polarization expansion coefficients $S_{\mu}$ and the polarization projection matrix $\Sigma_{\mu}$ can be determined according to the approach described in~\cite{Zhang:2024rbl}. For $S_{\mu}$, we adopt the following scheme:
\begin{align}
\frac{1}{S_{0}}\left\{\begin{array}{ccc}
S_{0} & S_{1} & S_{2}\\
S_{3} & S_{4} & S_{5}\\
S_{6} & S_{7} & S_{8}
\end{array}\right\}= & \left\{\begin{array}{ccc}
1 & S_{L} & S_{T}^{x}\\
S_{T}^{y} & S_{LL} & S_{LT}^{x}\\
S_{LT}^{y} & S_{TT}^{xx} & S_{TT}^{xy}
\end{array}\right\},\label{eq: Smu_1}
\end{align}
where $S_{L}$, ..., $S_{TT}^{xy}$ are the spin components of the spin-1 particle in Cartesian form, as detailed in Appendix~\ref{subsec: SDM_Spin1}. The explicit expression of $\Sigma_{\mu}$ in this scheme is provided in Appendix~\ref{subsec: pro_matr}.

\subsection{Production amplitudes for the subprocess $e^{+}e^{-}\rightarrow J/\psi$}

\begin{figure}[ht]
\begin{centering}
\includegraphics[width=0.6\textwidth]{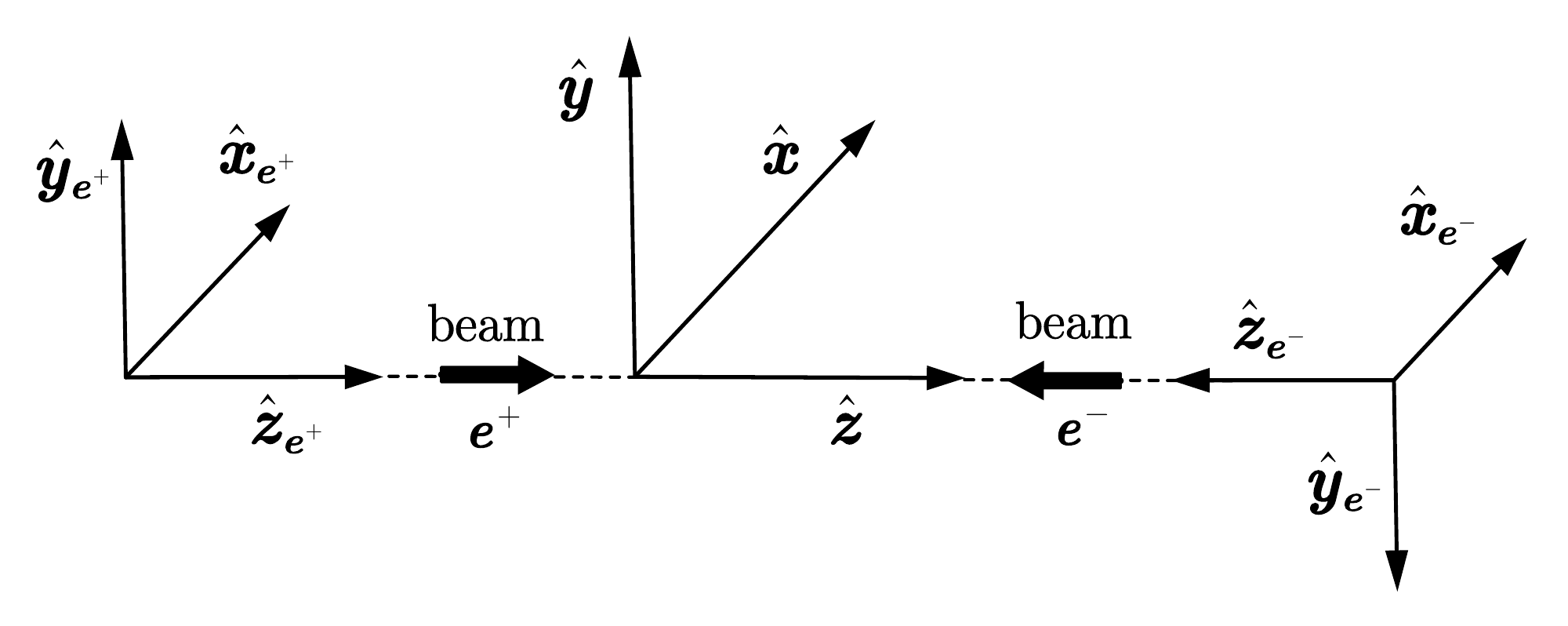}
\par\end{centering}
\caption{\label{fig: Jpsi_prod} The $e^{+}e^{-}\rightarrow J/\psi$ process in c.m. frame. The coordinate systems $x_{e^{+}}$-$y_{e^{+}}$-$z_{e^{+}}$, $x_{e^{-}}$-$y_{e^{-}}$-$z_{e^{-}}$ and $x$-$y$-$z$ are the helicity frames for positron, electron, and $J/\psi$. }
\end{figure}

In the helicity formalism, the helicity amplitude for the subprocess $e^{+}e^{-}\rightarrow J/\psi$ can be written as
\begin{align}
A_{\lambda_{1}\lambda_{2}}^{ee}= & \left(\begin{array}{cc}
h_{1}^{ee}+h_{3}^{ee} & h_{2}^{ee}-h_{4}^{ee}\\
h_{2}^{ee}+h_{4}^{ee} & h_{1}^{ee}-h_{3}^{ee}
\end{array}\right),\label{eq: Aij_ee}
\end{align}
where $\lambda_{1}$ and $\lambda_{2}$ are the helicities of the positron and electron, respectively.  Then, we have
\begin{align}
h_{1}^{ee}=&	\frac{1}{2}\left(A_{\frac{1}{2},\frac{1}{2}}^{ee}+A_{-\frac{1}{2},-\frac{1}{2}}^{ee}\right),\nonumber\\
h_{2}^{ee}=&	\frac{1}{2}\left(A_{-\frac{1}{2},\frac{1}{2}}^{ee}+A_{\frac{1}{2},-\frac{1}{2}}^{ee}\right),\nonumber\\
h_{3}^{ee}=&	\frac{1}{2}\left(A_{\frac{1}{2},\frac{1}{2}}^{ee}-A_{-\frac{1}{2},-\frac{1}{2}}^{ee}\right),\nonumber\\
h_{4}^{ee}=&	\frac{1}{2}\left(A_{-\frac{1}{2},\frac{1}{2}}^{ee}-A_{\frac{1}{2},-\frac{1}{2}}^{ee}\right).
\end{align}

These helicity amplitudes are correlated to the scattering amplitudes expressed in the (anomalous) forms factors of the electron, which are given by:
\begin{align}
\mathcal{M}_{\lambda_{1}\lambda_{2}}= & \bar{v}\left(p_{1},\lambda_{1}\right)\left(F_{1}(q^{2})\gamma^{\mu}+F_{2}(q^{2})\frac{i\sigma^{\mu\nu}q_{\nu}}{2m}+F_{3}(q^{2})\gamma^{\mu}\gamma_{5}\right.\nonumber \\
 & \left.+F_{4}(q^{2})\frac{\sigma^{\mu\nu}q_{\nu}}{2m}\gamma_{5}\right)u\left(p_{2},\lambda_{2}\right)\varepsilon_{\mu}\left(q,\kappa\right),\label{eq: Mij_ee}
\end{align}
where $\bar{v}$ and $u$ are the spinors of the positron and electron, respectively. $p_{1}\left(p_{2}\right)$, $\lambda_{1}\left(\lambda_{2}\right)$, and $m$ represent their momenta, helicities, and mass. $\varepsilon_{\mu}$ and $q=p_{1}+p_{2}$ are the polarization vector and momentum of $J/\psi$, and $\kappa=\lambda_{1}-\lambda_{2}$ is the helicity of $J/\psi$. The form factors $F_{1}$, $F_{2}$, and $F_{4}$ correspond to the charge, the anomalous magnetic dipole moment~\cite{Schwinger:1948iu}, and the electric dipole moment form factor of the electron, respectively, and $F_{3}$ represents the parity violation form factor. In the time-like region, these form factors can be complex. 

At one-loop order, the anomalous magnetic moment of the electron, see, for instance, in~\cite{Peskin:1995ev}, is given by:
\begin{align}
F_{2}(q^{2})= & \frac{\alpha m^{2}}{\pi}\int_{0}^{1}dx\int_{0}^{1-x}dy\frac{\left(x+y\right)\left(1-x-y\right)}{m^{2}(x+y)^{2}-Q^{2}xy}\nonumber\\
= & \frac{\alpha m^{2}}{\pi Q\sqrt{Q^{2}-4m^{2}}}\ln\frac{Q-\sqrt{Q^{2}-4m^{2}}}{Q+\sqrt{Q^{2}-4m^{2}}},
\end{align}
with $Q=\sqrt{q^2}$ and $\alpha=e^2/(4\pi)$. The parity violation form factor $F_{3}$ arises primarily from the contribution of $Z$-boson exchange,
\begin{align}
F_{3}(q^{2})= & \frac{E_{c}^{2}(4\cos2\theta_{W}-1)}{M_{z}^{2}\left(1-\cos4\theta_{W}\right)-3E_{c}^{2}\left(3-2\cos2\theta_{W}\right)},
\end{align}
where $E_{c}=\sqrt{q^{2}}/2$ represents the beam energy in the center-of-mass frame, $M_{z}$ is the mass of the $Z$ boson, and $\theta_{W}$ is the weak coupling angle. The electric dipole moment form factor $F_4$ is related to the electron’s dipole moment $d_{e}$ by
\begin{align}
d_{e}= & -\frac{F_{4}}{2m}.
\end{align}
According to the experimental constraint $d_{e}<4.1\times10^{-30}e\text{cm}$~\cite{ParticleDataGroup:2024cfk}, we obtain:
\begin{align}
\left|F_{4}\right|< & 6.43\times10^{-20}.
\end{align}
It is evident that the contribution from the electric dipole moment is much smaller than the contributions from the other terms. 

Since Eq.~\eqref{eq: Aij_ee} and Eq.~\eqref{eq: Mij_ee} are equivalent, i.e., $A_{\lambda_{1}\lambda_{2}}^{ee}=\mathcal{M}_{\lambda_{1}\lambda_{2}}$, we derive the relationship between the helicity amplitude and form factors,
\begin{align}
h_{1}^{ee}= & 2mF_{1}(q^{2})-\frac{2E_{c}^{2}}{m}F_{2}(q^{2}),\nonumber \\
h_{2}^{ee}= & 2\sqrt{2}E_{c}\left(F_{1}(q^{2})-F_{2}(q^{2})\right),\nonumber \\
h_{3}^{ee}= & i\frac{2E_{c}}{m}\sqrt{E_{c}^{2}-m^{2}}F_{4}(q^{2}),\nonumber \\
h_{4}^{ee}= & 2\sqrt{2}\sqrt{E_{c}^{2}-m^{2}}F_{3}(q^{2}).
\end{align}
We normalize these amplitudes with $h_{2}^{ee}$ and parametrize the helicity amplitudes in the following scheme:
\begin{align}
\frac{h_{4}^{ee}}{h_{2}^{ee}}= & \sqrt{a_{1}}e^{i\varphi_{1}},\nonumber \\
\frac{h_{1}^{ee}}{h_{2}^{ee}}= & \sqrt{a_{2}}e^{i\varphi_{2}},\nonumber \\
\frac{h_{3}^{ee}}{h_{2}^{ee}}= & \sqrt{a_{3}}e^{i\varphi_{3}}.\label{eq: ee_para}
\end{align}
At leading order, we obtain:
\begin{align}
\sqrt{a_{1}}= & \frac{E_{c}\sqrt{E_{c}^{2}-m^{2}}(4\cos2\theta_{W}-1)}{M_{z}^{2}\left(1-\cos4\theta_{W}\right)-3E_{c}^{2}\left(3-2\cos2\theta_{W}\right)}\approx2.33\times10^{-4},\nonumber \\
\sqrt{a_{2}}= & \frac{m}{\sqrt{2}E_{c}}\left(1-\frac{E_{c}^{2}}{m^{2}}F_{2}\left(q^{2}\right)\right)\approx2.36\times10^{-4},\nonumber \\
\sqrt{a_{3}}= &\left|i\frac{\sqrt{E_{c}^{2}-m^{2}}}{\sqrt{2}m}F_{4}\left(q^{2}\right)\right|<1.38\times10^{-16}.\label{eq: para_value}
\end{align}
We find that although the parameter $a_{3}$, associated with the electric dipole moment of the electron, is negligibly small and can be safely ignored, the parameters $a_{1}$ and $a_{2}$, which are related to the electron mass, anomalous magnetic moment, and $Z$-boson exchange, remain important for polarization analysis, especially in investigations of $CP$ violation.

\subsection{The polarization components of $J/\psi$ }

The production density matrix of $J/\psi$ for the subprocess $e^-e^+\rightarrow J/\psi$  can be expressed as
\begin{align}
\rho_{\kappa,\kappa^{\prime}}^{J/\psi}= & \;2\sum_{\lambda_{1},\lambda_{1},\lambda_{2},\lambda_{2}}\mathcal{D}_{\kappa,\lambda_{1}-\lambda_{2}}^{1*}\left(0,0,0\right)\mathcal{D}_{\kappa^{\prime},\lambda_{1}^{\prime}-\lambda_{2}^{\prime}}^{1}\left(0,0,0\right)\nonumber \\
 & \times\rho_{\lambda_{1},\lambda_{1}^{\prime}}^{+}\rho_{\lambda_{2},\lambda_{2}^{\prime}}^{-}A_{\lambda_{1},\lambda_{2}}A_{\lambda_{1}^{\prime},\lambda_{2}^{\prime}}^{*},
\end{align}
where $\mathcal{D}_{\kappa,\lambda}^{J}\left(\Omega\right)=\mathcal{D}_{\kappa,\lambda}^{1}\left(0,0,0\right)$ is the Wigner-$\mathcal{D}$ matrix. We can then obtain the polarization components of $J/\psi$, which are given by:
\begin{align}
S_{\mu}= & \frac{\text{Tr}\left[\rho_{\kappa,\kappa^{\prime}}^{J/\psi}\Sigma_{\mu}\right]}{\text{Tr}\left[\Sigma_{\mu}\Sigma_{\mu}\right]}.
\end{align}
Substituting Eqs.~\eqref{eq: Aij_ee} and~\eqref{eq: ee_para} into the above expression, we obtain:
\begin{align}
S_{0}= & 1+a_{1}+a_{2}+a_{3}-\left(1+a_{1}-a_{2}-a_{3}\right)\bar{S}_{L}S_{L}\nonumber \\
 & +\left(a_{2}-a_{3}\right)\left(\bar{S}_{T}^{x}S_{T}^{x}-\bar{S}_{T}^{y}S_{T}^{y}\right)-2\sqrt{a_{1}}\left(\bar{S}_{L}-S_{L}\right)\cos\varphi_{1}\nonumber \\
 & +2\sqrt{a_{2}a_{3}}\left[\left(\bar{S}_{L}+S_{L}\right)\cos\left(\varphi_{2}-\varphi_{3}\right)-\left(\bar{S}_{T}^{x}S_{T}^{y}+\bar{S}_{T}^{y}S_{T}^{x}\right)\sin\left(\varphi_{2}-\varphi_{3}\right)\right],\nonumber \\
S_{1}= & \left(1+a_{1}\right)\left(\bar{S}_{L}-S_{L}\right)-2\sqrt{a_{1}}\left(1-\bar{S}_{L}S_{L}\right)\cos\varphi_{1},\nonumber \\
S_{2}= & \sqrt{2}\left\{ \sqrt{a_{2}}\left[\left(\bar{S}_{T}^{x}+S_{T}^{x}\right)\cos\varphi_{2}+\left(\bar{S}_{L}S_{T}^{y}+S_{L}\bar{S}_{T}^{y}\right)\sin\varphi_{2}\right]\right.\nonumber \\
 & \left.+\sqrt{a_{3}}\left[\left(\bar{S}_{L}S_{T}^{x}+S_{L}\bar{S}_{T}^{x}\right)\cos\varphi_{3}+\left(\bar{S}_{T}^{y}+S_{T}^{y}\right)\sin\varphi_{3}\right]\right.\nonumber \\
 & \left.-\sqrt{a_{1}a_{2}}\left[\left(\bar{S}_{L}S_{T}^{x}-S_{L}\bar{S}_{T}^{x}\right)\cos\left(\varphi_{1}-\varphi_{2}\right)+\left(\bar{S}_{T}^{y}-S_{T}^{y}\right)\sin\left(\varphi_{1}-\varphi_{2}\right)\right]\right.\nonumber \\
 & \left.-\sqrt{a_{1}a_{3}}\left[\left(S_{T}^{x}-\bar{S}_{T}^{x}\right)\cos\left(\varphi_{1}-\varphi_{3}\right)+\left(S_{L}\bar{S}_{T}^{y}-\bar{S}_{L}S_{T}^{y}\right)\sin\left(\varphi_{1}-\varphi_{3}\right)\right]\right\} ,\nonumber \\
S_{3}= & \sqrt{2}\left\{ \sqrt{a_{2}}\left[\left(\bar{S}_{T}^{y}-S_{T}^{y}\right)\cos\varphi_{2}+\left(\bar{S}_{L}S_{T}^{x}-S_{L}\bar{S}_{T}^{x}\right)\sin\varphi_{2}\right]\right.\nonumber \\
 & \left.+\sqrt{a_{3}}\left[\left(S_{L}\bar{S}_{T}^{y}-\bar{S}_{L}S_{T}^{y}\right)\cos\varphi_{3}+\left(S_{T}^{x}-\bar{S}_{T}^{x}\right)\sin\varphi_{3}\right]\right.\nonumber \\
 & \left.+\sqrt{a_{1}a_{2}}\left[\left(\bar{S}_{L}S_{T}^{y}+S_{L}\bar{S}_{T}^{y}\right)\cos\left(\varphi_{1}-\varphi_{2}\right)+\left(\bar{S}_{T}^{x}+S_{T}^{x}\right)\sin\left(\varphi_{1}-\varphi_{2}\right)\right]\right.\nonumber \\
 & \left.+\sqrt{a_{1}a_{3}}\left[\left(\bar{S}_{T}^{y}+S_{T}^{y}\right)\cos\left(\varphi_{1}-\varphi_{3}\right)+\left(\bar{S}_{L}S_{T}^{x}+S_{L}\bar{S}_{T}^{x}\right)\sin\left(\varphi_{1}-\varphi_{3}\right)\right]\right\} ,\nonumber \\
S_{4}= & \frac{1}{2}\left\{ 1+a_{1}-2a_{2}-2a_{3}-\left(1+a_{1}+2a_{2}+2a_{3}\right)\bar{S}_{L}S_{L}\right.\nonumber \\
 & \left.-2\sqrt{a_{1}}\left(\bar{S}_{L}-S_{L}\right)\cos\varphi_{1}-2\left(a_{2}-a_{3}\right)\left(\bar{S}_{T}^{x}S_{T}^{x}-\bar{S}_{T}^{y}S_{T}^{y}\right)\right.\nonumber \\
 & \left.-4\sqrt{a_{2}a_{3}}\left[\left(\bar{S}_{L}+S_{L}\right)\cos\left(\varphi_{2}-\varphi_{3}\right)-\left(\bar{S}_{T}^{x}S_{T}^{y}+\bar{S}_{T}^{y}S_{T}^{x}\right)\sin\left(\varphi_{2}-\varphi_{3}\right)\right]\right\} ,\nonumber \\
S_{5}= & \sqrt{2}\left\{ \sqrt{a_{2}}\left[\left(\bar{S}_{L}S_{T}^{x}-S_{L}\bar{S}_{T}^{x}\right)\cos\varphi_{2}+\left(S_{T}^{y}-\bar{S}_{T}^{y}\right)\sin\varphi_{2}\right]\right.\nonumber \\
 & \left.+\sqrt{a_{3}}\left[\left(S_{T}^{x}-\bar{S}_{T}^{x}\right)\cos\varphi_{3}+\left(\bar{S}_{L}S_{T}^{y}-S_{L}\bar{S}_{T}^{y}\right)\sin\varphi_{3}\right]\right.\nonumber \\
 & \left.-\sqrt{a_{1}a_{2}}\left[\left(\bar{S}_{T}^{x}+S_{T}^{x}\right)\cos\left(\varphi_{1}-\varphi_{2}\right)-\left(\bar{S}_{L}S_{T}^{y}+S_{L}\bar{S}_{T}^{y}\right)\sin\left(\varphi_{1}-\varphi_{2}\right)\right]\right.\nonumber \\
 & \left.-\sqrt{a_{1}a_{3}}\left[\left(\bar{S}_{L}S_{T}^{x}+S_{L}\bar{S}_{T}^{x}\right)\cos\left(\varphi_{1}-\varphi_{3}\right)-\left(\bar{S}_{T}^{y}+S_{T}^{y}\right)\sin\left(\varphi_{1}-\varphi_{3}\right)\right]\right\} ,\nonumber \\
S_{6}= & \sqrt{2}\left\{ -\sqrt{a_{2}}\left[\left(\bar{S}_{L}S_{T}^{y}+S_{L}\bar{S}_{T}^{y}\right)\cos\varphi_{2}-\left(\bar{S}_{T}^{x}+S_{T}^{x}\right)\sin\varphi_{2}\right]\right.\nonumber \\
 & \left.-\sqrt{a_{3}}\left[\left(\bar{S}_{T}^{y}+S_{T}^{y}\right)\cos\varphi_{3}-\left(\bar{S}_{L}S_{T}^{x}+S_{L}\bar{S}_{T}^{x}\right)\sin\varphi_{3}\right]\right.\nonumber \\
 & \left.-\sqrt{a_{1}a_{2}}\left[\left(\bar{S}_{T}^{y}-S_{T}^{y}\right)\cos\left(\varphi_{1}-\varphi_{2}\right)+\left(S_{L}\bar{S}_{T}^{x}-\bar{S}_{L}S_{T}^{x}\right)\sin\left(\varphi_{1}-\varphi_{2}\right)\right]\right.\nonumber \\
 & \left.-\sqrt{a_{1}a_{3}}\left[\left(S_{L}\bar{S}_{T}^{y}-\bar{S}_{L}S_{T}^{y}\right)\cos\left(\varphi_{1}-\varphi_{3}\right)+\left(\bar{S}_{T}^{x}-S_{T}^{x}\right)\sin\left(\varphi_{1}-\varphi_{3}\right)\right]\right\} ,\nonumber \\
S_{7}= & \left(1-a_{1}\right)\left(\bar{S}_{T}^{x}S_{T}^{x}+\bar{S}_{T}^{y}S_{T}^{y}\right)-2\sqrt{a_{1}}\left(\bar{S}_{T}^{y}S_{T}^{x}-\bar{S}_{T}^{x}S_{T}^{y}\right)\sin\varphi_{1},\nonumber \\
S_{8}= & \left(1-a_{1}\right)\left(\bar{S}_{T}^{y}S_{T}^{x}-\bar{S}_{T}^{x}S_{T}^{y}\right)+2\sqrt{a_{1}}\left(\bar{S}_{T}^{x}S_{T}^{x}+\bar{S}_{T}^{y}S_{T}^{y}\right)\sin\varphi_{1}.
\end{align}
Based on Eq.~\eqref{eq: para_value}, and by neglecting the complex phase of the helicity amplitude and the small contributions from the terms related to the electric dipole moment, we obtain
\begin{align}
S_{0}= & 1+a_{1}+a_{2}-\left(1+a_{1}-a_{2}\right)\bar{S}_{L}S_{L}-2\sqrt{a_{1}}\left(\bar{S}_{L}-S_{L}\right)+a_{2}\left(\bar{S}_{T}^{x}S_{T}^{x}-\bar{S}_{T}^{y}S_{T}^{y}\right),\nonumber \\
S_{1}= & \left(1+a_{1}\right)\left(\bar{S}_{L}-S_{L}\right)-2\sqrt{a_{1}}\left(1-\bar{S}_{L}S_{L}\right),\nonumber \\
S_{2}= & \sqrt{2}\left[\sqrt{a_{2}}\left(\bar{S}_{T}^{x}+S_{T}^{x}\right)-\sqrt{a_{1}a_{2}}\left(\bar{S}_{L}S_{T}^{x}-S_{L}\bar{S}_{T}^{x}\right)\right],\nonumber \\
S_{3}= & \sqrt{2}\left[\sqrt{a_{2}}\left(\bar{S}_{T}^{y}-S_{T}^{y}\right)+\sqrt{a_{1}a_{2}}\left(\bar{S}_{L}S_{T}^{y}+S_{L}\bar{S}_{T}^{y}\right)\right],\nonumber \\
S_{4}= & \frac{1}{2}\left[1+a_{1}-2a_{2}-\left(1+a_{1}+2a_{2}\right)\bar{S}_{L}S_{L}-2\sqrt{a_{1}}\left(\bar{S}_{L}-S_{L}\right)-2a_{2}\left(\bar{S}_{T}^{x}S_{T}^{x}-\bar{S}_{T}^{y}S_{T}^{y}\right)\right],\nonumber \\
S_{5}= & \sqrt{2}\left[\sqrt{a_{2}}\left(\bar{S}_{L}S_{T}^{x}-S_{L}\bar{S}_{T}^{x}\right)-\sqrt{a_{1}a_{2}}\left(\bar{S}_{T}^{x}+S_{T}^{x}\right)\right],\nonumber \\
S_{6}= & \sqrt{2}\left[-\sqrt{a_{2}}\left(\bar{S}_{L}S_{T}^{y}+S_{L}\bar{S}_{T}^{y}\right)-\sqrt{a_{1}a_{2}}\left(\bar{S}_{T}^{y}-S_{T}^{y}\right)\right],\nonumber \\
S_{7}= & \left(1-a_{1}\right)\left(\bar{S}_{T}^{x}S_{T}^{x}+\bar{S}_{T}^{y}S_{T}^{y}\right),\nonumber \\
S_{8}= & \left(1-a_{1}\right)\left(\bar{S}_{T}^{y}S_{T}^{x}-\bar{S}_{T}^{x}S_{T}^{y}\right).\label{eq: Jpsi_pol}
\end{align}

By substituting the expressions of the polarization components above into Eq.~\eqref{eq: Smu_1}, one can obtain the spin components of $J/\psi$ in Cartesian form. We retain the leading-order contributions from $a_{1}$ and $a_{2}$ to analyze these Cartesian polarization components. Notably, two polarization components remain non-zero even when the beam is unpolarized,
\begin{align}
S_{L}^{\psi}= & -2\sqrt{a_{1}},\nonumber \\
S_{LL}^{\psi}= & \frac{1}{2}.
\end{align}
When the beam is polarized, these components are modified to
\begin{align}
S_{L}^{\psi}= & \frac{\bar{S}_{L}-S_{L}}{1-\bar{S}_{L}S_{L}}-2\sqrt{a_{1}}\left[1-\left(\frac{\bar{S}_{L}-S_{L}}{1-\bar{S}_{L}S_{L}}\right)^2\right],\nonumber \\
S_{LL}^{\psi}= & \frac{1}{2}\left[1-\frac{3a_{2}\left(1+\bar{S}_{L}S_{L}+\bar{S}_{T}^{x}S_{T}^{x}-\bar{S}_{T}^{y}S_{T}^{y}\right)}{1-\bar{S}_{L}S_{L}}\right].
\end{align}
The longitudinal polarization component $S_{L}^{\psi}$ of $J/\psi$ is sensitive to both beam polarization and weak coupling at the electron–$J/\psi$ vertex. This makes it a potential observable for probing the weak mixing angle $\theta_{W}$.

The two components of the transverse tensor polarization are significantly affected by beam polarization:
\begin{align}
S_{TT}^{xx,\psi}= & \left(1+2\sqrt{a_{1}}\frac{\bar{S}_{L}-S_{L}}{1-\bar{S}_{L}S_{L}}\right)\frac{\bar{S}_{T}^{x}S_{T}^{x}+\bar{S}_{T}^{y}S_{T}^{y}}{1-\bar{S}_{L}S_{L}},\nonumber \\
S_{TT}^{xy,\psi}= & \left(1+2\sqrt{a_{1}}\frac{\bar{S}_{L}-S_{L}}{1-\bar{S}_{L}S_{L}}\right)\frac{\bar{S}_{T}^{y}S_{T}^{x}-\bar{S}_{T}^{x}S_{T}^{y}}{1-\bar{S}_{L}S_{L}}.
\end{align}
The remaining four polarization components are slightly influenced by the beam polarization. Their contributions are suppressed by the smallness of the electron mass and the anomalous magnetic moment form factor,
\begin{align}
S_{T}^{x,\psi}= & \frac{\sqrt{2a_{2}}\left(\bar{S}_{T}^{x}+S_{T}^{x}\right)}{1-\bar{S}_{L}S_{L}},\nonumber \\
S_{T}^{y,\psi}= & \frac{\sqrt{2a_{2}}\left(\bar{S}_{T}^{y}-S_{T}^{y}\right)}{1-\bar{S}_{L}S_{L}},\nonumber \\
S_{LT}^{x,\psi}= & \frac{\sqrt{2a_{2}}\left(\bar{S}_{L}S_{T}^{x}-S_{L}\bar{S}_{T}^{x}\right)}{1-\bar{S}_{L}S_{L}},\nonumber \\
S_{LT}^{y,\psi}= & -\frac{\sqrt{2a_{2}}\left(\bar{S}_{L}S_{T}^{y}+S_{L}\bar{S}_{T}^{y}\right)}{1-\bar{S}_{L}S_{L}}.
\end{align}
When the electron mass and form factors are neglected, the polarization of $J/\psi$ reduces to the form discussed in Refs.~\cite{Cao:2024tvz,Salone:2022lpt}. Furthermore, if beam polarization is also ignored, the $J/\psi$ polarization simplifies to the case presented in Ref.~\cite{Perotti:2018wxm}, where $S_{LL}^{\psi} = 1/2$ and all other polarization components vanish. Under conditions of beam transverse polarization, these specific polarization components of $J/\psi$ provide access to the anomalous magnetic moment of the electron in the time-like region. 

\section{Polarization analysis in decay of $J/\psi$ \label{sec: Jpsi_decay}}

In the helicity formalism, we present a complete polarization analysis of the decay of $J/\psi$ into baryon–antibaryon pairs. The analysis is divided into two types: baryon-antibaryon pair production and associated production of different baryons. By introducing helicity amplitudes that account for parity and $CP$ violation, we define observables that can be used to search for $CP$ violation signals. Finally, we propose a novel expression to represent the polarization transfer for the subprocess $J/\psi\rightarrow B_{1}\bar{B}_{2}$.

\subsection{Decay amplitudes for the subprocess $J/\psi\rightarrow B\bar{B}$}

\begin{figure}[ht]
\begin{centering}
\includegraphics[width=0.6\textwidth]{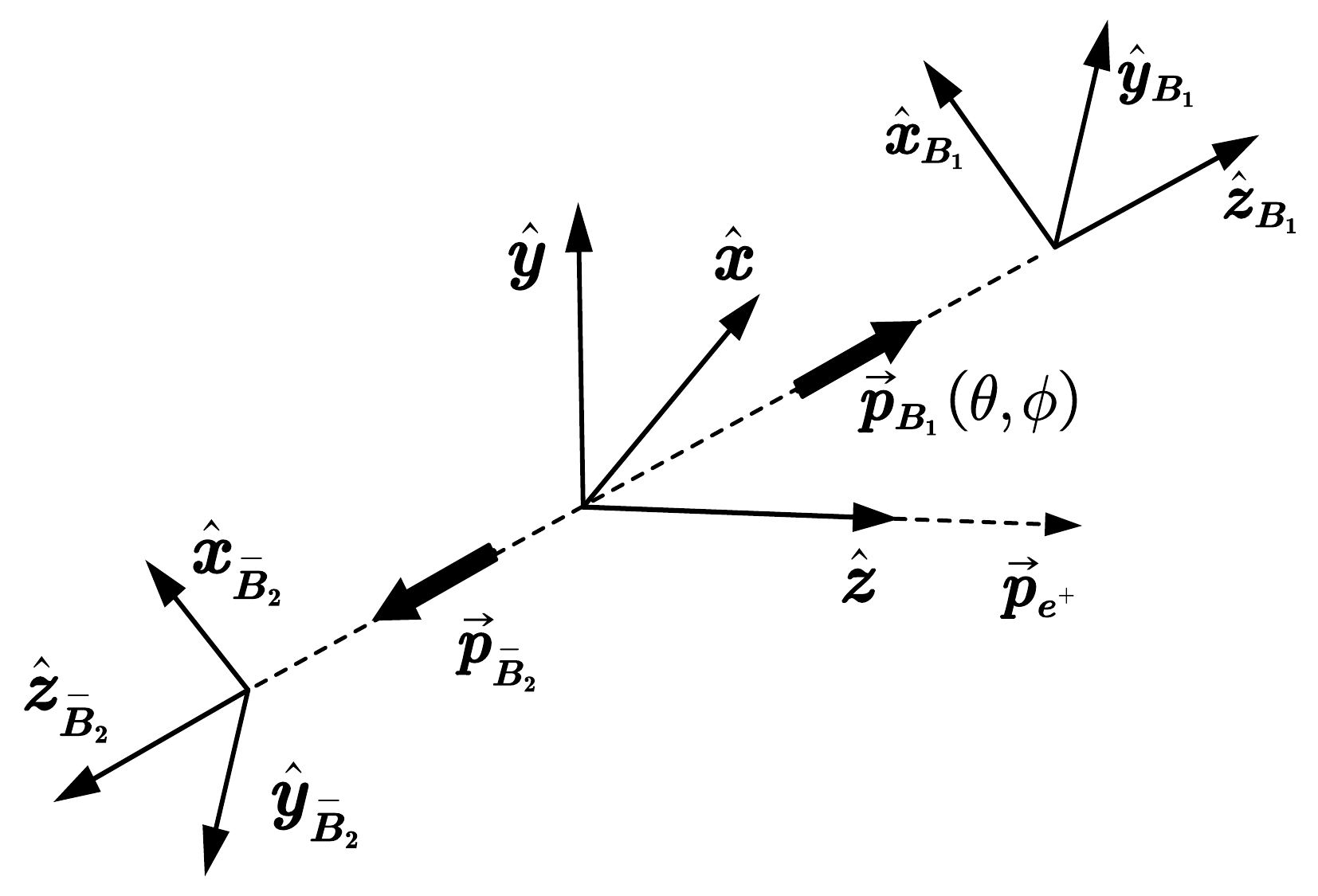}
\par\end{centering}
\caption{\label{fig: Jpsi_decay}The $J/\psi\rightarrow B_{1}\bar{B}_{2}$ process in $J/\psi$ rest frame (helicity frame). The coordinate systems $x_{B_{1}}$-$y_{B_{1}}$-$z_{B_{1}}$, $x_{\bar{B}_{2}}$-$y_{\bar{B}_{2}}$-$z_{\bar{B}_{2}}$, and $x$-$y$-$z$ represent the helicity frame for $B_{1}$, $\bar{B}_{2}$, and $J/\psi$, respectively. $\theta$ and $\phi$ denote the production angles of the $B_{1}$ baryon in the $J/\psi$ rest frame.}
\end{figure}

For the subprocess $J/\psi\rightarrow B\bar{B}$, where $\bar{B}$ is the antiparticle of $B$ (e.g., $J/\psi\rightarrow\Lambda\bar{\Lambda}$), the helicity amplitudes can be written as:
\begin{align}
A_{\lambda_{1},\lambda_{2}}= & \left(\begin{array}{cc}
h_{1}+h_{3} & h_{2}+h_{4}\\
h_{2}-h_{4} & h_{1}-h_{3}
\end{array}\right),\label{eq: Aij_BB}
\end{align}
where $\lambda_{1}$ and $\lambda_{2}$ denote the helicities of $B$ and $\bar{B}$, respectively. Then, we have
\begin{align}
h_{1}=&	\frac{1}{2}\left(A_{\frac{1}{2},\frac{1}{2}}+A_{-\frac{1}{2},-\frac{1}{2}}\right),\nonumber\\
h_{2}=&	\frac{1}{2}\left(A_{\frac{1}{2},-\frac{1}{2}}+A_{-\frac{1}{2},\frac{1}{2}}\right),\nonumber\\
h_{3}=&	\frac{1}{2}\left(A_{\frac{1}{2},\frac{1}{2}}-A_{-\frac{1}{2},-\frac{1}{2}}\right),\nonumber\\
h_{4}=&	\frac{1}{2}\left(A_{\frac{1}{2},-\frac{1}{2}}-A_{-\frac{1}{2},\frac{1}{2}}\right).
\end{align}
Under parity and $CP$ transformations, the helicity amplitude follows the transformation relations:
\begin{align}
A_{\lambda_{1},\lambda_{2}}^{B\bar{B}}&\xrightarrow{\text{Parity Transform}} P_{B}P_{\bar{B}}P_{J/\psi}\left(-1\right)^{J-s_{1}-s_{2}}A_{-\lambda_{1},-\lambda_{2}}^{B\bar{B}},\nonumber \\
A_{\lambda_{1},\lambda_{2}}^{B\bar{B}}&\xrightarrow{\text{Charge Conjugation}}\eta_{J/\psi}\left(-1\right)^{J}A_{\lambda_{1},\lambda_{2}}^{\bar{B}B},\nonumber\\
A_{\lambda_{1},\lambda_{2}}^{B\bar{B}} &\xrightarrow{\text{$CP$ Transform}} P_{B}P_{\bar{B}}P_{J/\psi}\left(-1\right)^{J-s_{1}-s_{2}}\eta_{J/\psi}\left(-1\right)^{J}A_{-\lambda_{1},-\lambda_{2}}^{\bar{B}B},
\end{align}
where $P_{B}$, $P_{\bar{B}}$, and $P_{J/\psi}$ represent the parity of $B$, $\bar{B}$, and $J/\psi$, and $J$, $s_{1}$, and $s_{2}$ represent their spins, $\eta_{J/\psi}$ is the charge parity of $J/\psi$, $\left(-1\right)^{J}$ is the charge parity of the fermion-antifermion system $B\bar{B}$. We obtain that $h_{1}$ and $h_{2}$ correspond to amplitudes that conserve both parity and $CP$, $h_{3}$ corresponds to an amplitude with both parity and $CP$ violation, and $h_{4}$ corresponds to an amplitude with parity violation but $CP$ conservation.

These helicity amplitudes are related to the form factors of the baryon $B$, which are introduced by:
\begin{align}
\mathcal{M}_{\lambda_{1}\lambda_{2}}= & \bar{u}\left(p_{1},\lambda_{1}\right)\left(F_{1}\left(q^{2}\right)\gamma^{\mu}+F_{2}\left(q^{2}\right)\frac{i\sigma^{\mu\nu}q_{\nu}}{2M}+F_{3}\left(q^{2}\right)\gamma^{\mu}\gamma_{5}\right.\nonumber \\
 & \left.+F_{4}\left(q^{2}\right)\frac{\sigma^{\mu\nu}q_{\nu}}{2M}\gamma_{5}\right)v\left(p_{2},\lambda_{2}\right)\varepsilon_{\mu}\left(q,\kappa\right),\label{eq: Mij_BB}
\end{align}
where $\bar{u}$ and $v$ are the spinors of $B$ and $\bar{B}$,  $p_1$ and $p_2$ denote their momenta, and $M$ is the baryon mass. $\varepsilon_{\mu}$ denotes the polarization vector of $J/\psi$,  and $q=p_{1}+p_{2}$, $\kappa=\lambda_{1}-\lambda_{2}$ are its momentum and helicity, respectively. The form factors $F_{1}$, $F_{2}$, and $F_{4}$ correspond to the charge, magnetic dipole moment, and electric dipole moment form factors of the baryon $B$, and $F_{3}$ represents the parity violation form factor. These form factors can be complex in the time-like region. The relation between the electric dipole moment form factor and the electric dipole moment $d_{B}$ is given by~\cite{He:1992ng,He:2022jjc}:
\begin{align}
\frac{1}{2M}F_{4}\left(q^{2}\right)= & \frac{2e}{3M_{J/\psi}^{2}}g_{V}d_{B},
\end{align}
where $M_{J/\psi}$ is the mass of $J/\psi$, and $g_{V}=1.286~\text{GeV}^{2}$ is the effective coupling constant between the charm quark and $J/\psi$~\cite{Du:2024jfc}.

According to Eqs.~\eqref{eq: Aij_BB} and~\eqref{eq: Mij_BB}, we derive the correspondence between the helicity amplitudes and the baryon form factors:
\begin{align}
h_{1}= & 2M\left(F_{1}\left(q^{2}\right)+\frac{E_{c}^{2}}{M^{2}}F_{2}\left(q^{2}\right)\right)=2MG_{E},\nonumber \\
h_{2}= & 2\sqrt{2}E_{c}\left(F_{1}\left(q^{2}\right)+F_{2}\left(q^{2}\right)\right)=2\sqrt{2}E_{c}G_{M},\nonumber \\
h_{3}= & i\frac{2E_{c}}{M}\sqrt{E_{c}^{2}-M^{2}}F_{4}\left(q^{2}\right),\nonumber \\
h_{4}= & 2\sqrt{2}\sqrt{E_{c}^{2}-M^{2}}F_{3}\left(q^{2}\right),\label{eq: BB_h1_Fi}
\end{align}
where $E_{c}=\sqrt{q^{2}}/2$ represents the beam energy in the center-of-mass frame. We parametrize these helicity amplitudes in the following scheme:
\begin{align}
h_{1}= & \frac{\sqrt{2}}{2}\sqrt{1-\alpha_{\psi}},\nonumber \\
h_{2}= & \sqrt{1+\alpha_{\psi}}e^{-i\Phi},\nonumber \\
h_{3}= & \sqrt{\alpha_{1}}e^{i\phi_{1}},\nonumber \\
h_{4}= & \sqrt{2\alpha_{2}}e^{i\phi_{2}}.\label{eq: BB_para}
\end{align}
The estimated sizes of the parity violation parameters for various baryon production processes are summarized in Table.~\ref{tab: p_vio}.
\begin{table}
\caption{\label{tab: p_vio}Based on the measurements in~\cite{BESIII:2022qax,BESIII:2025fre,BESIII:2023drj,BESIII:2023jhj,BESIII:2024nif} and the calculations in ~\cite{Du:2024jfc}, along with Eqs.~\eqref{eq: BB_h1_Fi} and~\eqref{eq: BB_para}, we calculate the size of the parity-violating parameters for the $J/\psi\rightarrow B\bar{B}$ processes.}
\begin{centering}
$\begin{array}{ccc}
\hline\hline 
\text{Process} & \quad\alpha_{2}\quad& \quad\phi_{2}\quad\\
\hline 
J/\psi\rightarrow p\bar{p} & 2.39\times10^{-7} & \pi\\
J/\psi\rightarrow n\bar{n} & 2.91\times10^{-7} & 0\\
J/\psi\rightarrow\Lambda\bar{\Lambda} & 1.22\times10^{-7} & 0\\
J/\psi\rightarrow\Sigma^{0}\bar{\Sigma}^{0} & 1.98\times10^{-13} & 0\\
J/\psi\rightarrow\Sigma^{+}\bar{\Sigma}^{-} & 1.02\times10^{-7} & \pi\\
J/\psi\rightarrow\Sigma^{-}\bar{\Sigma}^{+} & - & 0\\
J/\psi\rightarrow\Xi^{0}\bar{\Xi}^{0} & 1.23\times10^{-7} & 0\\
J/\psi\rightarrow\Xi^{-}\bar{\Xi}^{+} & 1.90\times10^{-7} & 0\\
\hline\hline \end{array}$
\par\end{centering}
\end{table}

\subsection{Decay amplitudes for the subprocess $J/\psi\rightarrow B_{1}\bar{B}_{2}$}

For subprocesses $J/\psi\rightarrow B_{1}\bar{B}_{2}$, where $\bar{B}_{2}$ is not the antiparticle of $B_{1}$ (e.g., $J/\psi\rightarrow\Lambda\bar{\Sigma}^{0}$), the helicity amplitudes can be expressed as
\begin{align}
A_{\lambda_{1},\lambda_{2}}= & \left(\begin{array}{cc}
h_{1}+h_{3} & h_{2}+h_{4}\\
h_{2}-h_{4} & h_{1}-h_{3}
\end{array}\right),\label{eq: Aij_B1B2}
\end{align}
where $\lambda_{1}$ and $\lambda_{2}$ denote the helicities of $B_{1}$ and $\bar{B}_{2}$, respectively. Under parity transformations, the helicity amplitudes follow the transformation relations:
\begin{align}
A_{\lambda_{1},\lambda_{2}}^{B_{1}\bar{B}_{2}}&\xrightarrow{\text{Parity Transform}} P_{B_{1}}P_{\bar{B}_{2}}P_{J/\psi}\left(-1\right)^{J-s_{1}-s_{2}}A_{-\lambda_{1},-\lambda_{2}}^{B_{1}\bar{B}_{2}},
\end{align}
where $P_{B_{1}}$, $P_{\bar{B}_{2}}$, and $P_{J/\psi}$ are the parities of $B_{1}$, $\bar{B}_{2}$, and $J/\psi$, and $J$, $s_{1}$, and $s_{2}$ are their spins. For the process $J/\psi\rightarrow\Lambda\bar{\Sigma}^{0}$, we obtain that $h_{1}$ and $h_{2}$ correspond to the amplitudes where parity is conserved, and $h_{3}$ and $h_{4}$ correspond to amplitudes with parity violation. 

For subprocesses $J/\psi\rightarrow B_{1}\bar{B}_{2}$, the helicity amplitudes are related to the transition form factors of $B_{1}\bar{B}_{2}$, which are introduced by~\cite{Korner:1976hv,Zhang:2024rbl}:
\begin{align}
\mathcal{M}_{\lambda_{1}\lambda_{2}}= & \bar{u}\left(p_{1},\lambda_{1}\right)\left[F_{1}\left(q^{2}\right)\gamma^{\mu}+F_{2}\left(q^{2}\right)\frac{i\sigma^{\mu\nu}q_{\nu}}{M_{1}+M_{2}}+F_{3}\left(q^{2}\right)\gamma^{\mu}\gamma_{5}\right.\nonumber \\
 & \left.+F_{4}\left(q^{2}\right)\frac{\sigma^{\mu\nu}q_{\nu}}{M_{1}+M_{2}}\gamma_{5}\right]v\left(p_{2},\lambda_{2}\right)\varepsilon_{\mu}\left(q,\kappa\right),\label{eq: Mij_B1B2}
\end{align}
where $\bar{u}$ and $v$ are the spinors of the baryons $B_{1}$ and $\bar{B}_{2}$, and $p_{1}$, $p_{2}$, $M_1$, $M_2$ are their momenta and masses. $\varepsilon_{\mu}$ is the polarization vector of $J/\psi$, and $q=p_{1}+p_{2}$, $\kappa=\lambda_{1}-\lambda_{2}$ are its momentum and helicity. $F_{1}$, $F_{2}$, $F_{3}$, and $F_{4}$ are the transition form factors for the production process of $J/\psi\rightarrow B_{1}\bar{B}_{2}$. Using the $CP$ transformation properties of fermion bilinears, we obtain the following relations between the form factors for the conjugate processes $J/\psi\rightarrow B_{1}\bar{B}_{2}$ and $J/\psi\rightarrow B_{2}\bar{B}_{1}$,
\begin{align}
F_{1}^{B_{1}\bar{B}_{2}}= & -P_{B_{1}}P_{\bar{B}_{2}}P_{J/\psi}\eta_{J/\psi}F_{1}^{B_{2}\bar{B}_{1}},\nonumber \\
F_{2}^{B_{1}\bar{B}_{2}}= & -P_{B_{1}}P_{\bar{B}_{2}}P_{J/\psi}\eta_{J/\psi}F_{2}^{B_{2}\bar{B}_{1}},\nonumber \\
F_{3}^{B_{1}\bar{B}_{2}}= & -P_{B_{1}}P_{\bar{B}_{2}}P_{J/\psi}\eta_{J/\psi}F_{3}^{B_{2}\bar{B}_{1}},\nonumber \\
F_{4}^{B_{1}\bar{B}_{2}}= & P_{B_{1}}P_{\bar{B}_{2}}P_{J/\psi}\eta_{J/\psi}F_{4}^{B_{2}\bar{B}_{1}}.
\end{align}

Since Eq.~\eqref{eq: Aij_B1B2} and Eq.~\eqref{eq: Mij_B1B2} are equivalent, i.e., $A_{\lambda_{1}\lambda_{2}} = \mathcal{M}_{\lambda_{1}\lambda_{2}}$, the correspondence between helicity amplitudes and form factors is given by
\begin{align}
h_{1}= & \frac{2\sqrt{1-\xi_{2}^{2}}}{\xi_{1}}E_{c}\left(\xi_{1}^{2}F_{1}+F_{2}\right),\nonumber \\
h_{2}= & 2\sqrt{2}\sqrt{1-\xi_{2}^{2}}E_{c}(F_{1}+F_{2}),\nonumber \\
h_{3}=&\frac{2\sqrt{1-\xi_{1}^{2}}}{\xi_{1}}E_{c}\left(iF_{4}\mp\xi_{1}\xi_{2}F_{3}\right),\nonumber\\
h_{4}=&\frac{2\sqrt{2}\sqrt{1-\xi_{1}^{2}}}{\xi_{1}}E_{c}\left(\xi_{1}F_{3}\mp i\xi_{2}F_{4}\right),\label{eq: B1B2_hi_Fi}
\end{align}
where $\xi_{1}=\left(M_{1}+M_{2}\right)/\sqrt{q^{2}}$ and $\xi_{2}=\left(M_{1}-M_{2}\right)/\sqrt{q^{2}}$, with $M_{1}$ and $M_{2}$ representing the masses of $B_{1}$ and $\bar{B}_{2}$, respectively. The sign $-$ before $\xi_{2}$ corresponds to the process $J/\psi \rightarrow B_{1}\bar{B}_{2}$, and the sign $+$ corresponds to the conjugate process $J/\psi \rightarrow B_{2}\bar{B}_{1}$. For specific reaction channels such as $J/\psi\rightarrow\Lambda\bar{\Sigma}^{0}$ and its conjugate process $J/\psi\rightarrow\Sigma^{0}\bar{\Lambda}$, the helicity amplitudes satisfy the following relations:
\begin{align}
h_{1}^{\Lambda\bar{\Sigma}^{0}}=&	h_{1}^{\Sigma^{0}\bar{\Lambda}}=h_{1}^{\bar{\Lambda}\Sigma^{0}},\nonumber \\
h_{2}^{\Lambda\bar{\Sigma}^{0}}=&	h_{2}^{\Sigma^{0}\bar{\Lambda}}=h_{2}^{\bar{\Lambda}\Sigma^{0}},\nonumber \\
h_{3}^{\Lambda\bar{\Sigma}^{0}}=&	-h_{3}^{\Sigma^{0}\bar{\Lambda}}=-h_{3}^{\bar{\Lambda}\Sigma^{0}},\nonumber \\
h_{4}^{\Lambda\bar{\Sigma}^{0}}=&	h_{4}^{\Sigma^{0}\bar{\Lambda}}=-h_{4}^{\bar{\Lambda}\Sigma^{0}}.\label{eq: B1B2_hi_CP0}
\end{align}
The helicity amplitudes are parametrized as follows
\begin{align}
h_{1}=&\frac{\sqrt{2}}{2}\sqrt{1-\alpha_{\psi}},\nonumber \\
h_{2}=&\sqrt{1+\alpha_{\psi}}e^{-i\Phi},\nonumber \\
h_{3}=&\sqrt{\alpha_{1}}e^{i\phi_{1}},\nonumber \\
h_{4}=&\sqrt{2\alpha_{2}}e^{i\phi_{2}}.\label{eq: para_LamSig0}
\end{align}
Under $CP$ conservation, the parameters satisfy the following relations
\begin{align}
\left\{ \alpha_{\psi}^{\Lambda\bar{\Sigma}^{0}},\alpha_{1}^{\Lambda\bar{\Sigma}^{0}},\alpha_{2}^{\Lambda\bar{\Sigma}^{0}}\right\} =&\left\{ \alpha_{\psi}^{\Sigma^{0}\bar{\Lambda}},\alpha_{1}^{\Sigma^{0}\bar{\Lambda}},\alpha_{2}^{\Sigma^{0}\bar{\Lambda}}\right\} =\left\{ \alpha_{\psi}^{\bar{\Lambda}\Sigma^{0}},\alpha_{1}^{\bar{\Lambda}\Sigma^{0}},\alpha_{2}^{\bar{\Lambda}\Sigma^{0}}\right\} ,\nonumber\\
\left\{ \Phi^{\Lambda\bar{\Sigma}^{0}},\phi_{1}^{\Lambda\bar{\Sigma}^{0}},\phi_{2}^{\Lambda\bar{\Sigma}^{0}}\right\} =&\left\{ \Phi^{\Sigma^{0}\bar{\Lambda}},\pi+\phi_{1}^{\Sigma^{0}\bar{\Lambda}},\phi_{2}^{\Sigma^{0}\bar{\Lambda}}\right\} =\left\{ \Phi^{\bar{\Lambda}\Sigma^{0}},\pi+\phi_{1}^{\bar{\Lambda}\Sigma^{0}},\pi+\phi_{2}^{\bar{\Lambda}\Sigma^{0}}\right\}.\label{eq: CP_LamSig0}
\end{align}

Notably, the amplitudes $h_{3}$ and $h_{4}$ involve mixtures of the $CP$-even and $CP$-odd form factors. To separate these, we introduce an alternative parametrization scheme. We define new helicity amplitudes
\begin{align}
h_{3}=&h_{3}^{\prime}\mp\xi_{1}\xi_{2}h_{4}^{\prime},\nonumber\\
h_{4}=&\sqrt{2}\left(\xi_{1}h_{4}^{\prime}\mp\xi_{2}h_{3}^{\prime}\right),
\end{align}
Then we have
\begin{align}
h_{3}^{\prime}= & i\frac{2\sqrt{1-\xi_{1}^{2}}}{\xi_{1}}E_{c}F_{4},\nonumber \\
h_{4}^{\prime}= & \frac{2\sqrt{1-\xi_{1}^{2}}}{\xi_{1}}E_{c}F_{3}.
\end{align}
For the specific channels $J/\psi\rightarrow\Lambda\bar{\Sigma}^{0}$ and its conjugate $J/\psi\rightarrow\Sigma^{0}\bar{\Lambda}$, the $CP$ conservation requires
\begin{align}
h_{1}^{\Lambda\bar{\Sigma}^{0}}= & h_{1}^{\Sigma^{0}\bar{\Lambda}},\nonumber \\
h_{2}^{\Lambda\bar{\Sigma}^{0}}= & h_{2}^{\Sigma^{0}\bar{\Lambda}},\nonumber \\
h_{3}^{\prime\Lambda\bar{\Sigma}^{0}}= & -h_{3}^{\prime\Sigma^{0}\bar{\Lambda}},\nonumber \\
h_{4}^{\prime\Lambda\bar{\Sigma}^{0}}= & h_{4}^{\prime\Sigma^{0}\bar{\Lambda}}.\label{eq: B1B2_hi_CP}
\end{align}
We then adopt the following parametrization scheme for the helicity amplitudes,
\begin{align}
h_{1}= & \frac{\sqrt{2}}{2}\sqrt{1-\alpha_{\psi}},\nonumber \\
h_{2}= & \sqrt{1+\alpha_{\psi}}e^{-i\Phi},\nonumber \\
h_{3}^{\prime}= & \sqrt{\alpha_{1}^{\prime}}e^{i\phi_{1}^{\prime}},\nonumber \\
h_{4}^{\prime}= & \sqrt{\alpha_{2}^{\prime}}e^{i\phi_{2}^{\prime}}.
\end{align}
we have 
\begin{align}
\alpha_{1}=&\alpha_{1}^{\prime}+\xi_{1}^{2}\xi_{2}^{2}\alpha_{2}^{\prime}\mp2\xi_{1}\xi_{2}\sqrt{\alpha_{1}^{\prime}\alpha_{2}^{\prime}}\cos(\phi_{1}^{\prime}-\phi_{2}^{\prime}),\nonumber\\
\alpha_{2}=&\xi_{2}^{2}\alpha_{1}^{\prime}+\xi_{1}^{2}\alpha_{2}^{\prime}\mp2\xi_{1}\xi_{2}\sqrt{\alpha_{1}^{\prime}\alpha_{2}^{\prime}}\cos(\phi_{1}^{\prime}-\phi_{2}^{\prime}),\nonumber\\
\sin\phi_{1}=&\frac{\sqrt{\alpha_{1}^{\prime}}\sin\phi_{1}^{\prime}\mp\xi_{1}\xi_{2}\sqrt{\alpha_{2}^{\prime}}\sin\phi_{2}^{\prime}}{\sqrt{\alpha_{1}}},\nonumber\\
\cos\phi_{1}=&\frac{\sqrt{\alpha_{1}^{\prime}}\cos\phi_{1}^{\prime}\mp\xi_{1}\xi_{2}\sqrt{\alpha_{2}^{\prime}}\cos\phi_{2}^{\prime}}{\sqrt{\alpha_{1}}},\nonumber\\
\sin\phi_{2}=&\frac{\xi_{1}\sqrt{\alpha_{2}^{\prime}}\sin\phi_{2}^{\prime}\mp\xi_{2}\sqrt{\alpha_{1}^{\prime}}\sin\phi_{1}^{\prime}}{\sqrt{\alpha_{2}}},\nonumber\\
\cos\phi_{2}=&\frac{\xi_{1}\sqrt{\alpha_{2}^{\prime}}\cos\phi_{2}^{\prime}\mp\xi_{2}\sqrt{\alpha_{1}^{\prime}}\cos\phi_{1}^{\prime}}{\sqrt{\alpha_{2}}}.\label{eq: para_LamSig}
\end{align}

By comparing these conjugate production processes, one can search for signals of $CP$ violation. For example, for the processes $J/\psi\rightarrow\Lambda\bar{\Sigma}^{0}$ and $J/\psi\rightarrow\Sigma^{0}\bar{\Lambda}$, based on Eq.~\eqref{eq: B1B2_hi_CP}, $CP$ conservation requires,
\begin{align}
\left\{ \alpha_{\psi}^{\Lambda\bar{\Sigma}^{0}},\alpha_{1}^{\prime\Lambda\bar{\Sigma}^{0}},\alpha_{2}^{\prime\Lambda\bar{\Sigma}^{0}}\right\} = & \left\{ \alpha_{\psi}^{\Sigma^{0}\bar{\Lambda}},\alpha_{1}^{\prime\Sigma^{0}\bar{\Lambda}},\alpha_{2}^{\prime\Sigma^{0}\bar{\Lambda}}\right\} ,\nonumber \\
\left\{ \Phi^{\Lambda\bar{\Sigma}^{0}},\phi_{1}^{\prime\Lambda\bar{\Sigma}^{0}},\phi_{2}^{\prime\Lambda\bar{\Sigma}^{0}}\right\} = & \left\{ \Phi^{\Sigma^{0}\bar{\Lambda}},\pi+\phi_{1}^{\prime\Sigma^{0}\bar{\Lambda}},\phi_{2}^{\prime\Sigma^{0}\bar{\Lambda}}\right\} .\label{eq: CP_LamSig}
\end{align}

For the baryon-antibaryon pair production processes $J/\psi\rightarrow B\bar{B}$, the relationships between the form factors and the partial decay widths of $J/\psi$ have been provided in Ref.~\cite{Du:2024jfc}. Here, we extend a similar analysis to the $J/\psi \rightarrow B_{1}\bar{B}_{2}$ case. The partial decay width for $J/\psi\rightarrow B_{1}\bar{B}_{2}$ is given by:
\begin{align}
\Gamma_{J/\psi\rightarrow B_{1}\bar{B}_{2}}= & \frac{\left|\vec{p}\right|}{8\pi M_{J/\psi}^2}\frac{1}{3}\sum_{\lambda_{1},\lambda_{2}}\left|\mathcal{M}_{\lambda_{1}\lambda_{2}}\right|^{2},
\end{align}
where $|\vec{p}|$ is the momentum of $B_{1}(\bar{B}_{2})$ in the rest frame of $J/\psi$. By substituting Eq.~\eqref{eq: Mij_B1B2} into above equation, we obtain:
\begin{align}
\Gamma_{J/\psi\rightarrow B_{1}\bar{B}_{2}}= & \frac{M_{J/\psi}}{24\pi}\left(1-\xi_{1}^{2}\right)^{1/2}\left(1-\xi_{2}^{2}\right)^{3/2}\left(2|G_{1}|^{2}+\xi_{1}^{2}|G_{2}|^{2}+2|G_{3}|^{2}+|G_{4}|^{2}\right),\label{eq: width_LamSig}
\end{align}
where the recombined form factors $G_{i}$ are defined as:
\begin{align}
G_{1}= & F_{1}+F_{2},\nonumber \\
G_{2}= & F_{1}+\frac{F_{2}}{\xi_{1}^{2}},\nonumber \\
G_{3}= & \frac{\sqrt{1-\xi_{1}^{2}}\left(\xi_{1}F_{3}-i \xi_{2}F_{4}\right)}{\xi_{1}\sqrt{1-\xi_{2}^{2}}},\nonumber \\
G_{4}= & \frac{\sqrt{1-\xi_{1}^{2}}\left(i F_{4}-\xi_{1}\xi_{2}F_{3}\right)}{\xi_{1}\sqrt{1-\xi_{2}{}^{2}}}.
\end{align}
For the process $J/\psi\rightarrow\Lambda\bar{\Sigma}^{0}$, the form factors $G_{1}$ and $G_{2}$ account for parity conservation contributions, with the relative phase defined as $\Phi^{\Lambda\bar{\Sigma}^{0}}=\text{arg}\left(G_{2}/G_{1}\right)$. Since the parity-violating form factors $G_{3}$ and $G_{4}$ are expected to be small, and their contributions to the decay width are quadratically suppressed, they can be safely neglected. Based on measurements from Ref.~\cite{BESIII:2023cvk}, we obtain:
\begin{align}
G_{1}= & 1.41*10^{-4},\nonumber \\
G_{2}= & 1.21*10^{-4},\nonumber \\
\Phi= & 1.011~\text{rad}.
\end{align}

\subsection{The polarization transfer matrix for subprocess $J/\psi\rightarrow B_{1}\bar{B}_{2}$}

The polarization correlations of $B_{1}\bar{B}_{2}$ for the subprocess $J/\psi\rightarrow B_{1}\bar{B}_{2}$ can be expressed as:
\begin{align}
\rho_{\lambda_{1},\lambda_{2};\lambda_{1}^{\prime},\lambda_{2}^{\prime}}^{B_{1}\bar{B}_{2}}= & \sum_{\kappa,\kappa^{\prime}}\mathcal{D}_{\kappa,\lambda_{1}-\lambda_{2}}^{1*}\left(0,\theta,\phi\right)\mathcal{D}_{\kappa^{\prime},\lambda_{1}^{\prime}-\lambda_{2}^{\prime}}^{1}\left(0,\theta,\phi\right)\nonumber \\
 & \times\rho_{\kappa,\kappa^{\prime}}^{J/\psi}A_{\lambda_{1},\lambda_{2}}A_{\lambda_{1}^{\prime},\lambda_{2}^{\prime}}^{*},
\end{align}
where $\theta$ and $\phi$ are the helicity angles of $B_{1}$ in the rest frame of $J/\psi$. $\rho_{\kappa,\kappa^{\prime}}^{J/\psi}$ is the spin density matrix of $J/\psi$, and $\rho_{\lambda_{1},\lambda_{2};\lambda_{1}^{\prime},\lambda_{2}^{\prime}}^{B_{1}\bar{B}_{2}}$ represents the polarization correlation density matrix (production density matrix) for $B_{1}\bar{B}_{2}$. In the helicity formalism, this density matrix can be written as:
\begin{align}
\rho_{\lambda_{1},\lambda_{2};\lambda_{1}^{\prime},\lambda_{2}^{\prime}}^{B_{1}\bar{B}_{2}}= & \sum_{\mu,\nu=0}S_{\mu\nu}\Sigma_{\mu}^{B_{1}}\otimes\Sigma_{\nu}^{\bar{B}_{2}},
\end{align}
where $S_{\mu\nu}$ is the polarization correlation matrix for $B_{1}\bar{B}_{2}$, and $\Sigma_{\mu}^{B_{1}}$, $\Sigma_{\nu}^{\bar{B}_{2}}$ are their polarization expansion matrices, given by Eq.~\eqref{eq: Sigma_1h}. We introduce a novel form to describe the decay process,
\begin{align}
S_{\mu\nu}= & S_{\rho}^{\psi}a_{\rho,\mu\nu},
\end{align}
where $S_{\rho}^{\psi}$ represents the polarization expansion coefficients for $J/\psi$, as given in Eq.~\eqref{eq: Smu_1}. $a_{\rho\mu\nu}$ represents the polarization transfer matrix from $J/\psi$ to $B_{1}\bar{B}_{2}$, which is defined by
\begin{align}
a_{\rho,\mu\nu}= & 4\sum_{\lambda_{1},\lambda_{1}^{\prime}}\sum_{\lambda_{2},\lambda_{2}^{\prime}}\left(\Sigma_{\mu}^{B_{1}}\right)_{\lambda_{1}^{\prime},\lambda_{1}}\left(\Sigma_{\mu}^{\bar{B}_{2}}\right)_{\lambda_{2}^{\prime},\lambda_{2}}A_{\lambda_{1},\lambda_{2}}A_{\lambda_{1}^{\prime},\lambda_{2}^{\prime}}^{*}\nonumber \\
 & \times\sum_{\kappa,\kappa^{\prime}}\left(\Sigma_{\rho}^{J/\psi}\right)_{\kappa,\kappa^{\prime}}\mathcal{D}_{\kappa,\lambda_{1}-\lambda_{2}}^{1*}\left(0,\theta,\phi\right)\mathcal{D}_{\kappa^{\prime},\lambda_{1}^{\prime}-\lambda_{2}^{\prime}}^{1}\left(0,\theta,\phi\right),\label{eq: Jpsi_decay}
\end{align}
where $\Sigma_{\rho}^{J/\psi}$ is the polarization projection matrix for $J/\psi$, as outlined in Appenda~\ref{subsec: pro_matr}. By substituting Eqs.~\eqref{eq: Aij_BB} and~\eqref{eq: BB_para} into the above equation, we obtain the specific form of $a_{\rho,\mu\nu}$ for $J/\psi\rightarrow B\bar{B}$, as detailed in Appenda~\ref{subsec: spin_trans} . Then then full polarization of $B\bar{B}$ can be expressed as 
\begin{align}
S_{00}= & \frac{1}{3}\left(3+\alpha_{\psi}+2\alpha_{1}+4\alpha_{2}\right)+2\sqrt{2}D_{3}^{c}\left[S_{L}^{\psi}\cos\theta+\left(S_{T}^{x,\psi}\cos\phi+S_{T}^{y,\psi}\sin\phi\right)\sin\theta\right]\nonumber \\
 & +\left(\alpha_{\psi}-\alpha_{1}+\alpha_{2}\right)\left[\frac{1}{3}S_{LL}\left(1+3\cos2\theta\right)+\left(S_{LT}^{x,\psi}\cos\phi+S_{LT}^{y,\psi}\sin\phi\right)\sin2\theta\right.\nonumber \\
 & \left.+\left(S_{TT}^{xx,\psi}\cos2\phi+S_{TT}^{xy,\psi}\sin2\phi\right)\sin^{2}\theta\right],\label{eq: S00}\\
S_{10}= & -\left(D_{0}^{c}-2D_{5}^{c}\right)\left[S_{L}^{\psi}\sin\theta-\left(S_{T}^{x,\psi}\cos\phi+S_{T}^{y,\psi}\sin\phi\right)\cos\theta\right]\nonumber \\
 & -\sqrt{2}\left(D_{1}^{s}+D_{4}^{s}\right)\left(S_{T}^{x,\psi}\sin\phi-S_{T}^{y,\psi}\cos\phi\right)+\sqrt{2}\left(D_{1}^{c}-D_{4}^{c}\right)\left[S_{LL}\sin2\theta\right.\nonumber \\
 & \left.-\left(S_{LT}^{x,\psi}\cos\phi+S_{LT}^{y,\psi}\sin\phi\right)\cos2\theta-\frac{1}{2}\left(S_{TT}^{xx,\psi}\cos2\phi+S_{TT}^{xy,\psi}\sin2\phi\right)\sin 2\theta\right]\nonumber \\
 & +\left(D_{0}^{s}-2D_{5}^{s}\right)\left[\left(S_{LT}^{x,\psi}\sin\phi-S_{LT}^{y,\psi}\cos\phi\right)\cos\theta+\left(S_{TT}^{xx,\psi}\sin2\phi-S_{TT}^{xy,\psi}\cos2\phi\right)\sin\theta\right],\\
S_{20}= & \sqrt{2}\left(D_{1}^{s}+D_{4}^{s}\right)\left[S_{L}^{\psi}\sin\theta-\left(S_{T}^{x,\psi}\cos\phi+S_{T}^{y,\psi}\sin\phi\right)\cos\theta\right]\nonumber \\
 & -\left(D_{0}^{c}-2D_{5}^{c}\right)\left(S_{T}^{x,\psi}\sin\phi-S_{T}^{y,\psi}\cos\phi\right)-\left(D_{0}^{s}-2D_{5}^{s}\right)\left[S_{LL}\sin2\theta\right.\nonumber \\
 & \left.-\left(S_{LT}^{x,\psi}\cos\phi+S_{LT}^{y,\psi}\sin\phi\right)\cos2\theta-\frac{1}{2}\left(S_{TT}^{xx,\psi}\cos2\phi+S_{TT}^{xy,\psi}\sin2\phi\right)\sin2\theta\right]\nonumber \\
 & +\sqrt{2}\left(D_{1}^{c}-D_{4}^{c}\right)\left[\left(S_{LT}^{x,\psi}\sin\phi-S_{LT}^{y,\psi}\cos\phi\right)\cos\theta+\left(S_{TT}^{xx,\psi}\sin2\phi-S_{TT}^{xy,\psi}\cos2\phi\right)\sin\theta\right],\\
S_{30}= & \frac{2\sqrt{2}}{3}\left(D_{2}^{c}+2D_{3}^{c}\right)+\left(1+\alpha_{\psi}+2\alpha_{2}\right)\left[S_{L}^{\psi}\cos\theta+\left(S_{T}^{x,\psi}\cos\phi+S_{T}^{y,\psi}\sin\phi\right)\sin\theta\right]\nonumber \\
 & -\sqrt{2}\left(D_{2}^{c}-D_{3}^{c}\right)\left[\frac{1}{3}S_{LL}^{\psi}\left(1+3\cos2\theta\right)+\left(S_{LT}^{x,\psi}\cos\phi+S_{LT}^{y,\psi}\sin\phi\right)\sin2\theta\right.\nonumber \\
 & \left.+\left(S_{TT}^{xx,\psi}\cos2\phi+S_{TT}^{xy,\psi}\sin2\phi\right)\sin^{2}\theta\right],
\end{align}
\begin{align}
S_{01}= & -\left(D_{0}^{c}+2D_{5}^{c}\right)\left[S_{L}^{\psi}\sin\theta-\left(S_{T}^{x,\psi}\cos\phi+S_{T}^{y,\psi}\sin\phi\right)\cos\theta\right]\nonumber \\
 & +\sqrt{2}\left(D_{1}^{s}-D_{4}^{s}\right)\left(S_{T}^{x,\psi}\sin\phi-S_{T}^{y,\psi}\cos\phi\right)-\sqrt{2}\left(D_{1}^{c}+D_{4}^{c}\right)\left[S_{LL}^{\psi}\sin2\theta\right.\nonumber \\
 & \left.-\left(S_{LT}^{x,\psi}\cos\phi+S_{LT}^{y,\psi}\sin\phi\right)\cos2\theta-\frac{1}{2}\left(S_{TT}^{xx,\psi}\cos2\phi+S_{TT}^{xy,\psi}\sin2\phi\right)\sin2\theta\right]\nonumber \\
 & +\left(D_{0}^{s}+2D_{5}^{s}\right)\left[\left(S_{LT}^{x,\psi}\sin\phi-S_{LT}^{y,\psi}\cos\phi\right)\cos\theta+\left(S_{TT}^{xx,\psi}\sin2\phi-S_{TT}^{xy,\psi}\cos2\phi\right)\sin\theta\right],\\
S_{11}= & \frac{1}{3}\left(1-\alpha_{\psi}-2\alpha_{1}\right)-S_{LL}^{\psi}\left[\frac{2}{3}\left(1-\alpha_{\psi}-2\alpha_{1}\right)-2\left(1-\alpha_{1}-\alpha_{2}\right)\sin^{2}\theta\right]\nonumber \\
 & -\left(1-\alpha_{1}-\alpha_{2}\right)\left(S_{LT}^{x,\psi}\cos\phi+S_{LT}^{y,\psi}\sin\phi\right)\sin2\theta\nonumber \\
 & +2\sqrt{2}D_{3}^{s}\left[\left(S_{LT}^{x,\psi}\sin\phi-S_{LT}^{y,\psi}\cos\phi\right)\sin\theta-\left(S_{TT}^{xx,\psi}\sin2\phi-S_{TT}^{xy,\psi}\cos2\phi\right)\cos\theta\right]\nonumber \\
 & +\left[\left(1+\alpha_{\psi}-2\alpha_{2}\right)-\left(1-\alpha_{1}-\alpha_{2}\right)\sin^{2}\theta\right]\left(S_{TT}^{xx,\psi}\cos2\phi+S_{TT}^{xy,\psi}\sin2\phi\right),\\
S_{21}= & -\frac{2\sqrt{2}}{3}D_{2}^{s}+\sqrt{2}S_{LL}^{\psi}\left[\frac{1}{3}D_{2}^{s}\left(1+3\cos2\theta\right)-2D_{3}^{s}\sin^{2}\theta\right]\nonumber \\
 & +\sqrt{2}\left(D_{2}^{s}+D_{3}^{s}\right)\left(S_{LT}^{x,\psi}\cos\phi+S_{LT}^{y,\psi}\sin\phi\right)\sin2\theta\nonumber \\
 & +\left(1+\alpha_{\psi}-2\alpha_{2}\right)\left[\left(S_{LT}^{x,\psi}\sin\phi-S_{LT}^{y,\psi}\cos\phi\right)\sin\theta-\left(S_{TT}^{xx,\psi}\sin2\phi-S_{TT}^{xy,\psi}\cos2\phi\right)\cos\theta\right]\nonumber \\
 & +\frac{\sqrt{2}}{2}\left[2D_{2}^{s}\sin^{2}\theta-D_{3}^{s}\left(3+\cos2\theta\right)\right]\left(S_{TT}^{xx,\psi}\cos2\phi+S_{TT}^{xy,\psi}\sin2\phi\right),\\
S_{31}= & -\sqrt{2}\left(D_{1}^{c}+D_{4}^{c}\right)\left[S_{L}^{\psi}\sin\theta-\left(S_{T}^{x,\psi}\cos\phi+S_{T}^{y,\psi}\sin\phi\right)\cos\theta\right]\nonumber \\
 & +\left(D_{0}^{s}+2D_{5}^{s}\right)\left(S_{T}^{x,\psi}\sin\phi-S_{T}^{y,\psi}\cos\phi\right)-\left(D_{0}^{c}+2D_{5}^{c}\right)\left[S_{LL}^{\psi}\sin2\theta\right.\nonumber \\
 & \left.-\left(S_{LT}^{x,\psi}\cos\phi+S_{LT}^{y,\psi}\sin\phi\right)\cos2\theta-\frac{1}{2}\left(S_{TT}^{xx,\psi}\cos2\phi+S_{TT}^{xy,\psi}\sin2\phi\right)\sin2\theta\right]\nonumber \\
 & +\sqrt{2}\left(D_{1}^{s}-D_{4}^{s}\right)\left[\left(S_{LT}^{x,\psi}\sin\phi-S_{LT}^{y,\psi}\cos\phi\right)\cos\theta+\left(S_{TT}^{xx,\psi}\sin2\phi-S_{TT}^{xy,\psi}\cos2\phi\right)\sin\theta\right],
\end{align}
 \begin{align}
S_{02}= & \sqrt{2}\left(D_{1}^{s}-D_{4}^{s}\right)\left[S_{L}^{\psi}\sin\theta-\left(S_{T}^{x,\psi}\cos\phi+S_{T}^{y,\psi}\sin\phi\right)\cos\theta\right]\nonumber \\
 & +\left(D_{0}^{c}+2D_{5}^{c}\right)\left(S_{T}^{x,\psi}\sin\phi-S_{T}^{y,\psi}\cos\phi\right)+\left(D_{0}^{s}+2D_{5}^{s}\right)\left[S_{LL}^{\psi}\sin2\theta\right.\nonumber \\
 & \left.-\left(S_{LT}^{x,\psi}\cos\phi+S_{LT}^{y,\psi}\sin\phi\right)\cos2\theta-\frac{1}{2}\left(S_{TT}^{xx,\psi}\cos2\phi+S_{TT}^{xy,\psi}\sin2\phi\right)\sin2\theta\right]\nonumber \\
 & +\sqrt{2}\left(D_{1}^{c}+D_{4}^{c}\right)\left[\left(S_{LT}^{x,\psi}\sin\phi-S_{LT}^{y,\psi}\cos\phi\right)\cos\theta+\left(S_{TT}^{xx,\psi}\sin2\phi-S_{TT}^{xy,\psi}\cos2\phi\right)\sin\theta\right],\\
S_{12}= & -\frac{2\sqrt{2}}{3}D_{2}^{s}+\sqrt{2}S_{LL}^{\psi}\left[\frac{1}{3}D_{2}^{s}\left(1+3\cos2\theta\right)+2D_{3}^{s}\sin^{2}\theta\right]\nonumber \\
 & +\sqrt{2}\left(D_{2}^{s}-D_{3}^{s}\right)\left(S_{LT}^{x,\psi}\cos\phi+S_{LT}^{y,\psi}\sin\phi\right)\sin2\theta\nonumber \\
 & -\left(1+\alpha_{\psi}-2\alpha_{2}\right)\left[\left(S_{LT}^{x,\psi}\sin\phi-S_{LT}^{y,\psi}\cos\phi\right)\sin\theta-\left(S_{TT}^{xx,\psi}\sin2\phi-S_{TT}^{xy,\psi}\cos2\phi\right)\cos\theta\right]\nonumber \\
 & +\frac{\sqrt{2}}{2}\left(2D_{2}^{s}\sin^{2}\theta+D_{3}^{s}\left(3+\cos2\theta\right)\right)\left(S_{TT}^{xx,\psi}\cos2\phi+S_{TT}^{xy,\psi}\sin2\phi\right),\\
S_{22}= & -\frac{1}{3}\left(1-\alpha_{\psi}-2\alpha_{1}\right)+S_{LL}^{\psi}\left[\frac{2}{3}\left(1-\alpha_{\psi}-2\alpha_{1}\right)+2\left(\alpha_{\psi}+\alpha_{1}-\alpha_{2}\right)\sin^{2}\theta\right]\nonumber \\
 & -\left(\alpha_{\psi}+\alpha_{1}-\alpha_{2}\right)\left(S_{LT}^{x,\psi}\cos\phi+S_{LT}^{y,\psi}\sin\phi\right)\sin2\theta\nonumber \\
 & +2\sqrt{2}D_{3}^{s}\left[\left(S_{LT}^{x,\psi}\sin\phi-S_{LT}^{y,\psi}\cos\phi\right)\sin\theta-\left(S_{TT}^{xx,\psi}\sin2\phi-S_{TT}^{xy,\psi}\cos2\phi\right)\cos\theta\right]\nonumber \\
 & +\left[\left(1+\alpha_{\psi}-2\alpha_{2}\right)-\left(\alpha_{\psi}+\alpha_{1}-\alpha_{2}\right)\sin^{2}\theta\right]\left(S_{TT}^{xx,\psi}\cos2\phi+S_{TT}^{xy,\psi}\sin2\phi\right),\\
S_{32}= & \left(D_{0}^{s}+2D_{5}^{s}\right)\left[S_{L}^{\psi}\sin\theta-\left(S_{T}^{x,\psi}\cos\phi+S_{T}^{y,\psi}\sin\phi\right)\cos\theta\right]\nonumber \\
 & +\sqrt{2}\left(D_{1}^{c}+D_{4}^{c}\right)\left(S_{T}^{x,\psi}\sin\phi-S_{T}^{y,\psi}\cos\phi\right)+\sqrt{2}\left(D_{1}^{s}-D_{4}^{s}\right)\left[S_{LL}^{\psi}\sin2\theta\right.\nonumber \\
 & \left.-\left(S_{LT}^{x,\psi}\cos\phi+S_{LT}^{y,\psi}\sin\phi\right)\cos2\theta-\frac{1}{2}\left(S_{TT}^{xx,\psi}\cos2\phi+S_{TT}^{xy,\psi}\sin2\phi\right)\sin2\theta\right]\nonumber \\
 & +\left(D_{0}^{c}+2D_{5}^{c}\right)\left[\left(S_{LT}^{x,\psi}\sin\phi-S_{LT}^{y,\psi}\cos\phi\right)\cos\theta+\left(S_{TT}^{xx,\psi}\sin2\phi-S_{TT}^{xy,\psi}\cos2\phi\right)\sin\theta\right],
\end{align}
\begin{align}
S_{03}= & \frac{2\sqrt{2}}{3}\left(D_{2}^{c}-2D_{3}^{c}\right)-\left(1+\alpha_{\psi}+2\alpha_{2}\right)\left[S_{L}^{\psi}\cos\theta+\left(S_{T}^{x,\psi}\cos\phi+S_{T}^{y,\psi}\sin\phi\right)\sin\theta\right]\nonumber \\
 & -\sqrt{2}\left(D_{2}^{c}+D_{3}^{c}\right)\left[\frac{1}{3}S_{LL}^{\psi}\left(1+3\cos2\theta\right)+\left(S_{LT}^{x,\psi}\cos\phi+S_{LT}^{y,\psi}\sin\phi\right)\sin2\theta\right.\nonumber \\
 & \left.+\left(S_{TT}^{xx,\psi}\cos2\phi+S_{TT}^{xy,\psi}\sin2\phi\right)\sin^{2}\theta\right],\\
S_{13}= & -\sqrt{2}\left(D_{1}^{c}-D_{4}^{c}\right)\left[S_{L}^{\psi}\sin\theta-\left(S_{T}^{x,\psi}\cos\phi+S_{T}^{y,\psi}\sin\phi\right)\cos\theta\right]\nonumber \\
 & -\left(D_{0}^{s}-2D_{5}^{s}\right)\left(S_{T}^{x,\psi}\sin\phi-S_{T}^{y,\psi}\cos\phi\right)+\left(D_{0}^{c}-2D_{5}^{c}\right)\left[S_{LL}^{\psi}\sin2\theta\right.\nonumber \\
 & \left.-\left(S_{LT}^{x,\psi}\cos\phi+S_{LT}^{y,\psi}\sin\phi\right)\cos2\theta-\frac{1}{2}\left(S_{TT}^{xx,\psi}\cos2\phi+S_{TT}^{xy,\psi}\sin2\phi\right)\sin2\theta\right]\nonumber \\
 & +\sqrt{2}\left(D_{1}^{s}+D_{4}^{s}\right)\left[\left(S_{LT}^{x,\psi}\sin\phi-S_{LT}^{y,\psi}\cos\phi\right)\cos\theta+\left(S_{TT}^{xx,\psi}\sin2\phi-S_{TT}^{xy,\psi}\cos2\phi\right)\sin\theta\right],\\
S_{23}= & \left(D_{0}^{s}-2D_{5}^{s}\right)\left[S_{L}^{\psi}\sin\theta-\left(S_{T}^{x,\psi}\cos\phi+S_{T}^{y,\psi}\sin\phi\right)\cos\theta\right]\nonumber \\
 & -\sqrt{2}\left(D_{1}^{c}-D_{4}^{c}\right)\left(S_{T}^{x,\psi}\sin\phi-S_{T}^{y,\psi}\cos\phi\right)-\sqrt{2}\left(D_{1}^{s}+D_{4}^{s}\right)\left[S_{LL}^{\psi}\sin2\theta\right.\nonumber \\
 & \left.-\left(S_{LT}^{x,\psi}\cos\phi+S_{LT}^{y,\psi}\sin\phi\right)\cos2\theta-\frac{1}{2}\left(S_{TT}^{xx,\psi}\cos2\phi+S_{TT}^{xy,\psi}\sin2\phi\right)\sin2\theta\right]\nonumber \\
 & +\left(D_{0}^{c}-2D_{5}^{c}\right)\left[\left(S_{LT}^{x,\psi}\sin\phi-S_{LT}^{y,\psi}\cos\phi\right)\cos\theta+\left(S_{TT}^{xx,\psi}\sin2\phi-S_{TT}^{xy,\psi}\cos2\phi\right)\sin\theta\right],\\
S_{33}= & -\frac{1}{3}\left(1+3\alpha_{\psi}-2\alpha_{1}+4\alpha_{2}\right)-2\sqrt{2}D_{3}^{c}\left[S_{L}^{\psi}\cos\theta+\left(S_{T}^{x,\psi}\cos\phi+S_{T}^{y,\psi}\sin\phi\right)\sin\theta\right]\nonumber \\
 & -\left(1+\alpha_{1}+\alpha_{2}\right)\left[\frac{1}{3}S_{LL}^{\psi}\left(1+3\cos2\theta\right)+\left(S_{LT}^{x,\psi}\cos\phi+S_{LT}^{y,\psi}\sin\phi\right)\sin2\theta\right.\nonumber \\
 & \left.+\left(S_{TT}^{xx,\psi}\cos2\phi+S_{TT}^{xy,\psi}\sin2\phi\right)\sin^{2}\theta\right].\label{eq: S33}
\end{align}
For convenience, we introduce the following $D$-type functions,
\begin{align}
D_{0}^{s}= & \sqrt{1-\alpha_{\psi}^{2}}\sin\Phi,\nonumber \\
D_{0}^{c}= & \sqrt{1-\alpha_{\psi}^{2}}\cos\Phi,\nonumber \\
D_{1}^{s}= & \sqrt{\alpha_{1}\left(1+\alpha_{\psi}\right)}\sin\left(\Phi+\phi_{1}\right),\nonumber \\
D_{1}^{c}= & \sqrt{\alpha_{1}\left(1+\alpha_{\psi}\right)}\cos\left(\Phi+\phi_{1}\right),\nonumber \\
D_{2}^{s}= & \sqrt{\alpha_{1}\left(1-\alpha_{\psi}\right)}\sin\phi_{1},\nonumber \\
D_{2}^{c}= & \sqrt{\alpha_{1}\left(1-\alpha_{\psi}\right)}\cos\phi_{1},\nonumber \\
D_{3}^{s}= & \sqrt{\alpha_{2}\left(1+\alpha_{\psi}\right)}\sin\left(\Phi+\phi_{2}\right),\nonumber \\
D_{3}^{c}= & \sqrt{\alpha_{2}\left(1+\alpha_{\psi}\right)}\cos\left(\Phi+\phi_{2}\right),\nonumber \\
D_{4}^{s}= & \sqrt{\alpha_{2}\left(1-\alpha_{\psi}\right)}\sin\phi_{2},\nonumber \\
D_{4}^{c}= & \sqrt{\alpha_{2}\left(1-\alpha_{\psi}\right)}\cos\phi_{2},\nonumber \\
D_{5}^{s}= & \sqrt{\alpha_{1}\alpha_{2}}\sin\left(\phi_{1}-\phi_{2}\right),\nonumber \\
D_{5}^{c}= & \sqrt{\alpha_{1}\alpha_{2}}\cos\left(\phi_{1}-\phi_{2}\right).
\end{align}
The terms $D_{1}^{s}$, $D_{1}^{c}$, $D_{2}^{s}$, and $D_{2}^{c}$ arise from CP-violating effects in baryon production, specifically due to contributions from the baryon electric dipole moment. The terms $D_{3}^{s}$, $D_{3}^{c}$, $D_{4}^{s}$, and $D_{4}^{c}$ originate from parity-violating effects, which are associated with the contribution of the parity-violating form factor $F_{3}$. Finally, the terms $D_{5}^{s}$ and $D_{5}^{c}$ are affected by both CP-violating and parity-violating effects in the baryon production process. When neglecting the electron form factors and assuming unpolarized beams, $S_{LL}^\psi = 1/2$ and all other polarization components of $J/\psi$ can be set to zero. Furthermore, if we also neglect parity-violating helicity amplitudes for baryon-antibaryon pair production processes, the expression simplifies to the form given in Ref.~\cite{Perotti:2018wxm}. 

For the subprocess $J/\psi \rightarrow \Lambda \bar{\Sigma}^{0}$ and its conjugate channel $J/\psi \rightarrow \Sigma^{0} \bar{\Lambda}$, the polarization correlation expressions derived above remain applicable. The angles $\theta$ and $\phi$ represent the production angles of the first baryon appearing in the expression. For example, for the process $J/\psi \rightarrow \Lambda \bar{\Sigma}^{0}$, $\theta$ and $\phi$ denote the production angles of the $\Lambda$ in the $J/\psi$ helicity frame. Correspondingly, for the process $J/\psi \rightarrow \Sigma^{0} \bar{\Lambda}$, these angles describe the production of the $\Sigma^{0}$ in the same reference frame. The constraints imposed by $CP$ conservation on the parameters are given in Eq.~\eqref{eq: CP_LamSig0}. Alternatively, an equivalent parametrization can be adopted, in which the parameters $\alpha_{1}, \alpha_{2}, \phi_{1},$ and $\phi_{2}$ are replaced by $\alpha_{1}^{\prime}, \alpha_{2}^{\prime}, \phi_{1}^{\prime},$ and $\phi_{2}^{\prime}$, as defined in Eq.~\eqref{eq: para_LamSig}. The corresponding $CP$ conservation relations for these parameters in the conjugate processes are presented in Eq.~\eqref{eq: CP_LamSig}.

Moreover, certain polarization components of $J/\psi$ produced directly in electron-positron annihilation are small. However, for $J/\psi$ production via alternative mechanisms, significant polarization may be present across all independent components.  For example, the decay chain $e^{+}e^{-} \rightarrow \psi(2S) \rightarrow J/\psi\, X$ can lead to a more diverse polarization structure. Our formalism provides a complete framework for experimentally accessing all independent polarization components of $J/\psi$.

\subsection{symmetry analysis of baryon--antibaryon polarization correlations}

\begin{figure}[ht]
\begin{centering}
\includegraphics[width=0.4\textwidth]{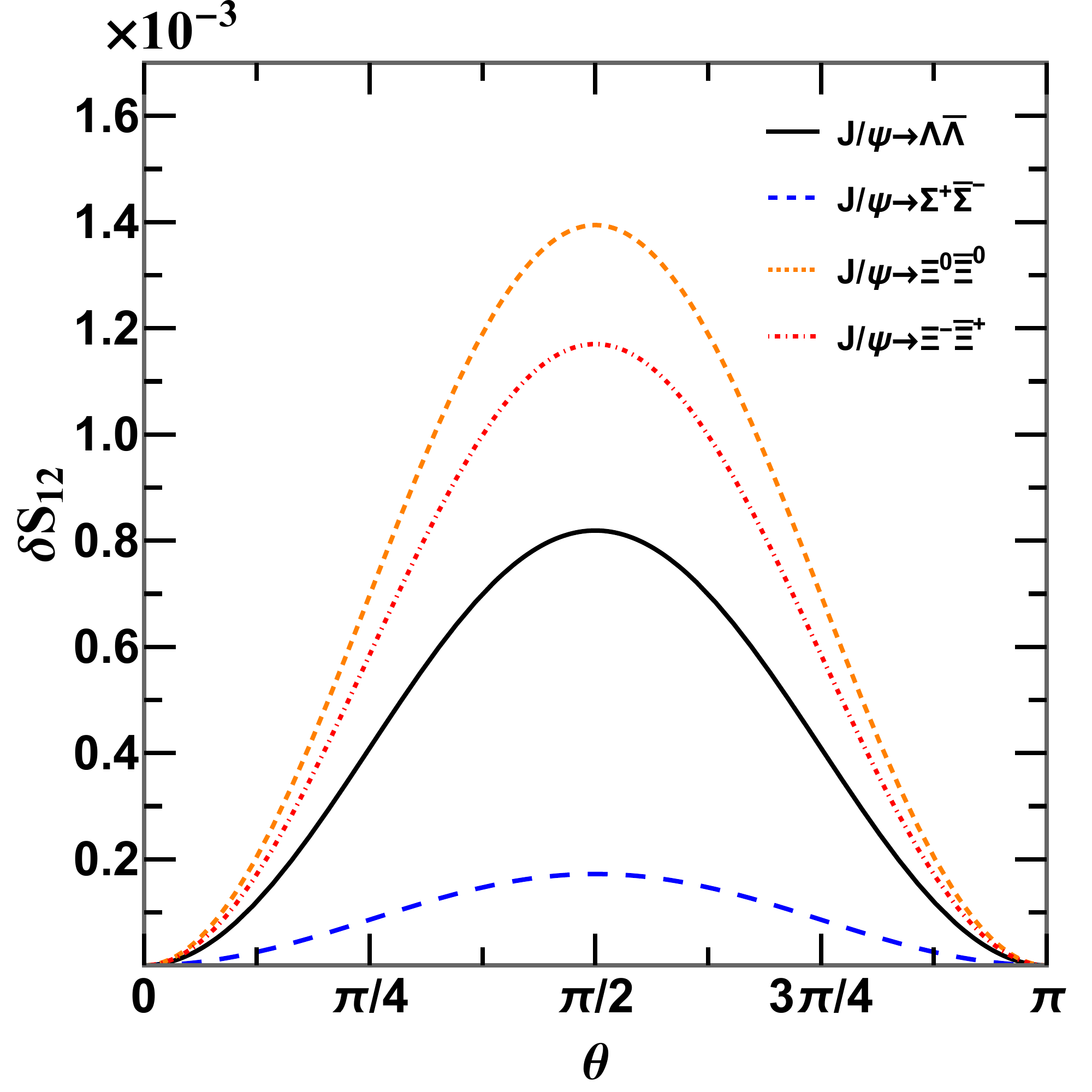}
\par\end{centering}
\caption{\label{fig: parity}Parity-violating effects in different hyperon production processes under unpolarized beam conditions.}
\end{figure}

In the absence of parity and $CP$-violating effects, the polarization correlations of baryon-antibaryon pairs satisfy the symmetry relation 
\begin{equation}
S_{\mu\nu}(\theta,\phi)=S_{\nu\mu}(\pi-\theta,\pi+\phi).
\end{equation}
This symmetry no longer holds when parity and $CP$-violating effects are included. To analyze the impact of parity and $CP$ violation in hyperon productions, we define the following observables:
\begin{align}
\Delta S_{ij}= & S_{ij}+S_{ji},\nonumber \\
\delta S_{ij}= & S_{ij}-S_{ji}.
\end{align}
When the azimuthal angle $\phi$ in hyperon production processes is not considered, i.e., by integrating over $\phi$, the baryon–antibaryon polarization correlations reduce to those that depend only on the purely longitudinal polarization components $S_L^\psi$ and $S_{LL}^\psi$. In this case, a parity-violating observable sensitive to baryon--antibaryon production is given by
\begin{align}
\delta S_{12}= & 4\sqrt{2}S_{LL}^{\psi}D_{3}^{s}\sin^{2}\theta.
\end{align}
According to Table~\ref{tab: p_vio}, the parity violation effects in different production channels are illustrated in Fig.~\ref{fig: parity}, assuming $S_{LL}^{\psi}=1/2$. For processes $J/\psi\rightarrow\Lambda\bar{\Lambda}$ and $J/\psi\rightarrow\Sigma^{+}\bar{\Sigma}^{-}$, the effects are of order $\mathcal{O}(10^{-4})$, and for $J/\psi\rightarrow\Xi^{0}\bar{\Xi}^{0}$ and $J/\psi\rightarrow\Xi^{-}\bar{\Xi}^{+}$, they reach $\mathcal{O}(10^{-3})$. 

Similarly, we define the following CP-violating observables:
\begin{align}
\Delta S_{02}= & 2\sqrt{2}D_{1}^{s}S_{L}^{\psi}\sin\theta+4D_{5}^{s}S_{LL}\sin2\theta,\nonumber\\
\Delta S_{03}= & \frac{4\sqrt{2}}{3}D_{2}^{c}-\frac{2\sqrt{2}}{3}D_{2}^{c}S_{LL}^{\psi}\left(1+3\cos2\theta\right),\nonumber\\
\Delta S_{12}= & -\frac{4\sqrt{2}}{3}D_{2}^{s}+\frac{2\sqrt{2}}{3}D_{2}^{s}S_{LL}^{\psi}\left(1+3\cos2\theta\right),\nonumber\\
\Delta S_{13}= & -2\sqrt{2}D_{1}^{c}S_{L}^{\psi}\sin\theta-4D_{5}^{c}S_{LL}^{\psi}\sin2\theta,\nonumber\\
\delta S_{01}= & -4D_{5}^{c}S_{L}^{\psi}\sin\theta-2\sqrt{2}D_{1}^{c}S_{LL}^{\psi}\sin2\theta,\nonumber\\
\delta S_{23}= & -4D_{5}^{s}S_{L}^{\psi}\sin\theta-2\sqrt{2}D_{1}^{s}S_{LL}^{\psi}\sin2\theta.
\end{align}
These observables serve as sensitive probes for detecting the violation of $CP$ in baryon production, which is contributed by the baryon's electric dipole moment.

\section{The joint angular distribution of products\label{sec: joint_ang}}

The characteristics of hyperon weak decays make them as self-analyzers for spin analysis. The expressions for baryon decays have been extensively discussed in Refs.~\cite{Donohue:1969fu,Chen:2007zzf,Perotti:2018wxm,Zhang:2023box,Zhang:2024rbl,Cao:2024tvz,Guo:2025bfn,Chung:1971ri,Hong:2023soc,Lee:1957qs,Kim:1992az,Byers:1963zz,Button-Shafer:1965, Lednicky:1975ry, Lednicky:1985zx, Ademollo:1964,Berman:1965rc,Xia:2019fjf,Shi:2025xkp}, with various forms of representation. To maintain the completeness of the article, we follow the decay expression scheme used in~\cite{Perotti:2018wxm,Zhang:2023box,Zhang:2024rbl} and provide a brief review of hadronic decays and radiative dacays of baryons. By combining the polarization analysis of the processes $e^{+}e^{-}\rightarrow J/\psi$, $J/\psi\rightarrow B_{1}\bar{B}_{2}$, and baryon decays, we present the full expression for the joint angular distribution of the final-state particles. Focusing on the process $J/\psi\rightarrow\Lambda\bar{\Sigma}^{0}$, we provide predictions for the statistical uncertainties of the helicity amplitude parameters and their corresponding transition form factors using the joint angular distribution.

\subsection{hadronic decay of baryon}

We focus on the most common hadronic decay process for baryons, $B_{1}\left(J^{P_{B_{1}}}\right)\rightarrow B_{3}\left(1/2^{+}\right)+\pi\left(0^{-}\right)$, in which the polarization of the daughter baryon $B_{3}$ can be expressed as:
\begin{align}
S_{\nu}^{B_{3}}= & \sum_{\mu}S_{\mu}^{B_{1}}a_{\mu\nu},
\end{align}
where $S_{\mu}^{B_{1}}$ and $S_{\nu}^{B_{3}}$ denote the polarization expansion coefficients for $B_{1}$ and $B_{3}$, respectively. The polarization transfer matrix $a_{\mu\nu}$ is defined as:
\begin{align}
a_{\mu\nu}^{h}= & \frac{2J+1}{2\pi}\sum_{\kappa,\kappa^{\prime}}\sum_{\lambda,\lambda^{\prime}}A_{\lambda}A_{\lambda^{\prime}}^{*}\left(\Sigma_{\mu}^{B_{1}}\right)_{\kappa,\kappa^{\prime}}\left(\Sigma_{\nu}^{B_{3}}\right)_{\lambda^{\prime}\lambda,}\mathcal{D}_{\kappa,\lambda}^{J*}\left(\Omega\right)\mathcal{D}_{\kappa^{\prime},\lambda^{\prime}}^{J}\left(\Omega\right),
\end{align}
where $\mathcal{D}_{\kappa,\lambda}^{J*}\left(\Omega\right)=\mathcal{D}_{\kappa,\lambda}^{J*}\left(0,\theta,\phi\right)$ is the Wigner-$\mathcal{D}$ matrix, with $J$ being the spin of the parent particle and $A_{\lambda}$ representing the helicity amplitude for the decay process. The indices $\kappa$ and $\lambda$ correspond to the helicities of $B_{1}$ and $B_{3}$, respectively. For parity-conserving decay processes, the helicity amplitude satisfies the following relation:
\begin{align}
A_{1/2}= & \left(-1\right)^{J+1/2}P_{B_{1}}A_{-1/2}.
\end{align}

For the case $J^{B_{1}}=1/2^{+}$, the relationship between the helicity amplitude and the canonical amplitude is~\cite{Jacob:1959at,Perotti:2018wxm}:
\begin{align}
A_{1/2}= & \frac{\sqrt{2}}{2}\left(A_{S}+A_{P}\right),\nonumber \\
A_{-1/2}= & \frac{\sqrt{2}}{2}\left(A_{S}-A_{P}\right).
\end{align}
Under the normalization condition  $\left|A_{S}\right|^{2}+\left|A_{P}\right|^{2}=\left|A_{1/2}\right|^{2}+\left|A_{-1/2}\right|^{2}=1$, the amplitudes can be parametrized as~\cite{Lee:1957qs}:
\begin{align}
\alpha_{D}= & 2\text{Re}\left[A_{S}^{*}A_{P}\right]=\left|A_{1/2}\right|^{2}-\left|A_{-1/2}\right|^{2},\nonumber \\
\beta_{D}= & 2\text{Im}\left[A_{S}^{*}A_{P}\right]=2\text{Im}\left[A_{1/2}A_{-1/2}^{*}\right],\nonumber \\
\gamma_{D}= & \left|A_{S}\right|^{2}-\left|A_{P}\right|^{2}=2\text{Re}\left[A_{1/2}A_{-1/2}^{*}\right],
\end{align}
where $\beta_{D}=\sqrt{1-\alpha_{D}^{2}}\sin\phi_{D}$ and $\gamma_{D}=\sqrt{1-\alpha_{D}^{2}}\cos\phi_{D}$. With this parametrization scheme, the expression for the polarization transfer matrix $a_{\mu\nu}$ is given by~\cite{Perotti:2018wxm},
\begin{align}
a_{00}^{h} & =1,\nonumber \\
a_{03}^{h} & =\alpha_{D},\nonumber \\
a_{10}^{h} & =\alpha_{D}\cos\phi\sin\theta,\nonumber \\
a_{11}^{h} & =\gamma_{D}\cos\theta\cos\phi-\beta_{D}\sin\phi,\nonumber \\
a_{12}^{h} & =-\beta_{D}\cos\theta\cos\phi-\gamma_{D}\sin\phi,\nonumber \\
a_{13}^{h} & =\sin\theta\cos\phi,\nonumber \\
a_{20}^{h} & =\alpha_{D}\sin\theta\sin\phi,\nonumber \\
a_{21}^{h} & =\beta_{D}\cos\phi+\gamma_{D}\cos\theta\sin\phi,\nonumber \\
a_{22}^{h} & =\gamma_{D}\cos\phi-\beta_{D}\cos\theta\sin\phi,\nonumber \\
a_{23}^{h} & =\sin\theta\sin\phi,\nonumber \\
a_{30}^{h} & =\alpha_{D}\cos\theta,\nonumber \\
a_{31}^{h} & =-\gamma_{D}\sin\theta,\nonumber \\
a_{32}^{h} & =\beta_{D}\sin\theta,\nonumber \\
a_{33}^{h} & =\cos\theta.\label{eq: decay_hadr}
\end{align}
For parity-conserving decay processes, we have $\alpha_{D}=\beta_{D}=0$ and $\gamma_{D}\rightarrow\left(-1\right)^{J+1/2}P_{B_{1}}$.

Under the $CP$-conservation constraint, the decay parameters for the particle and antiparticle should satisfy $\alpha_{D}=-\alpha_{\bar{D}}$ and $\phi_{D}=-\phi_{\bar{D}}$. Therefore, one can define these observables to characterize $CP$ violation,
\begin{align}
A_{CP}= & \frac{\alpha_{D}+\alpha_{\bar{D}}}{\alpha_{D}-\alpha_{\bar{D}}},\nonumber \\
B_{CP}= & \frac{\beta_{D}+\beta_{\bar{D}}}{\alpha_{D}-\alpha_{\bar{D}}}.
\end{align}
These observables are directly related to the weak $CP$-violating phases and the strong interaction phases from the final-state interactions~\cite{Donoghue:1985ww,Donoghue:1986hh,Guo:2025bfn}.

In experimental measurements, $CP$ violation is sometimes directly represented by the decay parameter $\phi_{D}$, which serves as an observable for $CP$ violation,
\begin{align}
\Phi_{CP}= & \frac{\phi_{D}+\phi_{\bar{D}}}{2}.
\end{align}
At the leading-order approximation for $\alpha_{D}$ and $\phi_{D}$, we have~\cite{Salone:2022lpt}:
\begin{align}
B_{CP}= & \Phi_{CP}\frac{\sqrt{1-\left\langle \alpha_{D}\right\rangle ^{2}}}{\left\langle \alpha_{D}\right\rangle }\cos\left\langle \phi_{D}\right\rangle -A_{CP}\frac{\left\langle \alpha_{D}\right\rangle }{\sqrt{1-\left\langle \alpha_{D}\right\rangle ^{2}}}\sin\left\langle \phi_{D}\right\rangle ,
\end{align}
where
\begin{align}
\left\langle \alpha_{D}\right\rangle = & \frac{\alpha_{D}-\alpha_{\bar{D}}}{2},\nonumber\\
\left\langle \phi_{D}\right\rangle = & \frac{\phi_{D}-\phi_{\bar{D}}}{2}.
\end{align}

\subsection{Radiative decay of Baryon}

We focus on the most common radiative decay process for baryons, $B_{1}\left(J^{P_{B_{1}}}\right)\rightarrow B_{3}\left(1/2^{+}\right)+\gamma\left(1^{-}\right)$, in which the polarization of the daughter baryon $B_{3}$ can be expressed as:
\begin{align}
S_{\nu}^{B_{3}}= & \sum_{\mu}S_{\mu}^{B_{1}}a_{\mu\nu},
\end{align}
The polarization transfer matrix $a_{\mu\nu}$ is defined as:
\begin{align}
a_{\mu\nu}^{\gamma}= & \frac{2J+1}{2\pi}\sum_{\kappa,\kappa^{\prime}}\sum_{\lambda,\lambda^{\prime}}A_{\lambda_{1},\lambda_{2}}A_{\lambda_{1}^{\prime},\lambda_{2}^{\prime}}^{*}\left(\Sigma_{\mu}^{B_{1}}\right)_{\kappa,\kappa^{\prime}}\left(\Sigma_{\nu}^{B_{3}}\right)_{\lambda_{1}^{\prime},\lambda_{1}}\delta_{\lambda_{2}^{\prime},\lambda_{2}}D_{\kappa,\lambda_{1}-\lambda_{2}}^{J*}\left(\Omega\right)D_{\kappa^{\prime},\lambda_{1}^{\prime}-\lambda_{2}^{\prime}}^{J}\left(\Omega\right),
\end{align}
where $A_{\lambda_{1},\lambda_{2}}$ denotes the helicity amplitude of the decay process, with $\lambda_{1}$ and $\lambda_{2}$ the helicities of the $B_{3}$ and photon, respectively. Based on helicity conservation and the fact that the helicity of a real photon cannot be zero, we obtain:
\begin{align}
A_{\lambda_{1},\lambda_{2}}= & \left(\begin{array}{ccc}
A_{1/2,1} & 0 & 0\\
0 & 0 & A_{-1/2,-1}
\end{array}\right),
\end{align}
For parity-conserving decay processes, the helicity amplitudes satisfy the following relation:
\begin{align}
A_{1/2,1}= & \left(-1\right)^{J-1/2}P_{B_{1}}A_{-1/2,-1}.
\end{align}
Under the normalization condition $\left|A_{1/2,1}\right|^{2}+\left|A_{-1/2,-1}\right|^{2}=1$, the amplitudes can be parametrized as:
\begin{align}
\alpha_{D}= & \left|A_{-1/2,-1}\right|^{2}-\left|A_{1/2,1}\right|^{2}.
\end{align}
With this parametrization scheme, the polarization transfer matrix $a_{\mu\nu}$ is given by,
\begin{align}
a_{00}^{\gamma} & =1,\nonumber \\
a_{03}^{\gamma} & =-\alpha_{D},\nonumber \\
a_{10}^{\gamma} & =\alpha_{D}\sin\theta\cos\phi,\nonumber \\
a_{13}^{\gamma} & =-\sin\theta\cos\phi,\nonumber \\
a_{20}^{\gamma} & =\alpha_{D}\sin\theta\sin\phi,\nonumber \\
a_{23}^{\gamma} & =-\sin\theta\sin\phi,\nonumber \\
a_{30}^{\gamma} & =\alpha_{D}\cos\theta,\nonumber \\
a_{33}^{\gamma} & =-\cos\theta.\label{eq: decay_radi}
\end{align}
Under the $CP$-conservation constraint, the decay parameters for the particle and antiparticle should satisfy $\alpha_{D}=-\alpha_{\bar{D}}$, and thus we can define an observable sensitive to $CP$ violation:
\begin{align}
A_{CP}= & \frac{\alpha_{D}+\alpha_{\bar{D}}}{\alpha_{D}-\alpha_{\bar{D}}}.
\end{align}

\subsection{The joint angular distribution of final products}

Combining the polarization analysis of the production and decay subprocesses of $J/\psi$ discussed in Sec.~\ref{sec: Jpsi_Pro} and Sec.~\ref{sec: Jpsi_decay}, along with the discussion of baryon decays in this section, we can express the joint angular distribution of the final-state particles as:
\begin{align}
W\propto & S_{\mu\nu}a_{\mu0}^{\left(h/\gamma\right)B_{1}}a_{\nu0}^{\left(h/\gamma\right)\bar{B}_{2}},
\end{align}
where $S_{\mu\nu}$ is the polarization correlation matrix for the $B_{1}\bar{B}_{2}$ system, as detailed in Eqs.~\eqref{eq: S00}-~\eqref{eq: S33}. The matrices $a_{\mu0}^{\left(h/\gamma\right)B_{1}}$ and $a_{\nu0}^{\left(h/\gamma\right)\bar{B}_{2}}$ represent the polarization transfer matrices for the decays of $B_{1}$ and $\bar{B}_{2}$, which are given by Eqs.~\eqref{eq: decay_hadr} and~\eqref{eq: decay_radi} for hadronic and radiative decays, respectively. For cascade decay particles, the transfer chain can be extended, e.g., $a_{\mu0}^{\left(h/\gamma\right)B_{1}}\rightarrow\sum_{\mu_{2}\cdots\mu_{n}}a_{\mu\mu_{2}}^{\left(h/\gamma\right)B_{1}}\cdots a_{\mu_{n}0}^{\left(h/\gamma\right)B_{n}}$.

As shown in Eq.~\eqref{eq: Jpsi_pol}, the polarization of $J/\psi$ is determined by both beam polarization and electron form factors. In this paper, we estimate the electron form factors for the $e^{+}e^{-} \rightarrow J/\psi$ subprocess at leading order, as shown in Eq.~\eqref{eq: para_value}.  We recommend that the experimental analyzes fit the polarization expansion coefficients $S_{L}^{\psi}, \ldots, S_{TT}^{xy,\psi}$ directly as free parameters. Given the beam polarization, the electron form factors can be extracted using Eqs.~\eqref{eq: para_value} and~\eqref{eq: Jpsi_pol}.  Comparison between experimental measurements and our theoretical predictions allows for a precise test of the electron–$J/\psi$ interaction.

\subsection{statistical significance}

While $CP$ violation has been widely studied in subprocesses such as $J/\psi\rightarrow\Lambda\bar{\Lambda}$ and $J/\psi\rightarrow\Xi\bar{\Xi}$~\cite{Fu:2023ose,Salone:2022lpt,Guo:2025bfn}, fewer investigations have focused on associated baryon production, such as $J/\psi\rightarrow\Lambda\bar{\Sigma}^{0}$. These isospin-violating processes, including $J/\psi \rightarrow \Lambda\bar{\Sigma}^{0}$ and its conjugate channel $J/\psi\rightarrow\Sigma^{0}\bar{\Lambda}$,
proceed primarily via electroweak interactions and thus offer unique opportunities for probing $CP$ violation. In this work, we analyze the sensitivity to the parameters $\alpha_{1}^{\prime}$, $\alpha_{2}^{\prime}$, and the relevant transition form factors in this channel.

For the subprocess $J/\psi\rightarrow\Lambda\bar{\Sigma}^{0}$, the joint angular distribution of the final-state particles is given by
\begin{align}
\frac{d\sigma}{d\Omega}\propto & \mathcal{W}\left(\vec{\omega};\vec{\eta}\right)= \sum_{\mu,\nu,\nu^{\prime}}
S_{\mu\nu}^{\Lambda\bar{\Sigma}} \,
a_{\mu0}^{h\Lambda} \,
a_{\nu\nu^{\prime}}^{\gamma\bar{\Sigma}} \,
a_{\nu^{\prime}0}^{h\bar{\Lambda}},
\end{align}
where $\vec{\omega}=\left\{\theta,\phi,\theta_{p},\phi_{p},\theta_{\bar{\Lambda}},\phi_{\bar{\Lambda}},\theta_{\bar{p}},\phi_{\bar{p}}\right\}$ is the set of angular variables, and $\vec{\eta}=\{\alpha_{\psi},\Phi,\alpha_{1},\cdots,S_{L}^{\psi},\\\cdots,S_{TT}^{xy,\psi}\}$ includes parameters associated with baryon production, decays, and polarization of $J/\psi$.

For $J/\psi$ produced directly in electron-positron annihilation, the most beam-sensitive polarization components are $S_{L}^{\psi}$ (primarily affected by longitudinal polarization of the beam) and $S_{TT}^{xx,\psi}$, $S_{TT}^{xy,\psi}$ (primarily affected by transverse polarization of the beam). The effects of beam polarization on other polarization components of $J/\psi$ are generally suppressed. At BESIII, electron beams are typically unpolarized, although some spontaneous transverse polarization may arise during beam propagation due to the Sokolov–Ternov effect~\cite{Cao:2024tvz,Sokolov:1963zn}. Future experiments such as STCF aim to prepare polarized beams~\cite{Achasov:2023gey}. According to the analysis in Ref.~\cite{Guo:2025bfn}, longitudinal polarization of the beams significantly improves the sensitivity of the parameters compared to the transverse polarization. Therefore, our sensitivity analysis focuses primarily on the longitudinal case, setting $S_{LL}^{\psi} \rightarrow 1/2$, $S_{L}^{\psi} \rightarrow -P_{e}(S_{L})$, and all other components to zero.

However, it should be noted that although most of the polarization components of $J/\psi$ are only slightly influenced by beam polarization, they can still induce asymmetry effects in final-state particles. Such asymmetries can overlap with the genuine $CP$-violating  effects. Therefore, these components must be retained in experimental fits to ensure accurate isolation of $CP$-violating observables.

\begin{table}[ht]
\caption{\label{tab: sensi} The statistical uncertainties of the parity-violating helicity amplitude parameters $\alpha_{1}^{\prime}$, $\alpha_{2}^{\prime}$, and the corresponding form factors $F_{3}$, $F_{4}$ in the $J/\psi\rightarrow\Lambda\bar{\Sigma}^{0}$ process. The predictions on the statistics at BESIII are conducted with unpolarized beams, and those at STCF are conducted with 80\% beam longitudinal polarization.}
\begin{centering}
$\begin{array}{ccc}
\hline\hline 
J/\psi\rightarrow\Lambda\bar{\Sigma}^{0} &\quad\text{BESIII}\quad & \quad\text{STCF}\quad\\
\hline n_{tag}^{evt} & 1.3\times10^{4} & 4.4\times10^{6}\\
\sigma\left(\sqrt{\alpha_{1}^{\prime}}\right) & 2.3\times10^{-2} & 8.3\times10^{-4}\\
\sigma\left(\sqrt{\alpha_{2}^{\prime}}\right) & 2.1\times10^{-2} & 7.9\times10^{-4}\\
\sigma\left(F_{3}\right) & 4.01\times10^{-6} & 1.48\times10^{-7}\\
\sigma\left(F_{4}\right) & 4.22\times10^{-6} & 1.55\times10^{-7}\\
\hline\hline \end{array}$
\par\end{centering}
\end{table}

\begin{figure}[ht]
\begin{centering}
\includegraphics[width=0.4\textwidth]{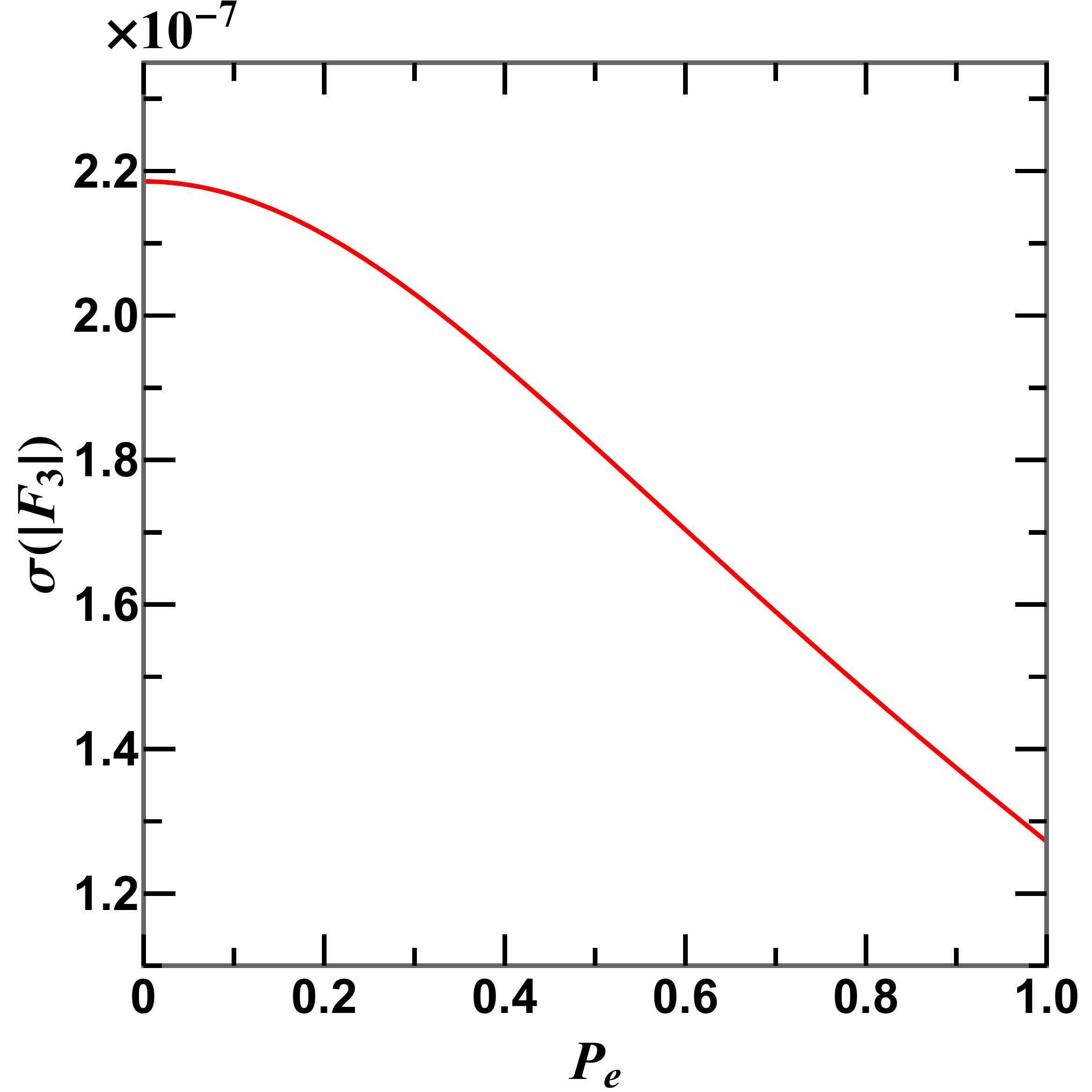}\quad
\includegraphics[width=0.4\textwidth]{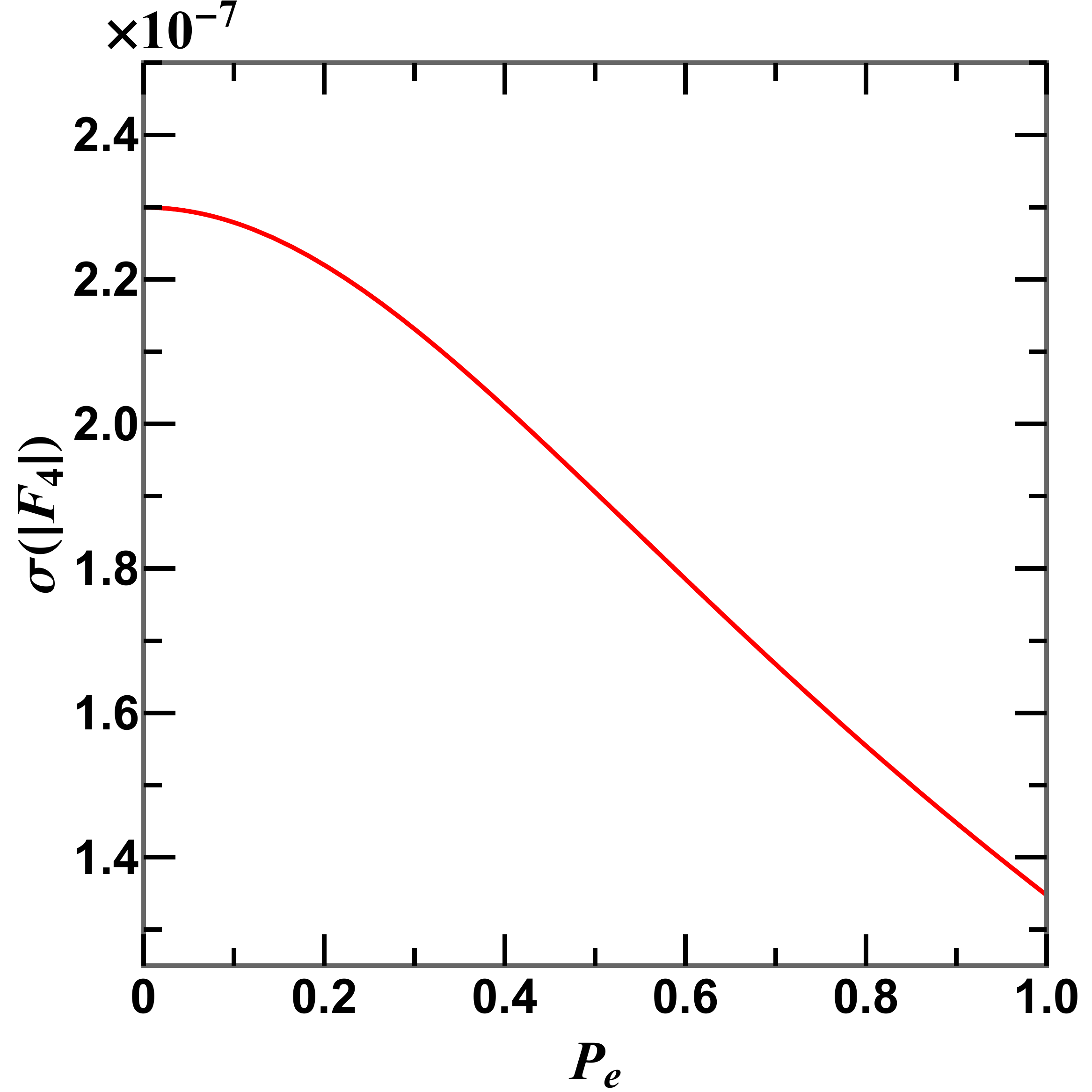}
\par\end{centering}
\caption{\label{fig: sensi}Predicted statistical uncertainties for the transition form factors $F_{3}$ and $F_{4}$ for the process $J/\psi\rightarrow\Lambda\bar{\Sigma}^{0}$ as a function of the degree of electron beam longitudinal polarization $P_{e}$.}
\end{figure}

To evaluate statistical uncertainties, we apply the maximum likelihood method to the joint angular distribution. The covariance matrix of the parameters can be obtained from the Fisher information matrix~\cite{Fisher:1922saa}:
\begin{align}
I_{ij}\left(\eta_{i},\eta_{j}\right)= & n\int\frac{1}{\tilde{\mathcal{W}}}\left(\frac{\partial\tilde{\mathcal{W}}}{\partial\eta_{i}}\right)\left(\frac{\partial\tilde{\mathcal{W}}}{\partial\eta_{j}}\right)d\vec{\omega},
\end{align}
where $n$ is the number of selected events and $\tilde{\mathcal{W}}=\mathcal{W}/\int\mathcal{W}d\vec{\omega}$ is the normalized joint angular distribution. The covariance matrix is the inverse of the Fisher information matrix, $V=I^{-1}$. The statistical uncertainty for a specific parameter $\eta_{i}$ is then given by
\begin{align}
\sigma\left(\eta_{i}\right)= & \sqrt{V_{ii}}.
\end{align}
In Table~\ref{tab: sensi}, we present the sensitivities for the parameters $\alpha_{1}^{\prime}$ and $\alpha_{2}^{\prime}$
for the process $J/\psi\rightarrow\Lambda\bar{\Sigma}^{0}$, based on projected statistics from BESIII~\cite{BESIII:2023cvk} and STCF~\cite{Achasov:2023gey}. The corresponding parameters in the conjugate channel have similar sensitivities. Using Eq.~\eqref{eq: B1B2_hi_Fi}, we also estimate the sensitivity to the transition form factors, whose values are constrained by Eq.~\eqref{eq: width_LamSig}. Although the statistical yield for $J/\psi\rightarrow\Lambda\bar{\Sigma}^{0}$ is relatively low, the sensitivities of the form factors reach levels comparable to those for processes such as $J/\psi\rightarrow\Lambda\bar{\Lambda}$~\cite{Guo:2025bfn,Salone:2022lpt}. Moreover, since the preparation of polarized beams is a key design objective of the planned STCF accelerator~\cite{Achasov:2023gey}, we illustrate in Fig.~\ref{fig: sensi} how parameter sensitivities vary with polarization degree of the electron beam. It is clear that beam polarization significantly enhances measurement precision.

\section{Summary\label{sec: Sum}}

In this paper, we present a comprehensive polarization analysis of the process $e^{+}e^{-}\rightarrow J/\psi\rightarrow B_{1}\bar{B}_{2}$. The analysis is divided into three parts: the subprocess $e^{+}e^{-}\rightarrow J/\psi$, the subprocess $J/\psi\rightarrow B_{1}\bar{B}_{2}$, and the subsequent decays of baryons.

As a spin-1 particle, the polarization state of $J/\psi$ is characterized by eight independent polarization components. For the subprocess $e^{+}e^{-}\rightarrow J/\psi$, we analyze how the polarization state of $J/\psi$ is influenced by both the beam polarization and the electron form factors. A detailed investigation of the initial beam polarization is provided, emphasizing the transformation effects from the laboratory frame to the center-of-mass frame. We present all possible helicity amplitudes for the annihilation process and explicitly establish their relationship to the electron form factors. By including both beam polarization and all relevant helicity amplitudes, we systematically analyze the resulting polarization of $J/\psi$ produced in $e^{+}e^{-}$ annihilation.

When parity-violating and $CP$-violating effects are included in the subprocess $J/\psi \rightarrow B_{1} \bar{B}{2}$, the polarization correlations of the baryon-antibaryon system are modified. This opens the possibility of searching for $CP$ violations in baryon production processes. We classify these subprocesses into two types: baryon–antibaryon pair production and associated production of different baryons. For the pair production process $J/\psi \rightarrow B \bar{B}$, we introduce two helicity amplitudes that conserve both parity and $CP$, one amplitude that violates parity but conserves $CP$, and one that violates both parity and $CP$. The explicit correspondence between these helicity amplitudes and the form factors of the baryon $B$ is provided. For the associated production process $J/\psi \rightarrow B_{1} \bar{B}_{2}$, we define two parity-conserving and two parity-violating helicity amplitudes, and establish their relationship to the transition form factors of the $B_{1} \bar{B}_{2}$ system. Based on the bilinear covariance of these transition form factors, we derive their transformation properties under $CP$ conservation. Subsequently, we determine the $CP$ constraints on the helicity amplitudes for the conjugate processes $J/\psi \rightarrow B_{1}\bar{B}_{2}$ and $J/\psi \rightarrow B_{2}\bar{B}_{1}$. To describe polarization transfer in these subprocesses, we introduce a polarization transfer matrix $a_{\rho,\mu\nu}$ and provide the complete expression for the polarization correlations of the $B_{1}\bar{B}_{2}$ system. Dedicated observables sensitive to parity and $CP$ violations in baryon production processes are proposed for future experimental investigation.

Due to the weak decay properties of hyperons, their polarization can be extracted experimentally from the angular distributions of their decay products. We express the polarization transfer matrices for both hadronic and radiative hyperon decays in terms of the corresponding decay parameters. Differences in the decay parameters between hyperons and their antiparticles serve as important observables for detecting violations of $CP$ in hyperon decays. By combining the polarization analyzes of the subprocesses $e^{+}e^{-} \rightarrow J/\psi$, $J/\psi \rightarrow B_{1} \bar{B}_{2}$, and subsequent baryon decays, we provide the complete joint angular distribution of all final-state particles. Focusing specifically on the process $J/\psi \rightarrow \Lambda \bar{\Sigma}^{0}$, we present sensitivity projections for the helicity amplitude parameters $\alpha_{1}^{\prime}$ and $\alpha_{2}^{\prime}$, along with their related transition form factors. These results offer essential theoretical guidance for future experimental measurements of $CP$ violations in associated baryon production channels.

At the mass threshold of $J/\psi$, the electron–positron annihilation cross section experiences significant enhancement, enabling high-statistics data samples in experiments. This provides a powerful opportunity to search for $CP$ violations in the production and decay of ordinary baryons through the $e^{+}e^{-} \rightarrow J/\psi \rightarrow B_{1}\bar{B}_{2}$ process. This paper presents a refined polarization analysis of these processes and provides the complete joint angular distribution of the final-state products, offering valuable theoretical support for future $CP$ violation studies involving ordinary baryons.

\acknowledgments{
The authors would like to thank Xu Cao, Shuangshi Fang, Mengjiao Guo, Jianbin Jiao, Yutie Liang, Liang Yan, and Wenjing Zheng for useful discussions. This work is Supported by the National Natural Science Foundation of China (Approvals Nos. 12175117, 12175244, 12247121, 12305085, 12321005, 12405103, and 12447132), the National Key R\&D Program of China No. 2024YFA1611004, and the Shandong Province Natural Science Foundation (Approvals Nos. 2023KJ146 and ZFJH202303).}

\appendix

\section{THE DEFINITION AND PROBABILITY OF THE SPIN COMPONENTS FOR SPIN ONE PARTICLES \label{subsec: SDM_Spin1}}

In the Cartesian coordinate system, the spin density matrix for a spin-1 particle can be expressed as~\cite{Bacchetta:2000jk}:
\begin{align}
\rho_{1}= & \frac{1}{3}\left(\mathbf{1}+\frac{3}{2}S^{i}\Sigma^{i}+3T^{ij}\Sigma^{ij}\right),
\end{align}
where $\Sigma^{i}$ and $\Sigma^{ij}$ represent independent, orthogonal, and Hermitian basis matrices. The matrix $\Sigma^{i}$, in the $S_{z}$ representation, can be written as:
\begin{align}
\Sigma^{x}= \frac{1}{\sqrt{2}}\left(\begin{array}{ccc}
0 & 1 & 0\\
1 & 0 & 1\\
0 & 1 & 0
\end{array}\right),\quad
\Sigma^{y}= \frac{i}{\sqrt{2}}\left(\begin{array}{ccc}
0 & -1 & 0\\
1 & 0 & -1\\
0 & 1 & 0
\end{array}\right),\quad
\Sigma^{z}= \left(\begin{array}{ccc}
1 & 0 & 0\\
0 & 0 & 0\\
0 & 0 & -1
\end{array}\right).
\end{align}
 The remaining basis matrices can be expressed in terms of $\Sigma^{i}$,
\begin{align}
\Sigma^{ij}= & \frac{1}{2}\left(\Sigma^{i}\Sigma^{j}+\Sigma^{j}\Sigma^{i}\right)-\frac{2}{3}\delta^{ij}\mathbf{1}.
\end{align}
For the polarization vector $S^{i}$, there are three independent polarization components, which can be represented as:
\begin{align}
S^{i}= & \left(S_{T}^{x},S_{T}^{y},S_{L}\right).
\end{align}
The definitions of these three polarization components are:
\begin{align}
S_{T}^{x}= & \left\langle \Sigma^{x}\right\rangle ,S_{T}^{y}=\left\langle \Sigma^{y}\right\rangle ,S_{L}=\left\langle \Sigma^{z}\right\rangle ,\label{eq: spin_vect}
\end{align}
where $\left\langle \Sigma^{a}\right\rangle =\text{Tr}\left[\Sigma^{a}\rho_{1}\right]$. For the rank-2 polarization tensor $T^{ij}$, it consists of five independent polarization tensors, expressed as:
\begin{align}
T^{ij}= & \frac{1}{2}\left(\begin{array}{ccc}
-\frac{2}{3}S_{LL}+S_{TT}^{xx} & S_{TT}^{xy} & S_{LT}^{x}\\
S_{TT}^{xy} & -\frac{2}{3}S_{LL}-S_{TT}^{xx} & S_{LT}^{y}\\
S_{LT}^{x} & S_{LT}^{y} & \frac{4}{3}S_{LL}
\end{array}\right).
\end{align}
The definitions of these five polarization components are:
\begin{align}
S_{LL}= & \frac{3}{2}\left\langle \Sigma^{zz}\right\rangle ,S_{LT}^{x}=2\left\langle \Sigma^{xz}\right\rangle ,S_{LT}^{y}=2\left\langle \Sigma^{yz}\right\rangle ,\nonumber \\
S_{TT}^{xx}= & \left\langle \Sigma^{xx}-\Sigma^{yy}\right\rangle ,S_{TT}^{xy}=2\left\langle \Sigma^{xy}\right\rangle .\label{eq: spin_tens}
\end{align}
To analyze the probabilistic interpretation of these polarization components, we introduce the eigenstate vector $\left|m_{\theta,\phi}\right\rangle$ along a specific direction, where $m$ represents the eigenvalue. Therefore, the probability of the system being in a specific polarization state is
\begin{align}
P\left(m_{\left(\theta,\phi\right)}\right)= & \text{Tr}\left[\rho_{1}\left|m_{\left(\theta,\phi\right)}\right\rangle \left\langle m_{\left(\theta,\phi\right)}\right|\right].
\end{align}
For convenience, we introduce the following notation:
\begin{align}
\left|m_{x}\right\rangle = & \left|m_{\left(\frac{\pi}{2},0\right)}\right\rangle ,\left|m_{y}\right\rangle =\left|m_{\left(\frac{\pi}{2},\frac{\pi}{2}\right)}\right\rangle ,\left|m_{z}\right\rangle =\left|m_{\left(0,0\right)}\right\rangle ,\nonumber \\
\left|m_{x+y}\right\rangle = & \left|m_{\left(\frac{\pi}{2},\frac{\pi}{4}\right)}\right\rangle ,\left|m_{x-y}\right\rangle =\left|m_{\left(\frac{\pi}{2},-\frac{\pi}{4}\right)}\right\rangle ,\left|m_{x+z}\right\rangle =\left|m_{\left(\frac{\pi}{4},0\right)}\right\rangle ,\nonumber \\
\left|m_{x-z}\right\rangle = & \left|m_{\left(-\frac{\pi}{4},0\right)}\right\rangle ,\left|m_{y+z}\right\rangle =\left|m_{\left(\frac{\pi}{4},\frac{\pi}{2}\right)}\right\rangle ,\left|m_{y-z}\right\rangle =\left|m_{\left(-\frac{\pi}{4},\frac{\pi}{2}\right)}\right\rangle .
\end{align}
Based on Eqs.~\eqref{eq: spin_vect} and~\eqref{eq: spin_tens}, one can interpret the probability of the polarization components as:
\begin{align}
S_{T}^{x}= & P_{x}\left(1\right)-P_{x}\left(-1\right),\nonumber \\
S_{T}^{y}= & P_{y}\left(1\right)-P_{y}\left(-1\right),\nonumber \\
S_{L}= & P_{z}\left(1\right)-P_{z}\left(-1\right),\nonumber \\
S_{LL}= & \frac{1}{2}P_{z}\left(1\right)+\frac{1}{2}P_{z}\left(-1\right)-P_{z}\left(0\right),\nonumber \\
S_{LT}^{x}= & P_{x-z}\left(0\right)-P_{x+z}\left(0\right),\nonumber \\
S_{LT}^{y}= & P_{y-z}\left(0\right)-P_{y+z}\left(0\right),\nonumber \\
S_{LT}^{xy}= & P_{x-y}\left(0\right)-P_{x+y}\left(0\right),\nonumber \\
S_{TT}^{xx}= & P_{y}\left(0\right)-P_{x}\left(0\right).
\end{align}
Thus, we can derive the threshold values for these polarization components as:
\begin{align}
S_{T}^{x}, & S_{T}^{y},S_{L}\in\left[-1,1\right],\nonumber \\
S_{LL}\in & \left[-1,\frac{1}{2}\right],\nonumber \\
S_{LT}^{x}, & S_{LT}^{y},S_{TT}^{xx},S_{TT}^{xy}\in\left[-1,1\right].
\end{align}
The total degree of polarization is:
\begin{align}
d= & \frac{1}{\sqrt{2s}}\sqrt{\left(2s+1\right)\text{Tr}\left[\rho^{2}\right]-1}\nonumber \\
= & \left[\frac{3}{4}S^{i}S^{i}+\frac{3}{2}T^{ij}T^{ij}\right]^{1/2}\nonumber \\
= & \left\{ \frac{3}{4}\left[(S_{T}^{x})^{2}+(S_{T}^{y})^{2}+S_{L}^{2}\right]+\frac{3}{2}\left[\frac{2}{3}S_{LL}^{2}\right.\right.\nonumber \\
 & \left.\left.+\frac{1}{2}\left((S_{LT}^{x})^{2}+(S_{LT}^{y})^{2}+(S_{TT}^{xx})^{2}+(S_{TT}^{xy})^{2}\right)\right]\right\} ^{1/2}.
\end{align}

\section{SPIN-ONE POLARIZATION PROJECTION MATRICS\label{subsec: pro_matr}}

Following the approach outlined in~\cite{Zhang:2024rbl}, we map the polarization expansion coefficients $S_{0}$,$S_{1}$,...,$S_{8}$ in the helicity framework for a spin-1 particle to the Cartesian spin components $1$,$S_{L}$,...,$S_{TT}^{xy}$, as detailed in Eq.~\eqref{eq: Smu_1}. In this scheme, the polarization expansion matrices in the helicity framework are chosen as follows:
\begin{align}
\Sigma_{0}= & \frac{1}{3}\mathbf{I},\Sigma_{1}=\frac{1}{2}\Sigma^{z},\Sigma_{2}=\frac{1}{2}\Sigma^{x},\nonumber \\
\Sigma_{3}= & \frac{1}{2}\Sigma^{y},\Sigma_{4}=\Sigma^{zz},\Sigma_{5}=\Sigma^{xz},\nonumber \\
\Sigma_{6}= & \Sigma^{yz},\Sigma_{7}=\frac{1}{2}\left(\Sigma^{xx}-\Sigma^{yy}\right),\Sigma_{8}=\Sigma^{xy}.
\end{align}
The explicit expressions for these matrices are:
\begin{align}
\Sigma_{0}= & \frac{1}{3}\left(\begin{array}{ccc}
1 & 0 & 0\\
0 & 1 & 0\\
0 & 0 & 1
\end{array}\right),\Sigma_{1}=\frac{1}{2}\left(\begin{array}{ccc}
1 & 0 & 0\\
0 & 0 & 0\\
0 & 0 & -1
\end{array}\right),\Sigma_{2}=\frac{\sqrt{2}}{4}\left(\begin{array}{ccc}
0 & 1 & 0\\
1 & 0 & 1\\
0 & 1 & 0
\end{array}\right),\nonumber \\
\Sigma_{3}= & \frac{i\sqrt{2}}{4}\left(\begin{array}{ccc}
0 & -1 & 0\\
1 & 0 & -1\\
0 & 1 & 0
\end{array}\right),\Sigma_{4}=\frac{1}{3}\left(\begin{array}{ccc}
1 & 0 & 0\\
0 & -2 & 0\\
0 & 0 & 1
\end{array}\right),\Sigma_{5}=\frac{1}{2\sqrt{2}}\left(\begin{array}{ccc}
0 & 1 & 0\\
1 & 0 & -1\\
0 & -1 & 0
\end{array}\right),\nonumber \\
\Sigma_{6}= & \frac{i}{2\sqrt{2}}\left(\begin{array}{ccc}
0 & -1 & 0\\
1 & 0 & 1\\
0 & -1 & 0
\end{array}\right),\Sigma_{7}=\frac{1}{2}\left(\begin{array}{ccc}
0 & 0 & 1\\
0 & 0 & 0\\
1 & 0 & 0
\end{array}\right),\Sigma_{8}=\frac{i}{2}\left(\begin{array}{ccc}
0 & 0 & -1\\
0 & 0 & 0\\
1 & 0 & 0
\end{array}\right).
\end{align}

\section{SPIN TRANSFER MATRIX FOR $J/\psi\rightarrow B\bar{B}$ \label{subsec: spin_trans}}

Based on Eq.~\eqref{eq: Jpsi_decay}, we present the complete expression for the polarization transfer matrix in the $J/\psi\rightarrow B\bar{B}$ subprocess in this section.

The components of the polarization transfer matrix associated with the production cross section $S_{0}$ of the $J/\psi$ are given by
\begin{align}
a_{0,00}= & \frac{1}{3}\left(3+\alpha_{\psi}+2\alpha_{1}+4\alpha_{2}\right),\nonumber \\
a_{0,30}= & \frac{2\sqrt{2}}{3}\left(D_{2}^{c}+2D_{3}^{c}\right),\nonumber \\
a_{0,11}= & \frac{1}{3}\left(1-2\alpha_{1}-\alpha_{\psi}\right),\nonumber \\
a_{0,21}= & -\frac{2\sqrt{2}}{3}D_{2}^{s},\nonumber \\
a_{0,12}= & -\frac{2\sqrt{2}}{3}D_{2}^{s},\nonumber \\
a_{0,22}= & -\frac{1}{3}\left(1-2\alpha_{1}-\alpha_{\psi}\right),\nonumber \\
a_{0,03}= & \frac{2\sqrt{2}}{3}\left(D_{2}^{c}-2D_{3}^{c}\right),\nonumber \\
a_{0,33}= & -\frac{1}{3}\left(1-2\alpha_{1}+4\alpha_{2}+3\alpha_{\psi}\right).\label{eq: Jpsi_decay_1}
\end{align}
The components of the polarization transfer matrix associated with the vector polarization component $S_{1}\left(S_{L}\right)$ are:
\begin{align}
a_{1,00}= & 2\sqrt{2}D_{3}^{c}\cos\theta,\nonumber \\
a_{1,10}= & -\left(D_{0}^{c}-2D_{5}^{c}\right)\sin\theta,\nonumber \\
a_{1,20}= & \sqrt{2}\left(D_{1}^{s}+D_{4}^{s}\right)\sin\theta,\nonumber \\
a_{1,30}= & \left(1+2\alpha_{2}+\alpha_{\psi}\right)\cos\theta,\nonumber \\
a_{1,01}= & -\left(D_{0}^{c}+2D_{5}^{c}\right)\sin\theta,\nonumber \\
a_{1,31}= & -\sqrt{2}\left(D_{1}^{c}+D_{4}^{c}\right)\sin\theta,\nonumber \\
a_{1,02}= & \sqrt{2}\left(D_{1}^{s}-D_{4}^{s}\right)\sin\theta,\nonumber \\
a_{1,32}= & \left(D_{0}^{s}+2D_{5}^{s}\right)\sin\theta,\nonumber \\
a_{1,03}= & -\left(1+\alpha_{\psi}+2\alpha_{2}\right)\cos\theta,\nonumber \\
a_{1,13}= & -\sqrt{2}\left(D_{1}^{c}-D_{4}^{c}\right)\sin\theta,\nonumber \\
a_{1,23}= & \left(D_{0}^{s}-2D_{5}^{s}\right)\sin\theta,\nonumber \\
a_{1,33}= & -2\sqrt{2}D_{3}^{c}\cos\theta.
\end{align}
 
The components of the polarization transfer matrix associated with the vector polarization component $S_{2}\left(S_{T}^{x}\right)$ are:
\begin{align}
a_{2,00}= & 2\sqrt{2}D_{3}^{c}\sin\theta\cos\phi,\nonumber \\
a_{2,10}= & \left(D_{0}^{c}-2D_{5}^{c}\right)\cos\theta\cos\phi-\sqrt{2}\left(D_{1}^{s}+D_{4}^{s}\right)\sin\phi,\nonumber \\
a_{2,20}= & -\left(D_{0}^{c}-2D_{5}^{c}\right)\sin\phi-\sqrt{2}\left(D_{1}^{s}+D_{4}^{s}\right)\cos\theta\cos\phi,\nonumber \\
a_{2,30}= & \left(1+\alpha_{\psi}+2\alpha_{2}\right)\sin\theta\cos\phi,\nonumber \\
a_{2,01}= & \left(D_{0}^{c}+2D_{5}^{c}\right)\cos\theta\cos\phi+\sqrt{2}\left(D_{1}^{s}-D_{4}^{s}\right)\sin\phi,\nonumber \\
a_{2,31}= & \left(D_{0}^{s}+2D_{5}^{s}\right)\sin\phi+\sqrt{2}\left(D_{1}^{c}+D_{4}^{c}\right)\cos\theta\cos\phi,\nonumber \\
a_{2,02}= & \left(D_{0}^{c}+2D_{5}^{c}\right)\sin\phi-\sqrt{2}\left(D_{1}^{s}-D_{4}^{s}\right)\cos\theta\cos\phi,\nonumber \\
a_{2,32}= & -\left(D_{0}^{s}+2D_{5}^{s}\right)\cos\theta\cos\phi+\sqrt{2}\left(D_{1}^{c}+D_{4}^{c}\right)\sin\phi,\nonumber \\
a_{2,03}= & -\left(1+\alpha_{\psi}+2\alpha_{2}\right)\sin\theta\cos\phi,\nonumber \\
a_{2,13}= & -\left(D_{0}^{s}-2D_{5}^{s}\right)\sin\phi+\sqrt{2}\left(D_{1}^{c}-D_{4}^{c}\right)\cos\theta\cos\phi,\nonumber \\
a_{2,23}= & -\left(D_{0}^{s}-2D_{5}^{s}\right)\cos\theta\cos\phi-\sqrt{2}\left(D_{1}^{c}-D_{4}^{c}\right)\sin\phi,\nonumber \\
a_{2,33}= & -2\sqrt{2}D_{3}^{c}\sin\theta\cos\phi.
\end{align}

The components of the polarization transfer matrix associated with the vector polarization component $S_{3}\left(S_{T}^{y}\right)$ are:
\begin{align}
a_{3,00}= & 2\sqrt{2}D_{3}^{c}\sin\theta\sin\phi,\nonumber \\
a_{3,10}= & \left(D_{0}^{c}-2D_{5}^{c}\right)\cos\theta\sin\phi+\sqrt{2}\left(D_{1}^{s}+D_{4}^{s}\right)\cos\phi,\nonumber \\
a_{3,20}= & \left(D_{0}^{c}-2D_{5}^{c}\right)\cos\phi-\sqrt{2}\left(D_{1}^{s}+D_{4}^{s}\right)\cos\theta\sin\phi,\nonumber \\
a_{3,30}= & \left(1+2\alpha_{2}+\alpha_{\psi}\right)\sin\theta\sin\phi,\nonumber \\
a_{3,01}= & \left(D_{0}^{c}+2D_{5}^{c}\right)\cos\theta\sin\phi-\sqrt{2}\left(D_{1}^{s}-D_{4}^{s}\right)\cos\phi,\nonumber \\
a_{3,31}= & -\left(D_{0}^{s}+2D_{5}^{s}\right)\cos\phi+\sqrt{2}\left(D_{1}^{c}+D_{4}^{c}\right)\cos\theta\sin\phi,\nonumber \\
a_{3,02}= & -\left(D_{0}^{c}+2D_{5}^{c}\right)\cos\phi-\sqrt{2}\left(D_{1}^{s}-D_{4}^{s}\right)\cos\theta\sin\phi,\nonumber \\
a_{3,32}= & -\left(D_{0}^{s}+2D_{5}^{s}\right)\cos\theta\sin\phi-\sqrt{2}\left(D_{1}^{c}+D_{4}^{c}\right)\cos\phi,\nonumber \\
a_{3,03}= & -\left(1+2\alpha_{2}+\alpha_{\psi}\right)\sin\theta\sin\phi,\nonumber \\
a_{3,13}= & \left(D_{0}^{s}-2D_{5}^{s}\right)\cos\phi+\sqrt{2}\left(D_{1}^{c}-D_{4}^{c}\right)\cos\theta\sin\phi,\nonumber \\
a_{3,23}= & -\left(D_{0}^{s}-2D_{5}^{s}\right)\cos\theta\sin\phi+\sqrt{2}\left(D_{1}^{c}-D_{4}^{c}\right)\cos\phi,\nonumber \\
a_{3,33}= & -2\sqrt{2}D_{3}^{c}\sin\theta\sin\phi.
\end{align}

The components of the polarization transfer matrix associated with the tensor polarization component $S_{4}\left(S_{LL}\right)$ are:
\begin{align}
a_{4,00}= & \frac{1}{3}\left(\alpha_{\psi}-\alpha_{1}+\alpha_{2}\right)\left(1+3\cos2\theta\right),\nonumber \\
a_{4,10}= & \sqrt{2}\left(D_{1}^{c}-D_{4}^{c}\right)\sin2\theta,\nonumber \\
a_{4,20}= & -\left(D_{0}^{s}-2D_{5}^{s}\right)\sin2\theta,\nonumber \\
a_{4,30}= & -\frac{\sqrt{2}}{3}\left(D_{2}^{c}-D_{3}^{c}\right)\left(1+3\cos2\theta\right),\nonumber \\
a_{4,01}= & -\sqrt{2}\left(D_{1}^{c}+D_{4}^{c}\right)\sin2\theta,\nonumber \\
a_{4,11}= & -\frac{2}{3}\left(1-\alpha_{\psi}-2\alpha_{1}\right)+2\sin^{2}\theta\left(1-\alpha_{1}-\alpha_{2}\right),\nonumber \\
a_{4,21}= & \frac{\sqrt{2}}{3}D_{2}^{s}\left(1+3\cos2\theta\right)-2\sqrt{2}D_{3}^{s}\sin^{2}\theta,\nonumber \\
a_{4,31}= & -\left(D_{0}^{c}+2D_{5}^{c}\right)\sin2\theta,\\
a_{4,02}= & \left(D_{0}^{s}+2D_{5}^{s}\right)\sin2\theta,\nonumber \\
a_{4,12}= & 2\sqrt{2}D_{3}^{s}\sin^{2}\theta+\frac{\sqrt{2}}{3}D_{2}^{s}\left(1+3\cos2\theta\right),\nonumber \\
a_{4,22}= & \frac{2}{3}\left(1-\alpha_{\psi}-2\alpha_{1}\right)+2\left(\alpha_{\psi}+\alpha_{1}-\alpha_{2}\right)\sin^{2}\theta,\nonumber \\
a_{4,32}= & \sqrt{2}\left(D_{1}^{s}-D_{4}^{s}\right)\sin2\theta,\nonumber \\
a_{4,03}= & -\frac{\sqrt{2}}{3}\left(D_{2}^{c}+D_{3}^{c}\right)\left(1+3\cos2\theta\right),\nonumber \\
a_{4,13}= & \left(D_{0}^{c}-2D_{5}^{c}\right)\sin2\theta,\nonumber \\
a_{4,23}= & -\sqrt{2}\left(D_{1}^{s}+D_{4}^{s}\right)\sin2\theta,\nonumber \\
a_{4,33}= & -\frac{1}{3}\left(1+\alpha_{1}+\alpha_{2}\right)\left(1+3\cos2\theta\right).
\end{align}

The components of the polarization transfer matrix associated with the tensor polarization component $S_{5}\left(S_{LT}^{x}\right)$ are:
\begin{align}
a_{5,00}= & \left(\alpha_{\psi}-\alpha_{1}+\alpha_{2}\right)\sin2\theta\cos\phi,\nonumber \\
a_{5,10}= & \left(D_{0}^{s}-2D_{5}^{s}\right)\cos\theta\sin\phi-\sqrt{2}\left(D_{1}^{c}-D_{4}^{c}\right)\cos2\theta\cos\phi,\nonumber \\
a_{5,20}= & \left(D_{0}^{s}-2D_{5}^{s}\right)\cos2\theta\cos\phi+\sqrt{2}\left(D_{1}^{c}-D_{4}^{c}\right)\cos\theta\sin\phi,\nonumber \\
a_{5,30}= & -\sqrt{2}\left(D_{2}^{c}-D_{3}^{c}\right)\sin2\theta\cos\phi,\nonumber \\
a_{5,01}= & \left(D_{0}^{s}+2D_{5}^{s}\right)\cos\theta\sin\phi+\sqrt{2}\left(D_{1}^{c}+D_{4}^{c}\right)\cos2\theta\cos\phi,\nonumber \\
a_{5,11}= & -\left(1-\alpha_{1}-\alpha_{2}\right)\sin2\theta\cos\phi+2\sqrt{2}D_{3}^{s}\sin\theta\sin\phi,\nonumber \\
a_{5,21}= & \left(1+\alpha_{\psi}-2\alpha_{2}\right)\sin\theta\sin\phi+\sqrt{2}\left(D_{2}^{s}+D_{3}^{s}\right)\sin2\theta\cos\phi,\nonumber \\
a_{5,31}= & \left(D_{0}^{c}+2D_{5}^{c}\right)\cos2\theta\cos\phi+\sqrt{2}\left(D_{1}^{s}-D_{4}^{s}\right)\cos\theta\sin\phi,\nonumber \\
a_{5,02}= & -\left(D_{0}^{s}+2D_{5}^{s}\right)\cos2\theta\cos\phi+\sqrt{2}\left(D_{1}^{c}+D_{4}^{c}\right)\cos\theta\sin\phi,\nonumber \\
a_{5,12}= & -\left(1+\alpha_{\psi}-2\alpha_{2}\right)\sin\theta\sin\phi+\sqrt{2}\left(D_{2}^{s}-D_{3}^{s}\right)\sin2\theta\cos\phi,\nonumber \\
a_{5,22}= & -\left(\alpha_{\psi}+\alpha_{1}-\alpha_{2}\right)\sin2\theta\cos\phi+2\sqrt{2}D_{3}^{s}\sin\theta\sin\phi,\nonumber \\
a_{5,32}= & \left(D_{0}^{c}+2D_{5}^{c}\right)\cos\theta\sin\phi-\sqrt{2}\left(D_{1}^{s}-D_{4}^{s}\right)\cos2\theta\cos\phi,\nonumber \\
a_{5,03}= & -\sqrt{2}\left(D_{2}^{c}+D_{3}^{c}\right)\sin2\theta\cos\phi,\nonumber \\
a_{5,13}= & -\left(D_{0}^{c}-2D_{5}^{c}\right)\cos2\theta\cos\phi+\sqrt{2}\left(D_{1}^{s}+D_{4}^{s}\right)\cos\theta\sin\phi,\nonumber \\
a_{5,23}= & \left(D_{0}^{c}-2D_{5}^{c}\right)\cos\theta\sin\phi+\sqrt{2}\left(D_{1}^{s}+D_{4}^{s}\right)\cos2\theta\cos\phi,\nonumber \\
a_{5,33}= & -\left(1+\alpha_{1}+\alpha_{2}\right)\sin2\theta\cos\phi.
\end{align}

The components of the polarization transfer matrix associated with the tensor polarization component $S_{6}\left(S_{LT}^{y}\right)$ are:
\begin{align}
a_{6,00}= & \left(\alpha_{\psi}-\alpha_{1}+\alpha_{2}\right)\sin2\theta\sin\phi,\nonumber \\
a_{6,10}= & -\left(D_{0}^{s}-2D_{5}^{s}\right)\cos\theta\cos\phi-\sqrt{2}\left(D_{1}^{c}-D_{4}^{c}\right)\cos2\theta\sin\phi,\nonumber \\
a_{6,20}= & \left(D_{0}^{s}-2D_{5}^{s}\right)\cos2\theta\sin\phi-\sqrt{2}\left(D_{1}^{c}-D_{4}^{c}\right)\cos\theta\cos\phi,\nonumber \\
a_{6,30}= & -\sqrt{2}\left(D_{2}^{c}-D_{3}^{c}\right)\sin2\theta\sin\phi,\nonumber \\
a_{6,01}= & -\left(D_{0}^{s}+2D_{5}^{s}\right)\cos\theta\cos\phi+\sqrt{2}\left(D_{1}^{c}+D_{4}^{c}\right)\cos2\theta\sin\phi,\nonumber \\
a_{6,11}= & -\left(1-\alpha_{1}-\alpha_{2}\right)\sin2\theta\sin\phi-2\sqrt{2}D_{3}^{s}\sin\theta\cos\phi,\nonumber \\
a_{6,21}= & -\left(1+\alpha_{\psi}-2\alpha_{2}\right)\sin\theta\cos\phi+\sqrt{2}(D_{2}^{s}+D_{3}^{s})\sin2\theta\sin\phi,\nonumber \\
a_{6,31}= & \left(D_{0}^{c}+2D_{5}^{c}\right)\cos2\theta\sin\phi-\sqrt{2}\left(D_{1}^{s}-D_{4}^{s}\right)\cos\theta\cos\phi,\nonumber \\
a_{6,02}= & -\left(D_{0}^{s}+2D_{5}^{s}\right)\cos2\theta\sin\phi-\sqrt{2}\left(D_{1}^{c}+D_{4}^{c}\right)\cos\theta\cos\phi,\nonumber \\
a_{6,12}= & \left(1+\alpha_{\psi}-2\alpha_{2}\right)\sin\theta\cos\phi+\sqrt{2}\left(D_{2}^{s}-D_{3}^{s}\right)\sin2\theta\sin\phi,\nonumber \\
a_{6,22}= & -\left(\alpha_{\psi}+\alpha_{1}-\alpha_{2}\right)\sin2\theta\sin\phi-2\sqrt{2}D_{3}^{s}\sin\theta\cos\phi,\nonumber \\
a_{6,32}= & -\left(D_{0}^{c}+2D_{5}^{c}\right)\cos\theta\cos\phi-\sqrt{2}\left(D_{1}^{s}-D_{4}^{s}\right)\cos2\theta\sin\phi,\nonumber \\
a_{6,03}= & -\sqrt{2}\left(D_{2}^{c}+D_{3}^{c}\right)\sin2\theta\sin\phi,\nonumber \\
a_{6,13}= & -\left(D_{0}^{c}-2D_{5}^{c}\right)\cos2\theta\sin\phi-\sqrt{2}\left(D_{1}^{s}+D_{4}^{s}\right)\cos\theta\cos\phi,\nonumber \\
a_{6,23}= & -\left(D_{0}^{c}-2D_{5}^{c}\right)\cos\theta\cos\phi+\sqrt{2}\left(D_{1}^{s}+D_{4}^{s}\right)\cos2\theta\sin\phi,\nonumber \\
a_{6,33}= & -\left(1+\alpha_{1}+\alpha_{2}\right)\sin2\theta\sin\phi.
\end{align}

The components of the polarization transfer matrix associated with the tensor polarization component $S_{7}\left(S_{TT}^{xx}\right)$ are:
\begin{align}
a_{7,00}= & \left(\alpha_{\psi}-\alpha_{1}+\alpha_{2}\right)\sin^{2}\theta\cos2\phi,\nonumber \\
a_{7,10}= & \left(D_{0}^{s}-2D_{5}^{s}\right)\sin\theta\sin2\phi-\frac{\sqrt{2}}{2}\left(D_{1}^{c}-D_{4}^{c}\right)\sin2\theta\cos2\phi,\nonumber \\
a_{7,20}= & \frac{1}{2}\left(D_{0}^{s}-2D_{5}^{s}\right)\sin2\theta\cos2\phi+\sqrt{2}\left(D_{1}^{c}-D_{4}^{c}\right)\sin\theta\sin2\phi,\nonumber \\
a_{7,30}= & -\sqrt{2}\left(D_{2}^{c}-D_{3}^{c}\right)\sin^{2}\theta\cos2\phi,\nonumber \\
a_{7,01}= & \left(D_{0}^{s}+2D_{5}^{s}\right)\sin\theta\sin2\phi+\frac{\sqrt{2}}{2}\left(D_{1}^{c}+D_{4}^{c}\right)\sin2\theta\cos2\phi,\nonumber \\
a_{7,11}= & \left[\left(1+\alpha_{\psi}-2\alpha_{2}\right)-\left(1-\alpha_{1}-\alpha_{2}\right)\sin^{2}\theta\right]\cos2\phi-2\sqrt{2}D_{3}^{s}\cos\theta\sin2\phi,\\
a_{7,21}= & -\left(1+\alpha_{\psi}-2\alpha_{2}\right)\cos\theta\sin2\phi+\frac{\sqrt{2}}{2}\left[2D_{2}^{s}\sin^{2}\theta-D_{3}^{s}\left(3+\cos2\theta\right)\right]\cos2\phi,\nonumber \\
a_{7,31}= & \frac{1}{2}\left(D_{0}^{c}+2D_{5}^{c}\right)\sin2\theta\cos2\phi+\sqrt{2}\left(D_{1}^{s}-D_{4}^{s}\right)\sin\theta\sin2\phi,\\
a_{7,02}= & -\frac{1}{2}\left(D_{0}^{s}+2D_{5}^{s}\right)\sin2\theta\cos2\phi+\sqrt{2}\left(D_{1}^{c}+D_{4}^{c}\right)\sin\theta\sin2\phi,\nonumber \\
a_{7,12}= & \left(1+\alpha_{\psi}-2\alpha_{2}\right)\cos\theta\sin2\phi+\frac{\sqrt{2}}{2}\left[2D_{2}^{s}\sin^{2}\theta+D_{3}^{s}\left(3+\cos2\theta\right)\right]\cos2\phi,\nonumber \\
a_{7,22}= & \left[\left(1+\alpha_{\psi}-2\alpha_{2}\right)-\left(\alpha_{\psi}+\alpha_{1}-\alpha_{2}\right)\sin^{2}\theta\right]\cos2\phi-2\sqrt{2}D_{3}^{s}\cos\theta\sin2\phi,\\
a_{7,32}= & \left(D_{0}^{c}+2D_{5}^{c}\right)\sin\theta\sin2\phi-\frac{\sqrt{2}}{2}\left(D_{1}^{s}-D_{4}^{s}\right)\sin2\theta\cos2\phi,\nonumber \\
a_{7,03}= & -\sqrt{2}\left(D_{2}^{c}+D_{3}^{c}\right)\sin^{2}\theta\cos2\phi,\nonumber \\
a_{7,13}= & -\frac{1}{2}\left(D_{0}^{c}-2D_{5}^{c}\right)\sin2\theta\cos2\phi+\sqrt{2}\left(D_{1}^{s}+D_{4}^{s}\right)\sin\theta\sin2\phi,\nonumber \\
a_{7,23}= & \frac{\sqrt{2}}{2}\left(D_{1}^{s}+D_{4}^{s}\right)\sin2\theta\cos2\phi+\left(D_{0}^{c}-2D_{5}^{c}\right)\sin\theta\sin2\phi,\nonumber \\
a_{7,33}= & -\left(1+\alpha_{1}+\alpha_{2}\right)\sin^{2}\theta\cos2\phi.
\end{align}

The components of the polarization transfer matrix associated with the tensor polarization component $S_{8}\left(S_{TT}^{xy}\right)$ are:
\begin{align}
a_{8,00}= & \left(\alpha_{\psi}-\alpha_{1}+\alpha_{2}\right)\sin^{2}\theta\sin2\phi,\nonumber \\
a_{8,10}= & -\left(D_{0}^{s}-2D_{5}^{s}\right)\sin\theta\cos2\phi-\frac{\sqrt{2}}{2}\left(D_{1}^{c}-D_{4}^{c}\right)\sin2\theta\sin2\phi,\nonumber \\
a_{8,20}= & \frac{1}{2}\left(D_{0}^{s}-2D_{5}^{s}\right)\sin2\theta\sin2\phi-\sqrt{2}\left(D_{1}^{c}-D_{4}^{c}\right)\sin\theta\cos2\phi,\nonumber \\
a_{8,30}= & -\sqrt{2}\left(D_{2}^{c}-D_{3}^{c}\right)\sin^{2}\theta\sin2\phi,\nonumber \\
a_{8,01}= & \frac{\sqrt{2}}{2}\left(D_{1}^{c}+D_{4}^{c}\right)\sin2\theta\sin2\phi-\left(D_{0}^{s}+2D_{5}^{s}\right)\sin\theta\cos2\phi,\nonumber \\
a_{8,11}= & \left[\left(1+\alpha_{\psi}-2\alpha_{2}\right)-\left(1-\alpha_{1}-\alpha_{2}\right)\sin^{2}\theta\right]\sin2\phi+2\sqrt{2}D_{3}^{s}\cos\theta\cos2\phi,\nonumber \\
a_{8,21}= & \left(1+\alpha_{\psi}-2\alpha_{2}\right)\cos\theta\cos2\phi+\frac{\sqrt{2}}{2}\left[2D_{2}^{s}\sin^{2}\theta-D_{3}^{s}\left(3+\cos2\theta\right)\right]\sin2\phi,\nonumber \\
a_{8,31}= & \frac{1}{2}\left(D_{0}^{c}+2D_{5}^{c}\right)\sin2\theta\sin2\phi-\sqrt{2}\left(D_{1}^{s}-D_{4}^{s}\right)\sin\theta\cos2\phi,\\
a_{8,02}= & -\frac{1}{2}\left(D_{0}^{s}+2D_{5}^{s}\right)\sin2\theta\sin2\phi-\sqrt{2}\left(D_{1}^{c}+D_{4}^{c}\right)\sin\theta\cos2\phi,\nonumber \\
a_{8,12}= & -\left(1+\alpha_{\psi}-2\alpha_{2}\right)\cos\theta\cos2\phi+\frac{\sqrt{2}}{2}\left[2D_{2}^{s}\sin^{2}\theta+D_{3}^{s}\left(3+\cos2\theta\right)\right]\sin2\phi,\nonumber \\
a_{8,22}= & \left[\left(1+\alpha_{\psi}-2\alpha_{2}\right)-\left(\alpha_{\psi}+\alpha_{1}-\alpha_{2}\right)\sin^{2}\theta\right]\sin2\phi+2\sqrt{2}D_{3}^{s}\cos\theta\cos2\phi,\\
a_{8,32}= & -\left(D_{0}^{c}+2D_{5}^{c}\right)\sin\theta\cos2\phi-\frac{\sqrt{2}}{2}\left(D_{1}^{s}-D_{4}^{s}\right)\sin2\theta\sin2\phi,\nonumber \\
a_{8,03}= & -\sqrt{2}\left(D_{2}^{c}+D_{3}^{c}\right)\sin^{2}\theta\sin2\phi,\nonumber \\
a_{8,13}= & -\frac{1}{2}\left(D_{0}^{c}-2D_{5}^{c}\right)\sin2\theta\sin2\phi-\sqrt{2}\left(D_{1}^{s}+D_{4}^{s}\right)\sin\theta\cos2\phi,\nonumber \\
a_{8,23}= & -\left(D_{0}^{c}-2D_{5}^{c}\right)\sin\theta\cos2\phi+\frac{\sqrt{2}}{2}\left(D_{1}^{s}+D_{4}^{s}\right)\sin2\theta\sin2\phi,\nonumber \\
a_{8,33}= & -\left(1+\alpha_{1}+\alpha_{2}\right)\sin^{2}\theta\sin2\phi.\label{eq: Jpsi_decay_2}
\end{align}
These polarization transfer matrices provide a complete description of polarization transfer in the $J/\psi\rightarrow B\bar{B}$ process. Experimentally, the polarization information of $J/\psi$ can be obtained by analyzing the polarization and angular distributions of $B\bar{B}$.

\section{CONVENTION FOR THE SPIN-1/2 AND SPIN-1 SPINORS}

In this section, we provide the conventions for the spin-1/2 and spin-1 field spinors in the Dirac-Pauli representation. For a particle with mass $m$, the four-momentum in spherical coordinates can be expressed as:
\begin{align}
p^{\mu}= & \left(E,\left|\vec{p}\right|\sin\theta\cos\phi,\left|\vec{p}\right|\sin\theta\sin\phi,\left|\vec{p}\right|\cos\theta\right).
\end{align}
For a spin-1/2 field, the Dirac equation must be satisfied~\cite{Dirac:1928hu}. The spinor field with $\lambda=\pm1$ can be written as:
\begin{align}
u\left(p,\pm\right)= & \left(\begin{array}{c}
\sqrt{E+m}\chi_{\pm}\\
\pm\sqrt{E-m}\chi_{\pm}
\end{array}\right),\nonumber \\
v\left(p,\pm\right)= & \left(\begin{array}{c}
\sqrt{E-m}\chi_{\mp}\\
\mp\sqrt{E+m}\chi_{\mp}
\end{array}\right),
\end{align}
where $\chi_{\pm}$ represents the two-component spinor,
\begin{align}
\chi_{+}= & \left(\begin{array}{c}
\cos\frac{\theta}{2}\\
\sin\frac{\theta}{2}e^{i\phi}
\end{array}\right),\nonumber \\
\chi_{-}= & \left(\begin{array}{c}
-\sin\frac{\theta}{2}\\
\cos\frac{\theta}{2}e^{i\phi}
\end{array}\right).
\end{align}
For a spin-1 particle, the field must satisfy the Proca equation~\cite{Proca:1936fbw}. The polarization vector with $\lambda=\pm1,0$ can be expressed as:
\begin{align}
\varepsilon^{\mu}\left(p,\pm\right)= & \left(\begin{array}{c}
0\\
\mp\cos\theta\cos\phi+i\sin\phi\\
\mp\cos\theta\sin\phi-i\cos\phi\\
\pm\sin\theta
\end{array}\right),\nonumber \\
\varepsilon^{\mu}\left(p,0\right)= & \frac{1}{m}\left(\begin{array}{c}
\left|\vec{p}\right|\\
E\sin\theta\cos\phi\\
E\sin\theta\sin\phi\\
E\cos\theta
\end{array}\right).
\end{align}


\begin{thebibliography}{0}%
\makeatletter
\providecommand \@ifxundefined [1]{%
 \@ifx{#1\undefined}
}%
\providecommand \@ifnum [1]{%
 \ifnum #1\expandafter \@firstoftwo
 \else \expandafter \@secondoftwo
 \fi
}%
\providecommand \@ifx [1]{%
 \ifx #1\expandafter \@firstoftwo
 \else \expandafter \@secondoftwo
 \fi
}%
\providecommand \natexlab [1]{#1}%
\providecommand \enquote  [1]{``#1''}%
\providecommand \bibnamefont  [1]{#1}%
\providecommand \bibfnamefont [1]{#1}%
\providecommand \citenamefont [1]{#1}%
\providecommand \href@noop [0]{\@secondoftwo}%
\providecommand \href [0]{\begingroup \@sanitize@url \@href}%
\providecommand \@href[1]{\@@startlink{#1}\@@href}%
\providecommand \@@href[1]{\endgroup#1\@@endlink}%
\providecommand \@sanitize@url [0]{\catcode `\\12\catcode `\$12\catcode `\&12\catcode `\#12\catcode `\^12\catcode `\_12\catcode `\%12\relax}%
\providecommand \@@startlink[1]{}%
\providecommand \@@endlink[0]{}%
\providecommand \url  [0]{\begingroup\@sanitize@url \@url }%
\providecommand \@url [1]{\endgroup\@href {#1}{\urlprefix }}%
\providecommand \urlprefix  [0]{URL }%
\providecommand \Eprint [0]{\href }%
\providecommand \doibase [0]{http://dx.doi.org/}%
\providecommand \selectlanguage [0]{\@gobble}%
\providecommand \bibinfo  [0]{\@secondoftwo}%
\providecommand \bibfield  [0]{\@secondoftwo}%
\providecommand \translation [1]{[#1]}%
\providecommand \BibitemOpen [0]{}%
\providecommand \bibitemStop [0]{}%
\providecommand \bibitemNoStop [0]{.\EOS\space}%
\providecommand \EOS [0]{\spacefactor3000\relax}%
\providecommand \BibitemShut  [1]{\csname bibitem#1\endcsname}%
\let\auto@bib@innerbib\@empty
\end{thebibliography}%


\begin{thebibliography}{}
\bibitem{Lee:1956qn}
T.~D.~Lee and C.~N.~Yang,
Question of parity conservation in weak interactions,
\href{https://doi.org/10.1103/PhysRev.104.254}{Phys. Rev. \textbf{104}, 254 (1956)}.

\bibitem{Wu:1957my}
C.~S.~Wu, E.~Ambler, R.~W.~Hayward, D.~D.~Hoppes and R.~P.~Hudson,
Experimental test of parity conservation in $\beta$ decay,
\href{https://doi.org/10.1103/PhysRev.105.1413}{Phys. Rev. \textbf{105}, 1413 (1957)}.

\bibitem{Garwin:1957hc}
R.~L.~Garwin, L.~M.~Lederman and M.~Weinrich,
Observations of the failure of conservation of parity and charge conjugation in meson decays: the magnetic moment of the free muon,
\href{https://doi.org/10.1103/PhysRev.105.1415}{Phys. Rev. \textbf{105}, 1415 (1957)}.

\bibitem{Friedman:1957mz}
J.~I.~Friedman and V.~L.~Telegdi,
Nuclear emulsion evidence for parity nonconservation in the secay chain $\pi^+ \to \mu^+ \to e^+$,
\href{https://doi.org/10.1103/PhysRev.106.1290}{Phys. Rev. \textbf{106}, 1290 (1957)}.

\bibitem{Sakharov:1967dj}
A.~D.~Sakharov,
Violation of $CP$ invariance, $C$ asymmetry, and baryon asymmetry of the universe,
\href{https://doi.org/10.1070/PU1991v034n05ABEH002497}{Pisma Zh. Eksp. Teor. Fiz. \textbf{5}, 32 (1967)}.

\bibitem{Christenson:1964fg}
J.~H.~Christenson, J.~W.~Cronin, V.~L.~Fitch and R.~Turlay,
Evidence for the $2\pi$ decay of the $K_2^0$ meson,
\href{https://doi.org/10.1103/PhysRevLett.13.138}{Phys. Rev. Lett. \textbf{13}, 138 (1964)}.

\bibitem{BaBar:2001pki}
B.~Aubert \textit{et al.} ($BABAR$ Collaboration),
Observation of $CP$ violation in the $B^0$ meson system,
\href{https://doi.org/10.1103/PhysRevLett.87.091801}{Phys. Rev. Lett. \textbf{87}, 091801 (2001)},
[\href{https://arxiv.org/abs/hep-ex/0107013}{\tt arXiv:hep-ex/0107013}].

\bibitem{Belle:2001zzw}
K.~Abe \textit{et al.} (Belle Collaboration),
Observation of large $CP$ violation in the neutral $B$ meson system,
\href{https://doi.org/10.1103/PhysRevLett.87.091802}{Phys. Rev. Lett. \textbf{87}, 091802 (2001)},
[\href{https://arxiv.org/abs/hep-ex/0107061}{\tt arXiv:hep-ex/0107061}].

\bibitem{LHCb:2019hro}
R.~Aaij \textit{et al.} (LHCb Collaboration),
Observation of $CP$ violation in charm decays,
\href{https://doi.org/10.1103/PhysRevLett.122.211803}{Phys. Rev. Lett. \textbf{122}, no.21, 211803 (2019)},
[\href{https://arxiv.org/abs/1903.08726}{\tt arXiv:1903.08726 [hep-ex]}].

\bibitem{Cabibbo:1963yz}
N.~Cabibbo,
Unitary symmetry and leptonic decays,
\href{https://doi.org/10.1103/PhysRevLett.10.531}{Phys. Rev. Lett. \textbf{10}, 531 (1963)}.

\bibitem{Kobayashi:1973fv}
M.~Kobayashi and T.~Maskawa,
$CP$ violation in the renormalizable theory of weak interaction,
\href{https://doi.org/10.1143/PTP.49.652}{Prog. Theor. Phys. \textbf{49}, 652 (1973)}.

\bibitem{Morrissey:2012db}
D.~E.~Morrissey and M.~J.~Ramsey-Musolf,
Electroweak baryogenesis,
\href{https://doi.org/10.1088/1367-2630/14/12/125003}{New J. Phys. \textbf{14}, 125003 (2012)},
[\href{https://arxiv.org/abs/1206.2942}{\tt arXiv:1206.2942 [hep-ph]}].

\bibitem{tHooft:1976rip}
G.~'t Hooft,
Symmetry breaking through Bell-Jackiw anomalies,
\href{https://doi.org/10.1103/PhysRevLett.37.8}{Phys. Rev. Lett. \textbf{37}, 8 (1976)}.

\bibitem{tHooft:1976snw}
G.~'t Hooft,
Computation of the quantum effects due to a four-dimensional pseudoparticle,
\href{https://doi.org/10.1103/PhysRevD.14.3432}{Phys. Rev. D \textbf{14}, 3432 (1976)};
\href{https://doi.org/10.1103/PhysRevD.18.2199.3}{Phys. Rev. D \textbf{18}, 2199(E) (1978)}.

\bibitem{Jackiw:1976pf}
R.~Jackiw and C.~Rebbi,
Vacuum periodicity in a Yang-Mills quantum theory,
\href{https://doi.org/10.1103/PhysRevLett.37.172}{Phys. Rev. Lett. \textbf{37}, 172 (1976)}.

\bibitem{Callan:1976je}
C.~G.~Callan, Jr., R.~F.~Dashen and D.~J.~Gross,
The structure of the gauge theory vacuum,
\href{https://doi.org/10.1016/0370-2693(76)90277-X}{Phys. Lett. B \textbf{63}, 334 (1976)}.

\bibitem{ParticleDataGroup:2024cfk}
S.~Navas \textit{et al.} (Particle Data Group),
Review of particle physics,
\href{https://doi.org/10.1103/PhysRevD.110.030001}{Phys. Rev. D \textbf{110}, no.3, 030001 (2024)},

\bibitem{Okubo:1958zza}
S.~Okubo,
Decay of the $\Sigma^+$ hyperon and its antiparticle,
\href{https://doi.org/10.1103/PhysRev.109.984}{Phys. Rev. \textbf{109}, 984 (1958)}.

\bibitem{Lee:1957qs}
T.~D.~Lee and C.~N.~Yang,
General partial wave analysis of the decay of a hyperon of spin 1/2,
\href{https://doi.org/10.1103/PhysRev.108.1645}{Phys. Rev. \textbf{108}, 1645 (1957)}.

\bibitem{Pais:1959zza}
A.~Pais,
Notes on antibaryon interactions,
\href{https://doi.org/10.1103/PhysRevLett.3.242}{Phys. Rev. Lett. \textbf{3}, 242 (1959)}.

\bibitem{Brown:1983wd}
T.~Brown, S.~F.~Tuan and S.~Pakvasa,
$CP$ nonconservation in hyperon decays,
\href{https://doi.org/10.1103/PhysRevLett.51.1823}{Phys. Rev. Lett. \textbf{51}, 1823 (1983)}.

\bibitem{Chau:1983ei}
L.~L.~Chau and H.~Y.~Cheng,
Partial rate differences from $CP$ violation in hyperon nonleptonic decays,
\href{https://doi.org/10.1016/0370-2693(83)91121-8}{Phys. Lett. B \textbf{131}, 202 (1983)}.

\bibitem{Donoghue:1985ww}
J.~F.~Donoghue and S.~Pakvasa,
Signals of $CP$ nonconservation in hyperon decay,
\href{https://doi.org/10.1103/PhysRevLett.55.162}{Phys. Rev. Lett. \textbf{55}, 162 (1985)}.

\bibitem{Donoghue:1986hh}
J.~F.~Donoghue, X.~G.~He and S.~Pakvasa,
Hyperon decays and $CP$ nonconservation,
\href{https://doi.org/10.1103/PhysRevD.34.833}{Phys. Rev. D \textbf{34}, 833 (1986)}.

\bibitem{LHCb:2024yzj}
R.~Aaij \textit{et al.} (LHCb Collaboration),
Study of $\Lambda_b^0$ and $\Xi_b^0$ decays to $\Lambda h^+h^{'-}$ and evidence for $CP$ violation in $\Lambda_b^0\rightarrow\Lambda K^+K^-$ decays,
\href{https://doi.org/10.1103/PhysRevLett.134.101802}{Phys. Rev. Lett. \textbf{134}, no.10, 101802 (2025)},
[\href{https://arxiv.org/abs/2411.15441}{\tt arXiv:2411.15441 [hep-ex]}].

\bibitem{LHCb:2025ray}
R.~Aaij \textit{et al.} (LHCb Collaboration),
Observation of charge-parity symmetry breaking in baryon decays,
\href{https://doi.org/10.1038/s41586-025-09119-3}{Nature \textbf{643}, 1223 (2025)},
[\href{https://arxiv.org/abs/2503.16954}{\tt arXiv:2503.16954 [hep-ex]}].

\bibitem{Kopke:1988cs}
L.~Kopke and N.~Wermes,
J/psi decays,
\href{https://doi.org/10.1016/0370-1573(89)90074-4}{Phys. Rept. \textbf{174}, 67 (1989)}.

\bibitem{Donohue:1969fu}
J.~T.~Donohue,
Spin and parity analysis for bosons decaying into spin 1/2 pairs,
\href{https://doi.org/10.1103/PhysRev.178.2288}{Phys. Rev. \textbf{178}, 2288 (1969)}.

\bibitem{Chen:2007zzf}
H.~Chen and R.~G.~Ping,
Helicity amplitude analysis of hyperon nonleptonic decays in $J/\psi$ or $\psi(2S)$ decays,
\href{https://doi.org/10.1103/PhysRevD.76.036005}{Phys. Rev. D \textbf{76}, 036005 (2007)}.

\bibitem{Perotti:2018wxm}
E.~Perotti, G.~F{\"a}ldt, A.~Kupsc, S.~Leupold and J.~J.~Song,
Polarization observables in $e^+e^-$ annihilation to a baryon-antibaryon pair,
\href{https://doi.org/10.1103/PhysRevD.99.056008}{Phys. Rev. D \textbf{99}, no.5, 056008 (2019)},
[\href{https://arxiv.org/abs/1809.04038}{\tt arXiv:1809.04038 [hep-ph]}].

\bibitem{Hong:2023soc}
P.~C.~Hong, R.~G.~Ping, T.~Luo, X.~R.~Zhou and H.~Li,
Phenomenological study of $J/\psi\to\Xi^0(\Lambda \pi^0) \bar{\Xi}^0(\bar{\Lambda} \gamma)$ decays*,
\href{https://doi.org/10.1088/1674-1137/ace354}{Chin. Phys. C \textbf{47}, no.9, 093103 (2023)},
[\href{https://arxiv.org/abs/2306.08517}{\tt arXiv:2306.08517 [hep-ph]}].

\bibitem{Zhang:2023wmd}
Z.~Zhang and J.~J.~Song,
Spin density matrix for $\Omega^-$ and its polarization alignment in $\psi(3686)\rightarrow\Omega^-\bar{\Omega}^+$,
\href{https://doi.org/10.1088/1674-1137/ace17e}{Chin. Phys. C \textbf{47}, no.9, 093101 (2023)},
[\href{https://arxiv.org/abs/2303.02629}{\tt arXiv:2303.02629 [hep-ex]}].

\bibitem{Zhang:2023box}
Z.~Zhang, J.~J.~Song and Y.~j.~Zhou,
Refined analysis of $\Omega^-\bar{\Omega}^+$ polarization in electron-positron annihilation process,
\href{https://doi.org/10.1103/PhysRevD.109.036005}{Phys. Rev. D \textbf{109}, no.3, 036005 (2024)},
[\href{https://arxiv.org/abs/2312.04363}{\tt arXiv:2312.04363 [hep-ph]}].

\bibitem{Zhang:2024rbl}
Z.~Zhang, R.~G.~Ping, T.~Liu, J.~J.~Song, W.~Yang and Y.~j.~Zhou,
Polarization analysis of two baryons with various spin combinations produced in electron-positron annihilation,
\href{https://doi.org/10.1103/PhysRevD.110.034034}{Phys. Rev. D \textbf{110}, no.3, 034034 (2024)},
[\href{https://arxiv.org/abs/2404.04787}{\tt arXiv:2404.04787 [hep-ph]}].

\bibitem{Dubnickova:1992ii}
A.~Z.~Dubnickova, S.~Dubnicka and M.~P.~Rekalo,
Investigation of the nucleon electromagnetic structure by polarization effects in $e^+ e^- \rightarrow B \bar{B}$ processes,
\href{https://doi.org/10.1007/BF02731012}{Nuovo Cim. A \textbf{109}, 241 (1996)}.

\bibitem{Tomasi-Gustafsson:2005svz}
E.~Tomasi-Gustafsson, F.~Lacroix, C.~Duterte and G.~I.~Gakh,
Nucleon electromagnetic form-factors and polarization observables in space-like and time-like regions,
\href{https://doi.org/10.1140/epja/i2005-10030-6}{Eur. Phys. J. A \textbf{24}, 419 (2005)},
[\href{https://arxiv.org/abs/nucl-th/0503001}{\tt arXiv:nucl-th/0503001}].

\bibitem{Czyz:2007wi}
H.~Czyz, A.~Grzelinska and J.~H.~Kuhn,
Spin asymmetries and correlations in $\Lambda$-pair production through the radiative return method,
\href{https://doi.org/10.1103/PhysRevD.75.074026}{Phys. Rev. D \textbf{75}, 074026 (2007)},
[\href{https://arxiv.org/abs/hep-ph/0702122}{\tt arXiv:hep-ph/0702122}].

\bibitem{Faldt:2017kgy}
G.~F{\"a}ldt and A.~Kupsc,
Hadronic structure functions in the $e^+ e^- \rightarrow \bar{\Lambda} \Lambda$ reaction,
\href{https://doi.org/10.1016/j.physletb.2017.06.011}{Phys. Lett. B \textbf{772}, 16 (2017)},
[\href{https://arxiv.org/abs/1702.07288}{\tt arXiv:1702.07288 [hep-ph]}].

\bibitem{BESIII:2018cnd}
M.~Ablikim \textit{et al.} (BESIII Collaboration),
Polarization and entanglement in baryon-antibaryon pair production in electron-positron annihilation,
\href{https://doi.org/10.1038/s41567-019-0494-8}{Nature Phys. \textbf{15}, 631 (2019)},
[\href{https://arxiv.org/abs/1808.08917}{\tt arXiv:1808.08917 [hep-ex]}].

\bibitem{BESIII:2022qax}
M.~Ablikim \textit{et al.} (BESIII Collaboration),
Precise measurements of decay parameters and $CP$ asymmetry with entangled $\Lambda-\bar{\Lambda}$ Pairs Pairs,
\href{https://doi.org/10.1103/PhysRevLett.129.131801}{Phys. Rev. Lett. \textbf{129}, no.13, 131801 (2022)},
[\href{https://arxiv.org/abs/2204.11058}{\tt arXiv:2204.11058 [hep-ex]}].

\bibitem{BESIII:2020fqg}
M.~Ablikim \textit{et al.} (BESIII Collaboration),
$\Sigma^{+}$ and $\bar{\Sigma}^-$ polarization in the $J/\psi$ and $\psi(3686)$ decays,
\href{https://doi.org/10.1103/PhysRevLett.125.052004}{Phys. Rev. Lett. \textbf{125}, no.5, 052004 (2020)},
[\href{https://arxiv.org/abs/2004.07701}{\tt arXiv:2004.07701 [hep-ex]}].

\bibitem{BESIII:2025fre}
M.~Ablikim \textit{et al.} (BESIII Collaboration),
Stringent test of $CP$ symmetry in $\Sigma^+$ hyperon decays,
[\href{https://arxiv.org/abs/2503.17165}{\tt arXiv:2503.17165 [hep-ex]}].

\bibitem{BESIII:2021ypr}
M.~Ablikim \textit{et al.} (BESIII Collaboration),
Probing $CP$ symmetry and weak phases with entangled double-strange baryons,
\href{https://doi.org/10.1038/s41586-022-04624-1}{Nature \textbf{606}, no.7912, 64 (2022)},
[\href{https://arxiv.org/abs/2105.11155}{\tt arXiv:2105.11155 [hep-ex]}].

\bibitem{BESIII:2023drj}
M.~Ablikim \textit{et al.} (BESIII Collaboration),
Tests of $CP$ symmetry in entangled $\Xi^0-\bar{\Xi}^0$ pairs,
\href{https://doi.org/10.1103/PhysRevD.108.L031106}{Phys. Rev. D \textbf{108}, no.3, L031106 (2023)},
[\href{https://arxiv.org/abs/2305.09218}{\tt arXiv:2305.09218 [hep-ex]}].

\bibitem{BESIII:2023jhj}
M.~Ablikim \textit{et al.} (BESIII Collaboration),
Investigating the $\Delta I = 1/2$ rule and CP violation through the measurement of decay asymmetry parameters in $\Xi^-$ decays, 
\href{https://doi.org/10.1103/PhysRevLett.132.101801}{Phys. Rev. Lett. \textbf{132}, no.10, 101801 (2024)},
[\href{https://arxiv.org/abs/2309.14667}{\tt arXiv:2309.14667 [hep-ex]}].

\bibitem{BESIII:2023cvk}
M.~Ablikim \textit{et al.} (BESIII Collaboration),
Extracting the femtometer structure of strange baryons using the vacuum polarization effect,
\href{https://doi.org/10.1038/s41467-024-51802-y}{Nature Commun. \textbf{15}, no.1, 8812 (2024)},
[\href{https://arxiv.org/abs/2309.04139 }{\tt arXiv:2309.04139 [hep-ex]}].

\bibitem{BESIII:2022rgl}
M.~Ablikim \textit{et al.} (BESIII Collaboration),
Measurement of the absolute branching fraction and decay asymmetry of $\Lambda\rightarrow n\gamma$,
\href{https://doi.org/10.1103/PhysRevLett.129.212002}{Phys. Rev. Lett. \textbf{129}, no.21, 212002 (2022)},
[\href{https://arxiv.org/abs/2206.10791}{\tt arXiv:2206.10791 [hep-ex]}].

\bibitem{BESIII:2023fhs}
M.~Ablikim \textit{et al.} (BESIII Collaboration),
Precision measurement of the decay $\Sigma^+\rightarrow p \gamma$ in the process $J/\psi\rightarrow\Sigma^+\Sigma^-$,
\href{https://doi.org/10.1103/PhysRevLett.130.211901}{Phys. Rev. Lett. \textbf{130}, no.21, 211901 (2023)},
[\href{https://arxiv.org/abs/2302.13568}{\tt arXiv:2302.13568 [hep-ex]}].

\bibitem{BESIII:2024nif}
M.~Ablikim \textit{et al.} (BESIII Collaboration),
Strong and weak $CP$ tests in sequential decays of polarized $\Sigma^0$ hyperons,
\href{https://doi.org/10.1103/PhysRevLett.133.101902}{Phys. Rev. Lett. \textbf{133}, no.10, 101902 (2024)},
[\href{https://arxiv.org/abs/2406.06118}{\tt arXiv:2406.06118 [hep-ex]}].

\bibitem{BESIII:2024lio}
M.~Ablikim \textit{et al.} (BESIII Collaboration),
Measurement of the decay $\Xi^0\rightarrow\Lambda\gamma$ with entangled $\Xi^0\Xi^0$ pairs,
\href{https://doi.org/10.1016/j.scib.2024.12.019}{Sci. Bull. \textbf{70}, 454 (2025)},
[\href{https://arxiv.org/abs/2408.16654}{\tt arXiv:2408.16654 [hep-ex]}].

\bibitem{Achasov:2023gey}
M.~Achasov,\textit{et al.}
STCF conceptual design report (Volume 1): Physics \& detector,
\href{https://doi.org/10.1007/s11467-023-1333-z}{Front. Phys. (Beijing) \textbf{19}, no.1, 14701 (2024)},
[\href{https://arxiv.org/abs/2303.15790}{\tt arXiv:2303.15790 [hep-ex]}].

\bibitem{He:1992ng}
X.~G.~He, J.~P.~Ma and B.~McKellar,
$CP$ violation in $J / \psi \rightarrow \Lambda \bar{\Lambda}$,
\href{https://doi.org/10.1103/PhysRevD.47.R1744}{Phys. Rev. D \textbf{47}, R1744 (1993)},
[\href{https://arxiv.org/abs/hep-ph/9211276}{\tt arXiv:hep-ph/9211276}].

\bibitem{He:2022jjc}
X.~G.~He and J.~P.~Ma,
Testing of P and $CP$ symmetries with $e^+e^-\rightarrow J/\psi\rightarrow\Lambda\bar{\Lambda}$,
\href{https://doi.org/10.1016/j.physletb.2023.137834}{Phys. Lett. B \textbf{839}, 137834 (2023)},
[\href{https://arxiv.org/abs/2212.08243}{\tt arXiv:2212.08243 [hep-ph]}].

\bibitem{Du:2024jfc}
Y.~Du, X.~G.~He, J.~P.~Ma and X.~Y.~Du,
Fundamental tests of P and $CP$ symmetries using octet baryons at the $J/\psi$ threshold,
\href{https://doi.org/10.1103/PhysRevD.110.076019}{Phys. Rev. D \textbf{110}, no.7, 076019 (2024)},
[\href{https://arxiv.org/abs/2405.09625}{\tt arXiv:2405.09625 [hep-ph]}].

\bibitem{Bondar:2019zgm}
A.~Bondar, A.~Grabovsky, A.~Reznichenko, A.~Rudenko and V.~Vorobyev,
Measurement of the weak mixing angle at a Super Charm-Tau factory with data-driven monitoring of the average electron beam polarization,
\href{https://doi.org/10.1007/JHEP03(2020)076}{JHEP \textbf{03}, 076 (2020)},
[\href{https://arxiv.org/abs/1912.09760}{\tt arXiv:1912.09760 [hep-ph]}].

\bibitem{Salone:2022lpt}
N.~Salone, P.~Adlarson, V.~Batozskaya, A.~Kupsc, S.~Leupold and J.~Tandean,
Study of $CP$ violation in hyperon decays at super-charm-tau factories with a polarized electron beam,
\href{https://doi.org/10.1103/PhysRevD.105.116022}{Phys. Rev. D \textbf{105}, no.11, 116022 (2022)},
[\href{https://arxiv.org/abs/2203.03035}{\tt arXiv:2203.03035 [hep-ph]}].

\bibitem{Zeng:2023wqw}
S.~Zeng, Y.~Xu, X.~R.~Zhou, J.~J.~Qin and B.~Zheng,
Prospects of $CP$ violation in $\Lambda$ decay with a polarized electron beam at the STCF*,
\href{https://doi.org/10.1088/1674-1137/ace9c5}{Chin. Phys. C \textbf{47}, no.11, 113001 (2023)},
[\href{https://arxiv.org/abs/2306.15602}{\tt arXiv:2306.15602 [hep-ex]}].

\bibitem{Fu:2023ose}
J.~Fu, H.~B.~Li, J.~P.~Wang, F.~S.~Yu and J.~Zhang,
Probing hyperon electric dipole moments with a full angular analysis,
\href{https://doi.org/10.1103/PhysRevD.108.L091301}{Phys. Rev. D \textbf{108}, no.9, 9 (2023)},
[\href{https://arxiv.org/abs/2307.04364}{\tt arXiv:2307.04364 [hep-ex]}].

\bibitem{Cao:2024tvz}
X.~Cao, Y.~T.~Liang and R.~G.~Ping,
Production and decay of hyperons in a transversely polarized electron-positron collider,
\href{https://doi.org/10.1103/PhysRevD.110.014035}{Phys. Rev. D \textbf{110}, no.1, 014035 (2024)},
[\href{https://arxiv.org/abs/2404.00298}{\tt arXiv:2404.00298 [hep-ph]}].

\bibitem{Guo:2025bfn}
M.~Guo, Z.~Zhang, R.~Ping and J.~Jiao,
Search for $CP$ violations in the production and decay of the hyperon-antihyperon pairs,
[\href{https://arxiv.org/abs/2502.20274}{\tt arXiv:2502.20274 [hep-ph]}].

\bibitem{Jacob:1959at}
M.~Jacob and G.~C.~Wick,
On the general theory of collisions for particles with spin,
\href{https://doi.org/10.1006/aphy.2000.6022}{Annals Phys. \textbf{7}, 404 (1959)}.

\bibitem{Sokolov:1963zn}
A.~A.~Sokolov and I.~M.~Ternov,
On polarization and spin effects in the theory of synchrotron radiation,
Sov. Phys. Dokl. \textbf{8}, 1203 (1963).

\bibitem{BESIII:2009fln}
M.~Ablikim \textit{et al.} (BESIII Collaboration),
Design and construction of the BESIII detector,
\href{https://doi.org/10.1016/j.nima.2009.12.050}{Nucl. Instrum. Meth. A \textbf{614}, 345 (2010)}.
[\href{https://arxiv.org/abs/0911.4960}{\tt arXiv:0911.4960 [physics.ins-det]}].

\bibitem{Schwinger:1948iu}
J.~S.~Schwinger,
On quantum electrodynamics and the magnetic moment of the electron,
\href{https://doi.org/10.1103/PhysRev.73.416}{Phys. Rev. \textbf{73}, 416 (1948)}.


\bibitem{Peskin:1995ev}
M.~E.~Peskin and D.~V.~Schroeder,
\textit{An Introduction to Quantum Field Theory} (Addison-Wesley, Reading, USA, 1995).

\bibitem{Korner:1976hv}
J.~G.~Korner and M.~Kuroda,
$e^+ e^-$ annihilation into baryon-antibaryon pairs,
\href{https://doi.org/10.1103/PhysRevD.16.2165}{Phys. Rev. D \textbf{16}, 2165 (1977)}.


\bibitem{Chung:1971ri}
S.~U.~Chung,
Spin formalism,
\href{http://dx.doi.org/10.5170/CERN-1971-008}{CERN Report No. CERN-1971-008}.

\bibitem{Kim:1992az}
J.~Kim, J.~Lee, J.~S.~Shim and H.~S.~Song,
Polarization effects in spin 3/2 hyperon decay,
\href{https://doi.org/10.1103/PhysRevD.46.1060}{Phys. Rev. D \textbf{46}, 1060 (1992)}.

\bibitem{Byers:1963zz}
N.~Byers and S.~Fenster,
Determination of spin and decay parameters of fermion states,
\href{https://doi.org/10.1103/PhysRevLett.11.52}{Phys. Rev. Lett. \textbf{11}, 52 (1963)}.

\bibitem{Button-Shafer:1965}
J.~Button-Shafer,
Fermion decay into spin-$\frac{3}{2}$ fermion plus spin-0 boson,
\href{https://doi.org/10.1103/PhysRev.139.B607}{Phys. Rev. \textbf{139}, B607 (1965)}.

\bibitem{Lednicky:1975ry}
R.~Lednicky,
About spin parity of a $p\omega$-enhancement near 1800-MeV,
\href{https://doi.org/10.1016/0370-2693(75)90736-4}{Phys. Lett. B \textbf{58}, 89 (1975)}.

\bibitem{Lednicky:1985zx}
R.~Lednicky,
On evaluation of polarization of the charmed baryon $\Lambda_c^+$,
Sov. J. Nucl. Phys. \textbf{43}, 817 (1986).

\bibitem{Ademollo:1964}
M.~Ademollo, and R.~Gatto,
Complete spin tests for fermions,
\href{https://doi.org/10.1103/PhysRev.133.B531}{Phys. Rev. \textbf{133}, B531 (1964)}.

\bibitem{Berman:1965rc}
S.~M.~Berman, and M.~Jacob,
Spin and parity analysis in two-step decay processes,
\href{https://doi.org/10.2172/4581914}{SLAC Report No. 43 (1965)}.

\bibitem{Xia:2019fjf}
X.~L.~Xia, H.~Li, X.~G.~Huang and H.~Z.~Huang,
Feed-down effect on $\Lambda$ spin polarization,
\href{https://doi.org/10.1103/PhysRevC.100.014913}{Phys. Rev. C \textbf{100}, no.1, 014913 (2019)},
[\href{https://arxiv.org/abs/1905.03120}{\tt arXiv:1905.03120 [nucl-th]}].

\bibitem{Shi:2025xkp}
R.~X.~Shi, Z.~Jia, L.~S.~Geng, H.~Peng, Q.~Zhao and X.~Zhou,
Status and prospect of weak radiative hyperon decays,
\href{https://doi.org/10.1088/0256-307X/42/3/032401}{Chin. Phys. Lett. \textbf{42}, no.3, 032401 (2025)},
[\href{https://arxiv.org/abs/2502.15473}{\tt arXiv:2502.15473 [hep-ph]}].

\bibitem{Fisher:1922saa}
R.~A.~Fisher,
On the mathematical foundations of theoretical statistics,
\href{https://doi.org/10.1098/rsta.1922.0009}{Phil. Trans. Roy. Soc. Lond. A \textbf{222}, 309 (1922)},

\bibitem{Bacchetta:2000jk}
A.~Bacchetta and P.~J.~Mulders,
Deep inelastic leptoproduction of spin-one hadrons,
\href{https://doi.org/10.1103/PhysRevD.62.114004}{Phys. Rev. D \textbf{62}, 114004 (2000)},
[\href{https://arxiv.org/abs/hep-ph/0007120}{\tt arXiv:hep-ph/0007120}].

\bibitem{Dirac:1928hu}
P.~A.~M.~Dirac,
The quantum theory of the electron,
\href{https://doi.org/10.1098/rspa.1928.0023}{Proc. Roy. Soc. Lond. A \textbf{117}, 610 (1928)}.

\bibitem{Proca:1936fbw}
A.~Proca,
Sur la theorie ondulatoire des electrons positifs et negatifs,
\href{https://doi.org/10.1051/jphysrad:0193600708034700}{J. Phys. Radium \textbf{7}, 347 (1936)}.
\end{thebibliography}

\end{document}